\documentclass[11pt,a4paper]{article}
\DeclareMathAlphabet{\scr}{U}{rsfs}{m}{n}

\usepackage{latexsym}
\usepackage{epsfig}
\usepackage[mathscr]{eucal}
\usepackage{amsfonts}
\usepackage{amscd}
\usepackage{cite}
\usepackage{array}   
\usepackage{amssymb}
\usepackage{bbold}
\usepackage{colordvi}
\usepackage[centertags]{amsmath}
\usepackage{enumerate}
\usepackage{graphicx}
\usepackage{booktabs}
\usepackage{theorem}
\usepackage{multirow}
\usepackage[footnotesize]{caption}
\usepackage{bbm}
\usepackage[utf8x]{inputenc}
\usepackage[labelfont=bf,textfont={sl,bf}]{subfig}
\usepackage{soul}
 \usepackage{slashed}
\usepackage[obeyFinal]{todonotes}
\usepackage{color}
\usepackage{ulem}

\usepackage{mathtools}

\DeclarePairedDelimiter\abs{\lvert}{\rvert}%
\newcommand{\newc}{\newcommand}
\newc{\be}{\begin{equation}}
\newc{\ee}{\end{equation}}
\newc{\bea}{\begin{eqnarray}}
\newc{\eea}{\end{eqnarray}}
\newc{\ol}{\overline}
\newc{\wt}{\widetilde}
\newc{\bs}{\boldsymbol}
\newc{\m}{\mathcal}
\newc{\lan}{\langle}
\newc{\ra}{\rangle}
\newc{\pa}{\partial}

\newcommand{\sbeta}{{s_{\beta}}}
\newcommand{\cbeta}{{c_{\beta}}}
\newcommand{\tbeta}{{\tan\beta}}

\newcommand{\non}{\nonumber} 
\newcommand{\crn}{\nonumber \\}
\newcommand{\beq}{\begin{eqnarray}} 
\newcommand{\eeq}{\end{eqnarray}} 
\newcommand{\bpmatrix}{\begin{pmatrix}}
\newcommand{\epmatrix}{\end{pmatrix}}
\newcommand{\ba}{\begin{array}}
\newcommand{\ea}{\end{array}}
\newcommand{\braket}[1]{\left(#1\right)}
\newcommand{\sbraket}[1]{\left[#1\right]}

\newcommand{\fr}{\frac}

\newcommand{\diag}{\text{diag}}
\newcommand{\al}{\alpha}

\newcommand{\calM}{{\cal M}}

\newcommand{\calO}{{\cal O}}
\newcommand{\MH}{\mathcal{M}_{hh}}

\newcommand{\drbar}{\overline{\text{DR}}}

\newcommand\Tstrut{\rule{0pt}{2.6ex}}       
\newcommand\Bstrut{\rule[-0.9ex]{0pt}{0pt}} 
\newcommand{\TBstrut}{\Tstrut\Bstrut} 

\newcommand{\mueff}{\mu_{\text{eff}}}

\renewcommand{\eqref}[1]{Eq.~(\ref{#1})}

\newcommand{\sect}[1]{Section~\ref{#1}}

\newcommand{\DRb}{\overline{\text{DR}}}

\newcommand{\OS}{\text{OS}}
\newcommand{\MSb}{\overline{\text{MS}}}
\newcommand{\Retilde}{\widetilde{\text{Re}}}
\renewcommand{\Re}{\text{Re}\,}
\renewcommand{\Im}{\text{Im}\,}

\newcommand{\ie}{{\it i.e.\;}}

\newcommand{\bc}{\begin{center}}
\newcommand{\ec}{\end{center}}

\newcommand{\gev}{~\text{GeV}}

\newcommand{\ti}{\tilde}

\newcommand{\calR}{{\cal R}}

\newcommand{\de}{\delta}

\newcommand{\order}{\mathcal{O}(\alpha_t^2)}
\renewcommand{\ol}{\text{1l}}

\newcommand{\tol}{{\text{\tiny(}\!\text{\tiny1}\!\text{\tiny)}\!}}
\newcommand{\ttl}{{\text{\tiny(}\!\text{\tiny2}\!\text{\tiny)}\!}}

\newcommand{\deltatwo}{\delta^{ \text{\tiny(}\!\text{\tiny2}\!\text{\tiny)}}\!}
\newcommand{\deltaone}{\delta^{ \text{\tiny(}\!\text{\tiny1}\!\text{\tiny)}}\!}

\newcommand{\Deltatwo}{\Delta^{ \text{\tiny(}\!\text{\tiny2}\!\text{\tiny)}}\!}
\newcommand{\Deltaone}{\Delta^{ \text{\tiny(}\!\text{\tiny1}\!\text{\tiny)}}\!}

\newcommand{\dol}{\delta^{ \text{\tiny(}\!\text{\tiny1}\!\text{\tiny)}}\!}
\newcommand{\wave}{\dol\mathcal{Z}}
\newcommand{\wavetwo}{\deltatwo\mathcal{Z}}

\newcommand{\abslambda}{|\lambda |}
\newcommand{\abskappa}{|\kappa |}
\newcommand{\mhpm}{{M_{H^\pm}^2}}
\newcommand{\msq}{m_{\tilde{Q}_3}^2}
\newcommand{\thd}{t_{h_d}}
\newcommand{\thu}{t_{h_u}}
\newcommand{\ths}{t_{h_s}}
\newcommand{\tad}{t_{a_d}}

\newcommand{\tas}{t_{a_s}}
\newcommand{\sphiy}{s_{\varphi_y}}
\newcommand{\cphiy}{c_{\varphi_y}}
\newcommand{\phiy}{{\varphi_y}}

\newcommand{\vs}{v_{s}}
\newcommand{\tanks}{t_{\varphi_\omega}}
\newcommand{\ReAkappa}{\Re A_{\kappa}}
\newcommand{\ReAlambda}{\Re A_{\lambda}}
\newcommand{\cosks}{c_{\varphi_\omega}}
\newcommand{\sinks}{s_{\varphi_\omega}}
\newcommand{\mathcalM}{\mathcal{M}}
\newcommand{\mathcalR}{\mathcal{R}}
\newcommand{\CDN}{c_{\beta -\beta_n}}
\newcommand{\SBN}{s_{\beta_n}}
\newcommand{\CBN}{c_{\beta_n}}

\newcommand{\unit}{\mathbbm{1}}

\newcommand{\ImAkappa}{\Im A_{\kappa}}
\newcommand{\ImAlambda}{\Im A_{\lambda}}
\newcommand{\phiom}{{\varphi_\omega}}

\newcommand{\expeta}{e^{i\varphi_u }}
\newcommand{\expetam}{e^{-i\varphi_u }}

\newcommand{\gsim}{\raisebox{-0.13cm}{~\shortstack{$>$ \\[-0.07cm]
      $\sim$}}~}

\newcommand{\s}{\newline \vspace*{-3.5mm}}

\begin{document}
\title{
\vspace*{-3cm}
\phantom{h} \hfill\mbox{\small CP3-Origins-2019-9 DNRF90}
\\[-1.1cm]
\phantom{h} \hfill\mbox{\small HU-EP-18/41} 
\\[-1.1cm]
\phantom{h} \hfill\mbox{\small IFIRSE-TH-2019-1} 
\\[-1.1cm]
\phantom{h} \hfill\mbox{\small KA-TP-03-2019}
\\[-1.1cm]
\phantom{h} \hfill\mbox{\small P3H-19-005} 
\\[1cm]
\textbf{Two-Loop $\order$ Corrections to the Neutral Higgs Boson
  Masses in the CP-Violating NMSSM }}

\date{}
\author{
Thi Nhung Dao$^{1}$\footnote{E-mail: \texttt{dtnhung@ifirse.icise.vn}},
Ramona Gr\"ober$^{2\,}$\footnote{E-mail:
  \texttt{ramona.groeber@physik.hu-berlin.de}}, Marcel Krause$^{3\,}$\footnote{E-mail:
  \texttt{marcel.krause@kit.edu}},
Margarete M\"{u}hlleitner$^{3\,}$\footnote{E-mail:
  \texttt{margarete.muehlleitner@kit.edu}}, 
Heidi Rzehak$^{4}$\footnote{E-mail: \texttt{rzehak@cp3.sdu.dk}}
\\[9mm]
{\small\it
$^1$Institute For Interdisciplinary Research in Science and Education, ICISE,}\\
{\small\it 590000, Quy Nhon, Vietnam.}\\[3mm]
{\small\it$^2$Humboldt Universit\"at zu Berlin, Institut f\"ur Physik, }\\
{\small\it  Newtonstr.~15, 12489 Berlin, Germany.}\\[3mm]
{\small\it
$^3$Institute for Theoretical Physics, Karlsruhe Institute of Technology,} \\
{\small\it Wolfgang-Gaede-Str. 1, 76131 Karlsruhe, Germany.}\\[3mm]
{\small\it$^4$CP3-Origins, University of Southern Denmark, 
Campusvej 55, } \\
{\small\it 5230 Odense M, Denmark.}\\[3mm]
}
\maketitle

\begin{abstract}
\noindent
We present our calculation of the two-loop corrections of ${\cal O}(\alpha_t^2)$ to
the neutral Higgs boson masses of the CP-violating Next-to-Minimal
Supersymmetric extension of the Standard Model (NMSSM). The
calculation is performed in the Feynman diagrammatic approach in the
gaugeless limit at vanishing external momentum. We apply a mixed
$\DRb$-on-shell (OS) renormalization scheme for the NMSSM input
parameters. Furthermore, we exploit a $\DRb$ as well as an OS
renormalization in the top/stop sector. The corrections are
implemented in the Fortran code {\tt NMSSMCALC} for the calculation of
the Higgs spectrum both in the CP-conserving and CP-violating
NMSSM. The code also provides the Higgs boson decays including the
state-of-the-art higher-order corrections. The corrections computed in
this work improve the already available corrections in {\tt NMSSMCALC}
which are the full one-loop corrections without any approximation and
the two-loop ${\cal O}(\alpha_t \alpha_s)$ corrections in the
gaugeless limit and at vanishing external momentum. Depending on the
chosen parameter point, we find that the ${\cal O}(\alpha_t \alpha_s +
\alpha_t^2)$ corrections add about 4-7\% to the one-loop mass of the SM-like Higgs
boson for $\DRb$ renormalization in the top/stop sector and they reduce
the mass by about 6-9\% if OS renormalization is applied. 
For an estimate of the theoretical uncertainty we vary the renormalization scale and change the renormalization scheme and show that care has to be taken in the  corresponding interpretation.
\end{abstract}
\thispagestyle{empty}
\vfill
\newpage

\section{Introduction}
Supersymmetric theories \cite{Golfand:1971iw, Volkov:1973ix, Wess:1974tw, Fayet:1974pd,Fayet:1977yc, Fayet:1976cr, Nilles:1982dy,Nilles:1983ge, Frere:1983ag,Derendinger:1983bz,Haber:1984rc, Sohnius:1985qm,
Gunion:1984yn, Gunion:1986nh} belong to
the best motivated and most intensively studied extensions of the
Standard Model (SM). Supersymmetry (SUSY) between bosonic and fermionic
degrees of freedom solves the hierarchy problem and relates the Higgs
boson masses and self-couplings to the electroweak gauge
couplings. The tree-level mass value of the SM-like Higgs
boson is therefore bound to be of the order of the electroweak
scale. In the Minimal Supersymmetric extension of the SM (MSSM)
\cite{Gunion:1989we,Martin:1997ns,Dawson:1997tz,Djouadi:2005gj}
it is less than or equal to the $Z$-boson mass. This
upper bound is lifted to higher values after the inclusion of the
radiative corrections to the Higgs boson masses. In order to match the
measured mass value of 125~GeV of the discovered Higgs boson large
values of the stop masses and/or mixing are required. This challenges
the model from the point of view of fine-tuning. The situation is more relaxed
in the Next-to-Minimal SUSY
extension (NMSSM)
\cite{Barbieri:1982eh,Dine:1981rt,Ellis:1988er,Drees:1988fc,Ellwanger:1993xa,Ellwanger:1995ru,Ellwanger:1996gw,Elliott:1994ht,King:1995vk,Franke:1995tc,Maniatis:2009re,Ellwanger:2009dp}. The
introduction of a complex singlet superfield coupling with the strength
$\lambda$ to the two Higgs doublet superfields of the MSSM, adds new
contributions to the quartic coupling so that the tree-level mass of
the lightest CP-even MSSM-like Higgs boson is shifted to a higher
value. \s

The Higgs boson mass is a crucial input parameter in all Higgs boson
observables (see {\it e.g.}~\cite{deFlorian:2016spz}) and plays an
important role for the stability of the electroweak vacuum
\cite{Degrassi:2012ry,Buttazzo:2013uya,Bednyakov:2015sca}. A precise
measurement of the Higgs boson mass (the current experimental value is
$125.09\pm 0.24$ GeV \cite{Aad:2015zhl}) is hence indispensable. The
experimental accuracy has to be matched by 
the precision of the theory predictions in order to fully exploit the experimental
information to constrain the still viable parameter space of
beyond-the-SM (BSM) extensions and to 
distinguish between new physics models in case of
discovery.\footnote{For a discussion see {\it e.g.}~Ref.~\cite{Muhlleitner:2017dkd}
  comparing various SUSY and non-SUSY extensions of the Higgs
  sector.} In the recent years there has been a lot of activity, and
is still ongoing, in order to improve the precision of the Higgs mass predictions
within the NMSSM, both for the CP-conserving and the CP-violating
case. 
In the CP-conserving NMSSM, after the computation of the leading
one-loop contributions to the Higgs boson masses in
\cite{Ellwanger:1993hn,Elliott:1993ex,Elliott:1993uc,Elliott:1993bs,Pandita:1993tg,Ellwanger:2005fh},
the full one-loop corrections including the momentum dependence
became available in the
$\overline{\mbox{DR}}$~\cite{Degrassi:2009yq,Staub:2010ty}
as well as a mixed $\DRb$-OS scheme
\cite{Ender:2011qh,Drechsel:2016jdg}.\footnote{We note that a full
  one-loop renormalization of the NMSSM has also been
  presented in \cite{Belanger:2016tqb,Belanger:2017rgu}.} Two-loop 
corrections at the order ${\cal O} (\alpha_t \alpha_s + \alpha_b
\alpha_s)$ in the $\overline{\mbox{DR}}$ scheme have been provided in \cite{Degrassi:2009yq} in the
approximation of zero external momentum. Two-loop corrections beyond these have
been derived in \cite{Goodsell:2014pla}, again exploiting the $\overline{\mbox{DR}}$ scheme.
In the CP-violating case first results for the dominant one-loop
corrections~\cite{Ham:2001kf,Ham:2001wt,Ham:2003jf,Funakubo:2004ka,Ham:2007mt}
 have been followed by the full one-loop and logarithmically enhanced
 two-loop effects computed in the renormalization group
 approach~\cite{Cheung:2010ba}. Our group complemented these
 calculations by computing the full one-loop corrections in the Feynman diagrammatic
approach \cite{Graf:2012hh} (see also \cite{Domingo:2017rhb},
published recently). We subsequently calculated the ${\cal 
  O}{(\al_t\al_s)}$ corrections in the approximation of vanishing
external momentum in the gaugeless limit
\cite{Muhlleitner:2014vsa}.\footnote{We also provided the complete
  one-loop corrections to the trilinear Higgs self-couplings in the
  CP-conserving NMSSM \cite{Nhung:2013lpa} and extended these  in
  \cite{Muhlleitner:2015dua}  to the ${\cal O}(\alpha_t
  \alpha_s)$ corrections in the CP-violating NMSSM.} The adopted
renormalization scheme is a 
mixture between $\DRb$ and OS conditions with the possibility to
choose between the $\DRb$ or OS scheme for the renormalization of the
top/stop sector. An automatized two-loop calculation of the Higgs boson masses in
CP-violating SUSY theories, applying the $\DRb$
scheme, was provided by \cite{Goodsell:2016udb}. \s

The higher-order corrections to the NMSSM Higgs boson masses have been
implemented in publicly available tools that partly also compute the
Higgs boson decays. The program package {\tt NMSSMTools}
\cite{Ellwanger:2004xm,Ellwanger:2005dv,Ellwanger:2006rn} calculates
the masses and decay widths in the 
CP-conserving $\mathbb{Z}^3$ and can be interfaced with {\tt SOFTSUSY}
\cite{Allanach:2001kg,Allanach:2013kza}, which includes the
possibility of $\mathbb{Z}^3$ violation. Recently, a tool became
available to support the extension of {\tt NMSSMTools} to the
CP-violating case \cite{Domingo:2015qaa,Domingo:2017rhb,Domingo:2018uim}.
The spectrum generation of different SUSY models, including the NMSSM,
on the other hand, is possible through the interface of {\tt SARAH}
\cite{Staub:2010jh,Staub:2012pb,Staub:2013tta,Goodsell:2014bna,Goodsell:2014pla}
 with {\tt SPheno} \cite{Porod:2003um,Porod:2011nf}. 
In the same spirit, {\tt SARAH} has been
interfaced with the package {\tt FlexibleSUSY}
\cite{Athron:2014yba,Athron:2014wta}. All these programs include the
Higgs mass corrections up to two-loop order, where in particular the
two-loop corrections are obtained in the effective potential
approach.\footnote{With the exception of the recent {\tt
  NMSSMTools} extension to the complex NMSSM, which relies on a Feynman
diagrammatic calculation in the OS scheme \cite{Domingo:2017rhb}.} 
In {\tt FlexibleEFTHiggs} \cite{Athron:2016fuq}, an effective field theory approach
has been combined with a diagrammatic calculation to obtain the
SM-like Higgs pole mass in various models, including the NMSSM.
Our program package {\tt NMSSMCALC}
\cite{Baglio:2013iia} computes the NMSSM
Higgs boson masses and decay widths in the Feynman diagrammatic
approach both for the CP-conserving and
CP-violating case.\footnote{One-loop corrected decay widths are also
  included in the code {\tt SloopS}
  \cite{Belanger:2016tqb,Belanger:2017rgu,Boudjema:2017ozm}. In the
  {\tt SARAH} and {\tt SPheno} framework a generic
implementation of the two-body partial decays widths exists at the full
one-loop level \cite{Goodsell:2017pdq}.} It incorporates the complete
one-loop corrections at non-zero external
momentum and the two-loop ${\cal O}(\alpha_t \alpha_s)$ corrections in
the limit of vanishing external momentum in the gaugeless
approximation. The renormalization is performed in a mixed $\DRb$-OS
scheme with the possibility to choose between the $\DRb$ or OS scheme
for the renormalization of the top/stop sector.
For the CP-conserving NMSSM, a detailed comparison of the results of
the various tools for the Higgs boson masses in the $\DRb$ scheme was
performed in \cite{Staub:2015aea}. In \cite{Drechsel:2016htw}, a
comparison of the Higgs boson masses and mixing matrices in the OS
scheme up to ${\cal  O}(\alpha_t \alpha_s)$ was performed between the
two codes {\tt NMSSMCALC} and {\tt NMSSM-FeynHiggs}. \s

With this paper we take another step in improving our
predictions for the NMSSM Higgs boson masses. We compute the two-loop
corrections to the neutral NMSSM Higgs boson masses in the Feynman
diagrammatic approach at the order ${\cal
  O}(\alpha_t^2)$ at vanishing external momentum in the
gaugeless limit. Strictly speaking, we provide the contributions of ${\cal
  O}(m_t^2 \alpha_t^2)$ neglecting further terms beyond the
approximation of the gaugeless limit and vanishing external
momentum. For simplicity, we call our newly calculated
contributions ${\cal O}(\alpha_t^2)$ terms in the following. The
calculation is performed both for the 
CP-conserving and the CP-violating case. We apply a mixed $\DRb$-OS
renormalization scheme. Our corrections are included in the program
package {\tt NMSSMCALC}.\footnote{The program package can be downloaded
  from the url: https://www.itp.kit.edu/$\sim$maggie/NMSSMCALC/} \s

The paper is organized as follows. In Section~\ref{sec:nmssm} we introduce the
model and set our notation.
Section~\ref{sec:homasses} contains the detailed presentation of our
calculation including the description of the renormalization and the
checks that were performed. Section~\ref{sec:analysis} is dedicated to
the numerical presentation of our results. We conclude in
Section~\ref{sec:concl}. We include an extensive appendix that
contains the presentation of the two-loop self-energy diagrams, the
scalar one-loop integrals expanded up to ${\cal O}(\varepsilon)$, the
details on the computation of the running $\overline{\mbox{DR}}$ top quark mass
without and with the inclusion of the gauge coupling contributions,
the required NMSSM renormalization group equations for the
investigation of the scale dependence of our results, the neutral
tree-level and counterterm mass matrices as well as the charged Higgs
mass counterterms. 

\section{The CP-Violating NMSSM \label{sec:nmssm}}
In order to introduce the model and set our notation we briefly
review the Higgs sector of the CP-violating NMSSM. For more details on
the NMSSM, see the reviews \cite{Ellwanger:2009dp,Maniatis:2009re}. Since
  the two-loop diagrams appearing in our calculation also involve
  contributions from the stops and the electroweakinos, we briefly introduce them in this section as well.

\subsection{The Higgs Sector}
The framework for our computation is the CP-violating NMSSM with a
scale-invariant superpotential and a discrete $\mathbb{Z}^3$
symmetry. The Higgs potential is derived from the superpotential, the
soft SUSY breaking Lagrangian and the $D$-term contributions. The
superpotential reads, in terms of the two Higgs doublet superfields
$\hat{H}_d$ and $\hat{H}_u$, the singlet superfield $\hat{S}$, the
quark and lepton superfields and their charge conjugates, denoted by
the superscript $c$, $\hat{Q}$, $\hat{U}^c$, $\hat{D}^c$, $\hat{L}$,
$\hat{E}^c$,
\beq
W_{\text{NMSSM}} = \epsilon_{ij} [y_e \hat{H}^i_d \hat{L}^j \hat{E}^c + y_d
\hat{H}_d^i \hat{Q}^j \hat{D}^c - y_u \hat{H}_u^i \hat{Q}^j \hat{U}^c]
- \epsilon_{ij} \lambda \hat{S} \hat{H}^i_d 
\hat{H}^j_u + \frac{1}{3} \kappa \hat{S}^3 \;. 
\eeq
Here $\epsilon_{ij}$ ($i,j=1,2$) is the totally antisymmetric tensor
with $\epsilon_{12}=\epsilon^{12}=1$ and $i,j$ denoting the indices of
the fundamental $SU(2)_L$ representation. Here and in the following,
we sum over repeated indices. For simplicity, we neglect colour and
generation indices. We neglect flavour mixing so that the Yukawa
couplings $y_e, y_d$ and $y_u$ are diagonal, and therefore complex phases can be
reabsorbed by redefining the quark fields without effect on the
physical meaning \cite{Kobayashi:1973fv}. The dimensionless
NMSSM-specific parameters $\lambda$ and $\kappa$ are complex in the
CP-violating NMSSM. For the soft SUSY breaking Lagrangian we have in
terms of the scalar component fields $H_u$, $H_d$ and $S$,
\begin{align}
{\cal L}_{\text{soft},\text{ NMSSM}} =& -m_{H_d}^2 H_d^\dagger H_d - m_{H_u}^2
H_u^\dagger H_u -
m_{\tilde{Q}}^2 \tilde{Q}^\dagger \tilde{Q} - m_{\tilde{L}}^2 \tilde{L}^\dagger \tilde{L}
- m_{\tilde{u}_R}^2 \tilde{u}_R^* 
\tilde{u}_R - m_{\tilde{d}_R}^2 \tilde{d}_R^* \tilde{d}_R 
\nonumber \\\nonumber
& - m_{\tilde{e}_R}^2 \tilde{e}_R^* \tilde{e}_R - (\epsilon_{ij} [y_e A_e H_d^i
\tilde{L}^j \tilde{e}_R^* + y_d
A_d H_d^i \tilde{Q}^j \tilde{d}_R^* - y_u A_u H_u^i \tilde{Q}^j
\tilde{u}_R^*] + \mathrm{h.c.}) \\
& -\frac{1}{2}(M_1 \tilde{B}\tilde{B} + M_2
\tilde{W}_i\tilde{W}_i + M_3 \tilde{G}\tilde{G} + \mathrm{h.c.})\\ \nonumber
&- m_S^2 |S|^2 +
(\epsilon_{ij} \lambda 
A_\lambda S H_d^i H_u^j - \frac{1}{3} \kappa
A_\kappa S^3 + \mathrm{h.c.}) \;.
\label{eq:softnmssm}
\end{align}
The summation over all three quark and lepton generations is
implicit. By $\tilde{Q}$ and $\tilde{L}$ we denote the complex
scalar components of the corresponding quark and lepton superfields, so that
we have for the first generation {\it e.g.}~$\tilde{Q}= (\tilde{u}_L,
\tilde{d}_L)^T$ and $\tilde{L}=(\tilde{\nu}_L,\tilde{e}_L)^T$. Working
in the CP-violating case, the soft SUSY breaking trilinear couplings
$A_x$ ($x=\lambda,\kappa,d,u,e$) and the gaugino mass parameters $M_k$
($k=1,2,3$) of the bino, wino and gluino fields $\tilde{B}$,
$\tilde{W}_i$ ($i=1,2,3$) and $\tilde{G}$, are complex. Application
of the $R$-symmetry transformation allows to choose either $M_1$ or $M_2$ to be
real. However, we keep both $M_1$ and $M_2$ complex in {\tt NMSSMCALC}.
The soft SUSY breaking mass parameters of the scalar fields, 
$m_X^2$ ($X=S,H_d,H_u,\tilde{Q},\tilde{u}_R,\tilde{d}_R,\tilde{L},\tilde{e}_R$),
are real. The final Higgs potential at tree level reads
\beq
V_{H}  &=& (|\lambda S|^2 + m_{H_d}^2)H_d^\dagger H_d+ (|\lambda S|^2
+ m_{H_u}^2)H_u^\dagger H_u +m_S^2 |S|^2 \nonumber \\
&& + \frac{1}{8} (g_2^2+g_1^{2})(H_d^\dagger H_d-H_u^\dagger H_u )^2
+\frac{1}{2} g_2^2|H_d^\dagger H_u|^2 \label{eq:higgspotential} \\ 
&&   + |-\epsilon^{ij} \lambda  H_{d,i}  H_{u,j} + \kappa S^2 |^2+
\big[-\epsilon^{ij}\lambda A_\lambda S   H_{d,i}  H_{u,j}  +\frac{1}{3} \kappa
A_{\kappa} S^3+\mathrm{h.c.} \big] \;,
\nonumber
\eeq
where $g_1$ and $g_2$ denote the $U(1)_Y$ and $SU(2)_L$ gauge
couplings, respectively. Two more CP-violating phases, $\varphi_u$ and
$\varphi_s$, are introduced in the expansion of the two Higgs doublets
and the singlet field about their vacuum expectation values (VEVs)
$v_d$, $v_u$ and $v_s$,
\beq
H_d =
 \bpmatrix \fr{1}{\sqrt 2}(v_d + h_d +i a_d)\\ h_d^- \epmatrix,\;
H_u = e^{i\varphi_u}\bpmatrix
h_u^+ \\ \fr{1}{\sqrt 2}(v_u + h_u +i a_u)\epmatrix,\;
S= \fr{e^{i\varphi_s}}{\sqrt 2}(v_s + h_s +ia_s) .
\label{eq:Higgs_decomposition} 
\eeq
The VEV $v\approx 246$~GeV is related to $v_u$ and $v_d$ through
$v^2=v_d^2+v_u^2$, and the ratio between the two VEVs is denoted by $\tan\beta$,
\be 
\tan\beta = \fr{v_u}{v_d} \;.
\ee
The phase $\varphi_u$ enters the top quark mass term.\footnote{For simplicity we focus on the top quark mass but the discussion holds for all up-type quark mass terms.} However, we keep
the coupling $y_t \exp(i \varphi_u)$, appearing in the mass
term, real by absorbing this phase into the
left- and right-handed top quark fields through the replacements
\be  
t_L \to e^{-i\varphi_u/2}\,t_L \quad \mbox{and} \quad t_R \to
e^{i\varphi_u/2}\,t_R\,. \label{eq:topfieldrephase}
\ee
This affects all couplings involving one top quark. After inserting
Eq.~(\ref{eq:Higgs_decomposition}) into Eq.~(\ref{eq:higgspotential})
the Higgs potential can be cast into the form 
\beq
V_H & = & V_H^{\mbox{\scriptsize const}}  +  t_{h_d} h_d + t_{h_u} h_u +
t_{h_s} h_s  +  t_{a_{d}} a_d+  t_{a_{u}} a_u 
+  t_{a_{s}} a_s  \\ \non
&& + 
 \frac{1}{2} \phi^{0,T}  {\mathcal{M}_{\phi\phi}} \, \phi^0 +
 \phi^{c,\dagger} {\mathcal{M}_{h^+h^-}} \, \phi^c 
+V_H^{\phi^3,\phi^4} \;,
\eeq
with $\phi^0 \equiv (h_d, h_u, h_s, a_d, a_u, a_s)^T$ and $\phi^c \equiv
((h_d^-)^*,h_u^+)^T$. The six tadpole coefficients $t_{\phi}$
($\phi=h_d,h_u,h_s,a_d,a_u,a_s$) are defined via
\begin{equation}
	t_\phi \equiv \left. \frac{\partial V_\text{Higgs}}{\partial \phi } \right| _\text{Min.} = 0
\end{equation}
at tree-level. This definition yields the following
tree-level relations for the tadpole coefficients,
\begin{align}
\frac{t_{h_d}}{vc_\beta } &=m_{H_d}^2+\frac{c_{2\beta}
                            M_Z^2}{2}-\frac{\abslambda \tbeta \vs}{2}
                            \left(\abskappa \cphiy \vs-\sqrt{2}
                            \ImAlambda s_{\phiom-\phiy}+\sqrt{2}
                            \ReAlambda c_{\phiom-\phiy}\right)
\nonumber \\
&\hspace*{0.5cm} +\frac{1}{2} \abslambda^2 \left(\sbeta^2 v^2+\vs^2\right)\,\\
\frac{t_{h_u}}{vs_\beta }
&=m_{H_u}^2-\frac{c_{2\beta}M_Z^2}{2}-\frac{\abslambda  \vs}{2 \tbeta} \left(\abskappa \cphiy \vs-\sqrt{2} \ImAlambda s_{\phiom-\phiy}+\sqrt{2} \ReAlambda c_{\phiom-\phiy}\right) \nonumber\\
&\hspace*{0.5cm} +\frac{1}{2} \abslambda^2 \left(\cbeta^2 v^2+\vs^2\right)\\
\frac{t_{h_s}}{v_S} &= m_S^2+\abskappa^2 \vs^2+\frac{\abslambda^2
                      v^2}{2}+\abslambda \cbeta \sbeta v^2
                      \left(\frac{\ImAlambda
                      s_{\phiom-\phiy}-\ReAlambda c_{\phiom-\phiy}
                      }{\sqrt{2} \vs} -\abskappa \cphiy\right)
                      \nonumber \\
&\hspace*{0.5cm}+\frac{\abskappa \vs (\ReAkappa c_\phiom-\ImAkappa s_\phiom )}{\sqrt{2}}\\
\frac{t_{a_d}}{vs_\beta } &= 
\frac{1}{2} \abslambda \vs \left(-\abskappa \vs s_\phiy+\sqrt{2} \ImAlambda c_{\phiom-\phiy}+\sqrt{2} \ReAlambda s_{\phiom-\phiy}\right)\\
t_{a_u} &= \frac{1}{\tbeta} t_{a_d}  \\
t_{a_s} &= \frac{1}{2} \abslambda \cbeta \sbeta v^2 \left(2 \abskappa
          \vs s_\phiy+\sqrt{2} \ImAlambda c_{\phiom-\phiy}+\sqrt{2}
          \ReAlambda s_{\phiom-\phiy}\right) \nonumber \\
&\hspace*{0.5cm}-\frac{\abskappa \vs^2 (\ImAkappa c_{\phiom}+\ReAkappa
  s_\phiom)}{\sqrt{2}} \;,
\end{align}
with $M_Z$ being the $Z$ boson mass. Here and in the following we use
the short-hand notations 
$c_x\equiv\cos x$ and $s_x\equiv \sin x$, and we express the
complex parameters $A_\lambda$ and $A_\kappa$ by their imaginary and
real parts in order to comply with the SUSY Les Houches Accord (SLHA) \cite{Skands:2003cj,Allanach:2008qq} as well as $\lambda$ and $\kappa$ by their absolute values
and their phases. The phases enter in two combinations together with
$\varphi_u$ and $\varphi_s$, 
\begin{align}
\varphi_y &= \varphi_\kappa - \varphi_\lambda + 2\varphi_s - \varphi_u \label{eq:varphiy} \\
\phiom &= \varphi_\kappa + 3\varphi_s\;. \label{eq:varphiom} 
\end{align}
Since $t_{a_u}$ and $t_{a_d}$ are linearly dependent, only five out of
the six tadpole equations yield linearly independent conditions in the
Higgs sector. While the tadpole coefficients vanish at tree level due
to the minimization conditions of the Higgs potential, they affect the
higher-order corrections through the appearance of tadpole
counterterms. We therefore keep the tadpole coefficients explicitly
in all quantities that need to be renormalized and set them to zero
only after the renormalization is performed. Note that
at lowest order the two parameters $\ImAlambda$ and $\ImAkappa$ can be
eliminated by using the minimisation conditions $t_{a_d}=0$ and
$t_{a_s}=0$, and the soft-breaking mass parameters $m_{H_d}^2$,
$m_{H_u}^2$ and $m_S^2$ can be eliminated by using $t_{h_d}=0$,
$t_{h_u}=0$ and $t_{h_s}=0$. \s 

The $6\times 6$ mass matrix for the neutral Higgs bosons
is denoted by ${\mathcalM}_{\phi\phi}$ and the $2\times 2$ mass matrix
for  the charged Higgs bosons by ${\mathcalM}_{h^+h^-}$. The explicit
expressions for the mass matrix
${\mathcalM}_{\phi\phi}$ can be found in
App.\,\ref{append:Hmass}, while ${\mathcalM}_{h^+h^-}$ is given by
\begin{align}
			{\mathcalM}_{h^+h^-} &= \frac{1}{2} \begin{pmatrix}
				\tbeta && 1 \\ 1 && \frac{1}{\tbeta}
			\end{pmatrix} \left[ M_W^2 s_{2\beta}+\frac{\abslambda \vs \left(\abskappa \vs \cos(\phiom)+\sqrt{2} \ReAlambda\right)}{\cos(\phiom-\phiy)}-\frac{1}{2} \abslambda^2 s_{2\beta} v^2 \right] \nonumber \\
			&\hspace*{0.4cm} + \begin{pmatrix}
				\frac{t_{h_d} - t_{a_d}\tan(\phiom-\phiy)}{vc_\beta } && -\frac{\tad (\tan(\phiom-\phiy)+i)}{\sbeta v}\\ -\frac{\tad (\tan(\phiom-\phiy)-i)}{\sbeta v} && \frac{\sbeta \thu-\cbeta \tad \tan(\phiom-\phiy)}{\sbeta^2 v} 
			\end{pmatrix} 
\end{align}
where we explicitly keep the dependence on the tadpole parameters.
Constant terms and trilinear and quartic interactions are summarized in
$V_H^{\mbox{\scriptsize const}}$ and $V_H^{\phi^3,\phi^4}$,
respectively. The mass eigenstates $h_i$ ($i=1,...,5$) are obtained by
rotating from the interaction to the mass basis with two consecutive
rotations, where the first rotation ${\cal R}^G$ singles out the
would-be Goldstone boson, and the second one, ${\cal R}$, performs the
rotation to the mass eigenstates,
\beq
(h_d,h_u,h_s,a,a_s, G)^T &=&  \mathcalR^G~(
h_d,h_u,h_s,a_d,a_u,a_s)^T \non \\ 
(h_1,h_2,h_3,h_4,h_5, G)^T &=& \mathcalR ~(h_d,h_u,h_s,a,a_s,
G)^T\,, \label{eq:rotationtreelevel}
\eeq
with the diagonal mass matrix
\beq
\diag(m_{h_1}^2,m_{h_2}^2,m_{h_3}^2,m_{h_4}^2,m_{h_5}^2,0)&=&
\mathcalR \mathcalM_{hh} \mathcalR^T\,, \quad \mathcalM_{hh}=
\mathcalR^G\mathcalM_{\phi\phi}(\mathcalR^G)^T, 
\label{eq:massmatrix}
\eeq
and 
\be 
\mathcalR^G = \bpmatrix \unit_{3\times3} & 0\\
                             0 & \ti \mathcalR^G  \epmatrix, \quad \ti \mathcalR^G= \bpmatrix
 \SBN & \CBN &0 \\ 0& 0& 1\\ \CBN & -\SBN & 0\epmatrix \;.
\ee
The mass eigenstates $h_i$
are ordered by ascending mass, {\it i.e.}~$m_{h_1} \le ... \le m_{h_5}$.
At tree level, the rotation angle $\beta_n$ is equal to the angle
$\beta$ defining the ratio of the VEVs. However, we distinguish them
here because in the subsequently applied renormalization procedure only
$\beta$ receives a counterterm but not $\beta_n$. The charged Higgs and
Goldstone boson $H^-$ and $G^-$, respectively, are obtained through
the rotation 
\be 
\bpmatrix G^-\\ H^- \epmatrix = \mathcalR^{G^-} \bpmatrix h_d^-\\
h_u^- \epmatrix, \quad  
\diag(\mhpm, 0 ) =  \mathcalR^{G^-} \mathcalM_{h^+h^-}
(\mathcalR^{G^-})^T, 
\ee
with 
\be   
\mathcalR^{G^-} = \bpmatrix - \cos\beta_c & \sin \beta_c \\
\sin\beta_c & \cos\beta_c \epmatrix \;.
\ee
Like the mixing angle $\beta_n$, the rotation angle $\beta_c$ of the
charged sector is considered to be already renormalized and hence does
not receive a counterterm. At tree level $\beta_c = \beta_n =\beta$
holds. The relation between the charged Higgs mass and $\ReAlambda$ is
given by 
\begin{align}
\mhpm=&\frac{\abslambda c_{\beta-\beta_c}^2 \vs  \left(\abskappa \vs
        \cos (\varphi_w)+\sqrt{2} \ReAlambda\right)}{s_{2\beta}\cos
        (\phiy-\varphi_w)}-\frac{1}{2} \abslambda^2 
c_{\beta-\beta_c}^2 v^2+ c_{\beta-\beta_c}^2 M_W^2
\nonumber\displaybreak[0]\\[2mm]&
+\frac{\sbeta \left(\cbeta c_{\beta_c}^2 \thu+\sbeta s_{\beta_c}^2
                                  \thd\right)+
                                  c_{\beta-\beta_c}^2 \tad \tan (\phiy-\varphi_w)}{\cbeta \sbeta^2 v}
\;,
\displaybreak[0]\, \label{eq:charalam}
\end{align}
where $M_W$ is the $W$ boson mass. Note that we explicitly kept the
difference between $\beta$ and $\beta _c$ as well as the tadpole parameters, all of
  which would vanish at tree level, in the formula for the charged
  Higgs boson mass since its counterterm at one- and two-loop order receives contributions from the counterterms of $\beta$ and of the tadpole parameters. \s

The MSSM limit of the complex NMSSM is obtained by $\lambda, \kappa
\to 0$ {(and constant ratio of $\kappa/\lambda$ for a smooth
  approximation of the limit) and keeping $A_\lambda$, $A_\kappa$ and
  the effective $\mu_{\text{eff}}$ parameter,
\beq
\mu_{\text{eff}} = \frac{\lambda v_s e^{i\varphi_s}}{\sqrt{2}} \;, \label{eq:effectiveMuParameter}
\eeq
fixed. \s

We choose the set of independent parameters entering the Higgs potential to be
\be 
\left\{ t_{h_d},t_{h_u},t_{h_s},t_{a_d},t_{a_s},M_{H^\pm}^2,v,s_{\theta_W},
e,\tan\beta,|\lambda|,v_s,|\kappa|,\ReAkappa,\varphi_\lambda,\varphi_\kappa,\varphi_u,\varphi_s
\right\} \,, \label{eq:inputset1}
\ee
where $\theta_W$ denotes the weak mixing angle.
Alternatively, we can choose to use $\ReAlambda$ as input and
calculate $M_{H^\pm}^2$ by means of \eqref{eq:charalam}. In this case,
the set of independent parameters entering the Higgs potential is
given by 
\be 
\left\{ t_{h_d},t_{h_u},t_{h_s},t_{a_d},t_{a_s},v,s_{\theta_W},
e,\tan\beta,|\lambda|,v_s,|\kappa|,\ReAlambda,\ReAkappa,\varphi_\lambda,\varphi_\kappa,\varphi_u,\varphi_s
\right\} \,. \label{eq:inputset2}
\ee

\subsection{The Squark Sector}
The top, bottom, stop and sbottom particles appear in the $\order$
two-loop diagrams of the self-energies of the neutral and charged
Higgs bosons. The two-loop corrections to the charged Higgs boson need
to be computed for the definition of the two-loop counterterm of the
squared charged Higgs mass, since this counterterm explicitly appears
in the $\order$ two-loop corrections to the masses of the neutral
Higgs bosons. In the gaugeless approximation that we apply in our
calculation, \textit{i.e.}~$e\to 0$ (or, in other words, $g_1 \to 0$ and $g_2 \to 0$), the stop mass matrix reads  
\begin{equation}
\mathcal{M}_{\tilde t}=
\bpmatrix
 \msq+m_t^2 & m_t \left(A_t^* e^{-i\varphi_u}-\fr{\mueff}{\tan\beta}\right) \\[2mm]
m_t \left(A_t e^{i\varphi_u}- \fr{\mueff^*}{\tan\beta}\right) & m_{\tilde{t}_R}^2+m_t^2
\epmatrix\,,
\end{equation}
where $m_t$ denotes the top quark mass. 
The matrix is diagonalized by a unitary matrix
$\mathcal{U}_{\tilde t}$, rotating the interaction states $\tilde{t}_L$
and $\tilde{t}_R$ to the mass eigenstates $\tilde{t}_1$ and $\tilde{t}_2$,
\bea
  (\tilde t_1,\tilde t_2)^T &=& \mathcal{U}_{\tilde t}~(\tilde
  t_L,\tilde t_R)^T \\
 \text{diag}(m_{\tilde t_1}^2,m_{\tilde t_2}^2)&=&\mathcal{U}_{\tilde
   t}~\mathcal{M}_{\tilde t}~\mathcal{U}_{\tilde t}^\dagger\,. 
\eea
Throughout our calculation of the $\order$ two-loop corrections, we set the bottom quark mass to zero, {\it i.e.}~$m_b=0$. Therefore, the left- and right-handed
sbottom states will not mix and only the left-handed sbottom with a
mass of $m_{\tilde{Q}_3}$ will contribute. With the parameters
$\varphi_u, \tan\beta$ and those defining $\mueff$ already appearing
in the Higgs sector, we are hence left with the following set of
independent parameters for the third generation quark/squark sector
\be 
m_t, \; m_{\tilde{Q}_3}, \; m_{\tilde t_R} \quad \mbox{and} \quad A_t \;.
\label{eq:stopparset}
\ee  
With this parameter choice for the mass matrix in the interaction
basis, the rotation matrix $\mathcal{U}_{\tilde t}$ does not need to receive a counterterm.
 We follow here the same approach as in the Higgs sector,
where we assumed the  rotation matrices to be renormalized already. 

\subsection{The Chargino and Neutralino Sectors \label{sec:ewino}}
The computation of the ${\cal O} (\alpha_t^2)$ contributions to
the Higgs self-energies and tadpoles also involves the neutralino and
chargino sectors. In the gaugeless approximation, electroweak gauginos
do not mix with the fermionic superpartners of the Higgs bosons and do
not contribute to the $\order$ correction. Only the Higgsinos enter
the two-loop diagrams of the $\order$ neutral and charged Higgs
self-energies. The neutralino mass matrix in the Weyl spinor basis
$\psi^0 =  (\tilde{B}, \tilde{W}_3, \tilde{H}^0_d,\tilde{H}^0_u,  \tilde{S})^T$
simplifies to 
\beq
M_{\tilde{\chi}^0} = \begin{pmatrix} M_G & 0 \\ 0 & M_N \end{pmatrix} \;,
\eeq
with the neutral $2\times 2$ gaugino mass matrix $M_G$ given by 
\beq
M_G = \begin{pmatrix} M_1 & 0 \\ 0 & M_2 \end{pmatrix}
\eeq
whose eigenstates decouple and do not contribute to the $\order$ results in the gaugeless limit in our calculation. The neutral $3\times 3$ Higgsino mass matrix $M_N$ reads
\begin{align}
M_N =  \begin{pmatrix} 
 0  & - \mueff   & 
- \frac{v s_\beta  \lambda e^{i\varphi_u}}{\sqrt{2}}
  \\ - \mueff & 0 &
-  \frac{ v c_\beta \lambda}{\sqrt{2}}  
   \\
- \frac{v s_\beta  \lambda e^{i\varphi_u}}{\sqrt{2}} & 
                            -  \frac{ v c_\beta \lambda}{\sqrt{2}} &
                             \sqrt{2} \kappa v_s e^{i \varphi_s} 
\end{pmatrix}. \label{eq:neuMass}       
\end{align}  
The symmetric matrix $M_N$ can be diagonalized by a $3 \times 3$ matrix $N$, yielding   
\beq
\text{diag}(m_{\tilde{\chi}^0_3}, 
m_{\tilde{\chi}^0_4},m_{\tilde{\chi}^0_5}) = N^* M_N N^\dagger \;, 
\eeq
where we denoted the Higgsino mass eigenstates by
  $\tilde{\chi}_{3,4,5}^0$. The  neutral Higgsino mass eigenstates, expressed 
as a Majorana spinor ($i=3,4,5$)
\beq
\tilde{\chi}_i^0 = \left( \begin{array}{c} \tilde{\chi}_i^0 \\
                         \overline{\tilde{\chi}_i^0}^T \end{array}
                     \right) \;,
\eeq
are obtained from 
\beq
\label{eq:neuspinor}
\left( \begin{array}{c} \tilde{\chi}_3^0 \\ \tilde{\chi}_4^0 \\
         \tilde{\chi}_5^0 \end{array} \right) = N
  \left( \begin{array}{c} \tilde{H}^0_d \\ \tilde{H}^0_u \\
           \tilde{S} \end{array} \right) \;.
\eeq
They are ordered by ascending mass, \textit{i.e.}~$|m_{\tilde{\chi}^0_3}|\leq
|m_{\tilde{\chi}^0_4}| \leq |m_{\tilde{\chi}^0_5}|$. \s

In the gaugeless approximation, the mass matrix of the charginos in
the interaction basis $\psi_R^- =
( \tilde{W}^- , \tilde{H} ^- _d ) ^T$, respectively, $\psi_L^+ = (
\tilde{W}^+ , \tilde{H}^+_u )^T$, simplifies to  
\beq
M_{\psi^\pm} = \begin{pmatrix} M_2 & 0 \\ 0 & \mu _\text{eff} \end{pmatrix} 
\eeq
and is equal to the mass matrix $M_{\chi ^\pm} $ in the basis of the
  mass eigenstates $\chi^\pm$. The fermionic superpartner of the
  charged Higgs boson has a mass of  
\begin{align}
m_{\ti H^-} = \abs{\mueff}\,. \label{eq:chaMass} 
\end{align}
Here we reabsorbed the phase $\mueff$ into the mixing matrices,
hence it implicitly appears in the charged Higgsino couplings. The other superpartner with mass $M_2$ decouples and does
not give any contribution to our calculation at $\order$. \s

Since vertices and propagators
  with charginos and neutralinos enter at the two-loop level only, they
do not need to be renormalized. In the one-loop inserted counterterms
the effective Higgsino-mass parameter $\mu_{\text{eff}}$ enters via
the couplings of Higgs bosons to stops. The parameter is defined
in terms of $\lambda$ and $v_s$ whose renormalization
is specified in the Higgs sector already, and therefore no further
renormalization conditions need to be specified. 

\section{The Higher-Order NMSSM Higgs Boson Masses \label{sec:homasses}}
The loop-corrected Higgs boson masses are given by the real parts
of the poles of the propagators of the Higgs bosons. They are obtained
numerically as the zeroes of 
the determinant 
\begin{equation}
\mbox{det} \left( \hat{\Gamma} (p^2) \right) = 0
\end{equation}
of the renormalized two-point correlation function
\begin{equation}
	\hat{\Gamma } (p^2) = i \left( p^2 \mathbb{1} - \mathcal{M}^{(n)} \right) ~.
\end{equation}
The matrix part of the renormalized two-point correlation function is
given by the tree-level masses $m_{h_i}$ of the neutral Higgs bosons
and the renormalized self-energies $\hat{\Sigma}_{ij}(p^2)$ for the
$h_i \to h_j$ transition of the tree-level mass eigenstates $h_{i,j}$ at $p^2$. It reads
\begin{equation}
\left( \mathcal{M}^{(n)} \right) _{ij} = m_{h_i}^2 \delta _{ij} - \hat{\Sigma } _{ij} (p^2) ~~~\quad (i,j = 1,...,5) \label{eq:massMatrixPartOfCorrelator}
\end{equation}
and contains all contributions of one- and two-loop order, where for
the latter only the $\order$ and $\mathcal{O} (\alpha _t \alpha _s)$
corrections in the gaugeless limit and at vanishing external momentum
are taken into account. \s 

The numerical recipe to extract the zeroes of the determinant
proceeds along the same lines as in \cite{Ender:2011qh,Graf:2012hh},
which is based on an iterative procedure. While such an iterative
procedure automatically mixes different orders of perturbation theory
and is thus not of strict $\order$ any more in our present
case\footnote{Moreover, the iterative procedure can
    introduce intricate gauge dependences by mixing different
    orders of perturbation theory. Since our order $\order$
    contributions are calculated in the gaugeless limit, however, additional
    gauge dependences are only introduced beyond the considered two-loop corrections
    through the iterative procedure.}, it nevertheless gives an
  improvement 
of the numerical results when compared to a fixed-order calculation,
as argued for the one-loop case in \cite{Frank20033672003}. In the
first iteration step for the calculation of the $n^\text{th}$ neutral
Higgs boson mass, the square of the external momentum in the
renormalized self-energies is chosen to be at its tree-level mass
shell, \textit{i.e.}~$p^2 = m _{h_n} ^2$. The mass matrix
part of the two-point correlation function,
\eqref{eq:massMatrixPartOfCorrelator}, is then diagonalized, 
yielding the $n^\text{th}$ eigenvalue as a first approximation to the
squared mass of the $n^\text{th}$ neutral Higgs boson. In the next
step of the iteration, this value is taken as the new OS value for
$p^2$, and the matrix part of $\hat{\Gamma}$ is again
diagonalized. This procedure is repeated until the $n^\text{th}$
eigenvalue changes by less than $10^{-9}$ between two consecutive steps
of the iteration. This iterative algorithm is applied for the
calculation of all five neutral Higgs boson masses.
\s

The renormalized Higgs boson self-energies $\hat{\Sigma}_{ij}$ consist
of one-loop and two-loop contributions labeled by the superscript
$(1)$ and $(2)$, respectively,
\beq
\hat{\Sigma}_{ij} = \hat{\Sigma}_{ij}^{\text{\tiny{(1)}}} (p^2) +
\hat{\Sigma}^{\text{\tiny{(2)}}}_{ij} (0) \;.
\eeq
In Refs.~\cite{Ender:2011qh,Graf:2012hh} we computed the complete
one-loop corrections to the neutral NMSSM Higgs bosons in the
CP-conserving and CP-violating NMSSM, respectively. For details, we refer to these
papers. The renormalized two-loop self-energies
$\hat{\Sigma}_{ij}(0)$, which are evaluated in the approximation of vanishing external
momentum, \textit{i.e.}~$p^2=0$, comprise the ${\cal O}(\alpha_t
\alpha_s)$ corrections, which we 
computed in \cite{Muhlleitner:2014vsa}, and the
${\cal O}(\alpha_t^2)$ contributions computed in this paper, 
\beq
\hat{\Sigma}^{\text{\tiny{(2)}}} (0)_{ij} = \hat{\Sigma}^{\tiny{(2)},\alpha_t
  \alpha_s}_{ij} (0) + \hat{\Sigma}^{\tiny{(2)},\alpha_t^2}_{ij} (0) \;.
\eeq
In the following, we concentrate on the $\order$ corrections and
suppress the superscript $\alpha_t^2$. Products of one-loop terms
of $\calO(\al_t)$ give also contributions to the renormalized Higgs
self-energies at $\order$. Therefore, if not stated explicitly
otherwise, the following $\calO(\al_t)$ contributions are also 
evaluated in the gaugeless limit and at zero external momentum, just
as the $\order$ contributions\footnote{Note, however, that in the
  pure one-loop corrections to the Higgs boson masses we do not apply
  these approximations.}.
Although the renormalized Higgs self-energies
are evaluated in the approximation $p^2=0$, in the following formulae we keep the 
momentum dependence for the purpose of introducing our notation. \s

The renormalized one-loop Higgs self-energy for the transition $h_i\to
h_j$  ($i,j=1,\ldots,5$) at $\calO(\al_t)$ can be decomposed as 
\bea 
\hat \Sigma^\tol_{ij}(p^2) &=& \Sigma^\tol_{ij}(p^2) +\fr12  p^2
\left[\calR(  \wave ^\dagger +\wave)\calR^T\right]_{ij} \crn 
&&-\sbraket{ \calR\braket{ \fr12 \wave^\dagger  \calM_{hh} + \fr12
    \calM_{hh}\wave + \deltaone {\calM_{hh}} }\calR^T}_{ij } \;, 
\eea
where $\Sigma^{\text{\tiny{(1)}}}_{ij} (p^2)$ denotes the unrenormalized self-energy
at ${\cal O}(\alpha_t)$. The remaining part is the one-loop counterterm consisting of
the wave function renormalization constant matrix $\wave$ and the mass 
counterterm $\deltaone {\cal M}_{hh}$.
At two-loop $\order$, the renormalized self-energy is given by 
\bea 
\hat \Sigma^\ttl_{ij}(p^2)=\Sigma^\ttl_{ij}(p^2)  +\fr12  p^2
\left[\calR\braket{\fr 12  (\wave)^\dagger\wave +  \wavetwo^\dagger
    +\wavetwo }\calR^T\right]_{ij}- \left( \deltatwo  M^2 \right) _{ij}  \;, 
\eea
where $\Sigma^{(2)}_{ij} (p^2)$ is the unrenormalized two-loop
self-energy which is evaluated at $p^2=0$ and $\left( \deltatwo  M^2
\right) _{ij}$ is the $\order$ mass counterterm given by 
\begin{align} 
\left( \deltatwo  M^2 \right) _{ij}  &= \fr 12 \left[\calR\left(\fr12
(\wave)^\dagger \calM_{hh} \wave + \wave^\dagger \deltaone
\calM_{hh} +\deltaone \calM_{hh}\wave + \wavetwo^\dagger\calM_{hh}
\right. \right. \crn
& + \calM_{hh} \wavetwo\bigg)\calR^T\bigg]_{ij}
+\braket{ \calR \deltatwo {\calM_{hh}}
  \calR^T}_{ij} \;. \label{eq:ReHSelf}
\end{align}
The Higgs field renormalization constant matrix is diagonal and can be
expressed as
\be   
\delta^{(n)} {\cal Z}=\diag (\Delta^{(n)} Z_{H_d},\Delta^{(n)} Z_{H_u},\Delta^{(n)}
Z_{S},s_\beta^2\Delta^{(n)} Z_{H_d}+c_\beta^2\delta^{(n)} Z_{H_u},\Delta^{(n)}
Z_{S})\,, \quad n=1,2 \;,
\ee
in terms of the renormalization constants $\Delta ^{(n)} Z_{H_u,H_d,S}$ for
the doublet and singlet fields. The definition of $\Delta^{(n)}
Z_{H_u,H_d,S}$ is given subsequently in Eq.~(\ref{eq:del1zdef}) for $n=1$
and Eq.~(\ref{eq:del2zdef}) for $n=2$, respectively. We note once more
that in the above formulae we kept the momentum dependence for the purpose of
defining the Higgs field renormalization constants. Our results are
obtained, however, in the approximation of vanishing external
momentum, \ie for $p^2=0$. \s  

We remind the reader that in the mass matrix ${\cal M}_{hh}$, and
hence also in its counterterm matrix, we have dropped the Goldstone
component, {\it cf.}~Eq.~(\ref{eq:massmatrix}) with the rotation
matrix ${\cal R}$ defined in Eq.~(\ref{eq:rotationtreelevel}). 
Hence, we neglect higher-order corrections due to
  mixing of the Goldstone boson with the remaining neutral Higgs
  bosons, which we have verified to be numerically negligible.
The mass matrix counterterms $\delta {\cal M}_{hh}^{(1,2)}$
implicitly contains the counterterms of the parameters that need to be
renormalized. We explicitly specify the renormalization of these
parameters and the Higgs wave function renormalization constants in
Secs.~\ref{ssect:HiggsCTs} and \ref{ssect:TopStopCTs}.

\subsection{The Unrenormalized Self-Energies of the Neutral Higgs Bosons}
\begin{figure}[t]
\centering
\includegraphics[width=0.9\linewidth, trim=0cm 1.3cm 0cm 0.7cm, clip]{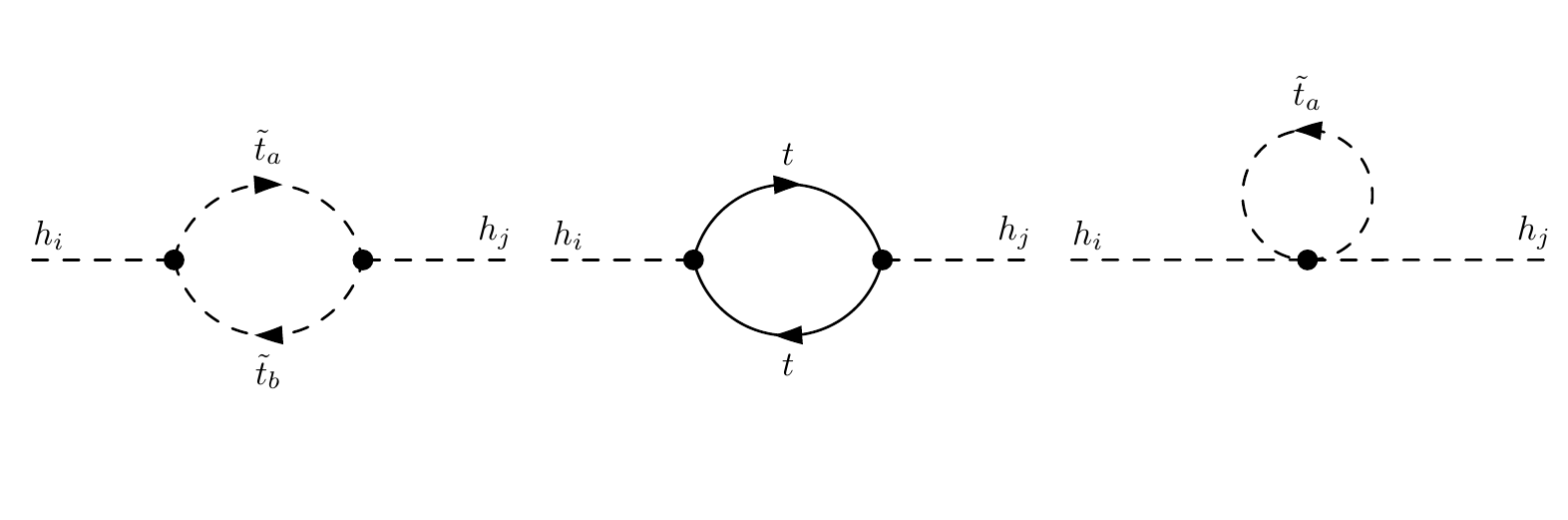}
\caption{Generic one-loop self-energies of the neutral Higgs bosons
  $h_{i,j}$ $(i,j=1,...,5)$ at $\mathcal{O} (\alpha _t)$. A summation
  over all internal particles with indices $a,b=1,2$ is implicit.}  
\label{fig:oneLoopSelfEnergiesNeutralHiggs}
\end{figure} 
The unrenormalized self-energies $\Sigma ^{(n)} _{ij} (p^2)$ of the
neutral Higgs bosons consist of all Feynman diagrammatic contributions
that are relevant at $\mathcal{O} (\alpha _t)$ and $\order$,
respectively. The one-loop diagrams contributing at $\mathcal{O}
(\alpha _t)$ are shown in
Fig.\,\ref{fig:oneLoopSelfEnergiesNeutralHiggs}, while the two-loop
diagrams of $\order$ are presented in
Fig.\,\ref{fig:twoLoopSelfEnergiesNeutralHiggs} in
App.\,\ref{append:twoLoopSelfEnergiesNeutralHiggs}. The two-loop
contributions consist of genuine two-loop diagrams as well as one-loop
diagrams with insertions of counterterm contributions from the
vertices and the particle masses relevant at $\order$, denoted by a
cross in Fig.\,\ref{fig:twoLoopSelfEnergiesNeutralHiggs}. The
definition of the counterterms necessary for the calculation of these
counterterm-inserted diagrams is discussed in
Sec.\,\ref{ssect:HiggsCTs}. \s

The calculation of the loop integrals is performed in the framework of
dimensional reduction \cite{SIEGEL1979193, 1126-6708-2005-03-076}
in $D=4-2\varepsilon$ dimensions, where $\varepsilon $ is the dimensional
regulator. For the MSSM, it has been proven that for two-loop
corrections of $\order$, SUSY is preserved by dimensional reduction
and no additional SUSY-restoring counterterms are needed
\cite{HOLLIK200663}. Since our $\order$ calculations in the NMSSM are
formally two-loop corrections calculated in the MSSM limit, this proof
is directly applicable to our two-loop $\order$ corrections in the
NMSSM as well, and no SUSY-restoring counterterms are needed. For the
calculation of the genuine two-loop integrals, we 
use the analytic results of the two-loop integrals, evaluated up to
$\mathcal{O} (\varepsilon ^0)$, as presented in
\cite{MARTIN2006133}. Some of the genuine two-loop integrals as well
as all integrals appearing in counterterm-induced Feynman diagrams can
be represented as the product 
of two one-loop one-point and two-point functions which are defined in
\cite{THOOFT1979365, Nierste1993}. For these, we derived and
implemented their expansion up to $\mathcal{O} (\varepsilon )$, given
in App.\,\ref{append:loopintegrals}. \s

The calculation of the renormalized two-loop self-energies is 
performed fully analytically. The implementation of the loop integrals
at $\mathcal{O} (\varepsilon^0)$ for the two-loop and at
$\mathcal{O} (\varepsilon)$ for the one-loop case allows for an explicit check of the
cancellation of all UV-divergent poles of $\mathcal{O} (\varepsilon ^{-2})$ and
$\mathcal{O} (\varepsilon ^{-1})$, which was confirmed explicitly at
$\order$ in the MSSM limit of the NMSSM, \textit{i.e.}~for vanishing
$\lambda$ and $\kappa$ while keeping $\mu _\text{eff}$ fixed at the
value given in the input file. Note that the calculation of the $\order$
two-loop contributions to the neutral Higgs masses is always evaluated
in the MSSM limit to ensure UV finiteness, while all other
contributions are not restricted to the MSSM limit. 

\subsection{The Renormalization of the Higgs Sector \label{ssect:HiggsCTs}}
Since we work in the gaugeless limit at $\order$, the Higgs
potential does not depend on $s_{\theta_W}$ and $e$ anymore 
so that we
restrict ourselves to one of the following two new sets of independent input
parameters entering the Higgs potential at $\order$,
\begin{align}
	 \left\{
  t_{h_d},t_{h_u},t_{h_s},t_{a_d},t_{a_s},M_{H^\pm}^2,v,\tan\beta,|\lambda|,v_s,|\kappa|,\ReAkappa,\varphi_\lambda,\varphi_\kappa,\varphi_u,\varphi_s\right\} 
\end{align}
or
\begin{align}
	  \left\{ t_{h_d},t_{h_u},t_{h_s},t_{a_d},t_{a_s},v,\tan\beta,|\lambda|,v_s,|\kappa|,\ReAlambda,\ReAkappa,\varphi_\lambda,\varphi_\kappa,\varphi_u,\varphi_s\right\} \,, 
\end{align}
depending on whether $M_{H^\pm}^2$ or $\ReAlambda$ is chosen as independent input. These are the parameters that need to be renormalized in order to
obtain a UV-finite result for the mass corrections. For the
renormalization we replace the parameters by the renormalized ones and
their corresponding counterterms as
\begin{align} 
t_{\phi} &\rightarrow t_{\phi} +\deltaone t_{\phi}  +\deltatwo
t_{\phi} \qquad \qquad\;\; \text{with}~~
           \phi=(h_d,h_u,h_s,a_d,a_s)  \label{eq:rencond1}\\  
M_{H^\pm}^2 &\rightarrow M_{H^\pm}^2+  \deltaone M_{H^\pm}^2+  \deltatwo M_{H^\pm}^2\\
v &\rightarrow v + \deltaone v +  \deltatwo v\\
\tan\beta &\rightarrow \tan\beta  +  \deltaone \tan\beta+  \deltatwo \tan\beta\\
v_s &\rightarrow v_s+  \deltaone v_s +  \deltatwo v_s\\
|\lambda| &\rightarrow |\lambda|+  \deltaone |\lambda| +  \deltatwo |\lambda|\\
|\kappa| &\rightarrow |\kappa|+  \deltaone |\kappa|+  \deltatwo |\kappa|\\
\ReAlambda &\rightarrow \ReAlambda +  \deltaone \ReAlambda  +  \deltatwo \ReAlambda \\
\ReAkappa &\rightarrow \ReAkappa+  \deltaone \ReAkappa +  \deltatwo \ReAkappa  \\
\varphi_p  & \rightarrow \varphi_p + \deltaone
              \varphi_p +  \deltatwo \varphi_p \qquad\qquad \quad \text{with}~~
              p=(u, s, \lambda, \kappa) \;,  
             \label{eq:rencond10}
\end{align}
where the superscript $(n)$ stands for the $n$-loop level. \s

For the consistent incorporation of the $\order$ corrections with
the previously computed one-loop corrections in \cite{Graf:2012hh} and two-loop
corrections of $\calO(\al_t\al_s)$ in \cite{Muhlleitner:2014vsa},
we use here also the mixed $\DRb$-OS renormalization scheme, {\it
  i.e.}~the parameters are renormalized as
\be
\underbrace{ t_{h_d},t_{h_u},t_{h_s},t_{a_d},t_{a_s},M_{H^\pm}^2,v}_{\mbox{on-shell
 scheme}},
\underbrace{\tan\beta,|\lambda|,v_s,|\kappa|,\ReAkappa,\varphi_\lambda,\varphi_\kappa,\varphi_u,\varphi_s}_{\overline{\mbox{DR}} \mbox{ scheme}}\,,
\label{eq:mixedcond1}
\ee
in case $M_{H^\pm}^2$ is used as independent input, or
\be
\underbrace{ t_{h_d},t_{h_u},t_{h_s},t_{a_d},t_{a_s},v}_{\mbox{on-shell
 scheme}},
\underbrace{\tan\beta,|\lambda|,v_s,|\kappa|,\ReAlambda,\ReAkappa,\varphi_\lambda,\varphi_\kappa,\varphi_u,\varphi_s}_{\overline{\mbox{DR}} \mbox{ scheme}}\,,
\label{eq:mixedcond2}
\ee
for $\ReAlambda$ as independent input. The counterterm matrix for the
neutral Higgs mass matrix $\mathcalM_{hh}$ is obtained by inserting the replacements
from Eqs.~(\ref{eq:rencond1})-(\ref{eq:rencond10}) in the tree-level mass
matrix and expanding order by order, yielding 
\be 
\mathcalM_{hh} \rightarrow  \mathcalM_{hh} +\deltaone\mathcalM_{hh}
+\deltatwo\mathcalM_{hh} \;. \label{eq:CTmassH}
\ee
In App.\,\ref{sec:dMH2loop}, we give the explicit expressions of
$\deltaone\mathcalM_{hh}$ and $\deltatwo\mathcalM_{hh}$ in terms of
all parameter counterterms. \s

Also the Higgs fields need to be renormalized. We
introduce the renormalization constants for the doublet and singlet
fields before rotating to the mass eigenstates as 
\bea 
H_d&\rightarrow& (1+ \fr{1}2 \deltaone Z_{H_d}+ \fr{1}2 \deltatwo
Z_{H_d} -\fr18 (\deltaone Z_{H_d})^2 )H_d \equiv (1+ \fr{1}2 \deltaone
Z_{H_d}+ \fr{1}2 \Deltatwo Z_{H_d})H_d\\ 
H_u&\rightarrow& (1+ \fr{1}2 \deltaone Z_{H_u}+ \fr{1}2 \deltatwo Z_{H_u}
 -\fr18 (\deltaone Z_{H_u})^2) H_u\equiv  (1+ \fr{1}2 \deltaone
 Z_{H_u}+ \fr{1}2 \Deltatwo Z_{H_u}) H_u\\ 
S&\rightarrow& (1+ \fr{1}2 \deltaone Z_{S}+ \fr{1}2 \deltatwo Z_{S}
-\fr18 (\deltaone Z_{H_d})^2)S\equiv  (1+ \fr{1}2 \deltaone Z_{S}+
\fr{1}2 \Deltatwo Z_{S})S \;, 
\eea
where we defined
\beq  
\Deltatwo Z_{i} \equiv  \deltatwo
Z_{i} - \fr14 (\deltaone Z_{i})^2 \;, \quad i=H_d,H_u,S \;. \label{eq:del2zdef} 
\eeq 
In the following, we discuss in detail the wave function
renormalization constants and parameter counterterms. \s

\noindent
\underline{\it Higgs wave function renormalization constants}\\
In analogy to our one-loop and two-loop $\calO(\al_t\al_s)$
calculation \cite{Ender:2011qh,Graf:2012hh,Muhlleitner:2014vsa} the
Higgs fields are renormalized through $\DRb$ conditions. The
renormalization conditions ($n=1,2$, $i=h_d, h_u, h_s$)
\be 
\left.\fr{\pa \hat \Sigma^{(n)}_{ii}(p^2)}{\pa p^2}
\right|_{\text{div}} =0
\ee
yield 
\bea \delta^{(n)} Z_{H_d} &=&- \left.\fr{\pa
    \Sigma^{(n)}_{h_dh_d}(p^2)}{\pa p^2} \right|_{\text{div}}   \\  
 \delta^{(n)} Z_{H_u} &=&- \left.\fr{\pa \Sigma^{(n)}_{h_uh_u}(p^2)}{\pa
     p^2} \right|_{\text{div}} \\
 \delta^{(n)} Z_{S} &=&- \left.\fr{\pa \Sigma^{(n)}_{h_sh_s}(p^2)}{\pa p^2} \right|_{\text{div}}
\,,  
 \eea
where the subscript '$\text{div}$' indicates that we
take the divergent part only. The wave function renormalization constants are
  defined through the Higgs self-energies in the gauge basis. They are
  obtained from the derivatives with respect to $p^2$
  of the diagrams shown in 
  Fig.\,\ref{fig:oneLoopSelfEnergiesNeutralHiggs} at one-loop and in
  Fig.\,\ref{fig:twoLoopSelfEnergiesNeutralHiggs} at two-loop level by
  choosing the mixing matrix of the neutral Higgs bosons to be
  diagonal. At ${\cal O}(\al_t)$ we find 
\begin{align}
	\deltaone Z_{H_d} &= 0  \\
	\deltaone Z_{H_u} &= \fr{-3 m_t^2}{8\pi^2 v^2 \sbeta^2}\fr{1}{\varepsilon}  \\
	\deltaone Z_{S} &= 0 
\end{align}
and at $\order$ we have, for the $\DRb$ renormalization scheme of the top/stop sector,
\begin{align}
\deltatwo Z_{H_d}^{\DRb} &= 0  \\
 \deltatwo Z_{H_u}^{\DRb} &= \fr{9(m_t^4)^{\DRb}}{128 \pi^4 v^4\sbeta^4 }\left( \fr{1}{\varepsilon} -\fr{1}{\varepsilon^2} \right)  \\
\deltatwo Z_{H_s}^{\DRb} &= 0 ~.
\end{align}
This result is in agreement with
\cite{Sperling:2013eva,Sperling:2013xqa} derived on the basis of the
renormalization group. In the OS scheme of the top/stop sector, we find
\begin{align}
\deltatwo Z_{H_d}^\OS &= 0  \\
\deltatwo Z_{H_u}^\OS &=\fr{9(m_t^4)^{\OS}}{128 \pi^4
  v^4\sbeta^4 }\left( \fr{1}{\varepsilon} -\fr{1}{\varepsilon^2} \right) -\fr{3(m_t^2)^{\OS} 
}{4 \pi^2 v^2 \sbeta^2 \varepsilon}\braket{\fr{dm_t^{\al_t} }{m_t^\OS}-
  \fr{ dv^{\al_t} }{v} } \\
\deltatwo Z_{H_s}^\OS &= 0  \;.
\end{align}
Note that the
superscripts $\DRb$ and $\OS$ on $\deltatwo Z_{H_u}$ 
refer to the renormalization of the top/stop sector. The expression
for $dm_t^{\al_t} $ is given 
in \eqref{eq:dmt} and  
for $dv^{\al_t} $ reads
\bea  
dv^{\al_t}&=& \fr{3}{32 \pi^2 s_{\theta_W}^2 v} \bigg(c_{2\theta_W}
|\mathcal{U}^{\tilde t}_{11}|^2 F_0(m_{\tilde t_1}^2,m_{\tilde Q_3}^2)
+  c_{2\theta_W} |\mathcal{U}_{\tilde t_{21}}|^2 F_0(m_{\tilde
  t_2}^2,m_{\tilde Q_3}^2)  \crn 
&&- c_{\theta_W}^2 |\mathcal{U}^{\tilde t}_{11}|^2 |\mathcal{U}^{\tilde 
  t}_{12}|^2 F_0(m_{\tilde t_1}^2,m_{\tilde t_2}^2) \bigg)\,,  
\eea
where 
\be 
F_0(x,y) = x+y -\fr{2xy}{x-y}\log\fr{x}{y} \;.
\ee 

\noindent
\underline{\it The tadpole counterterms}\\
\begin{figure}[t]
\centering
\includegraphics[width=0.5\linewidth, trim=0cm 0.9cm 0cm 0.2cm, clip]{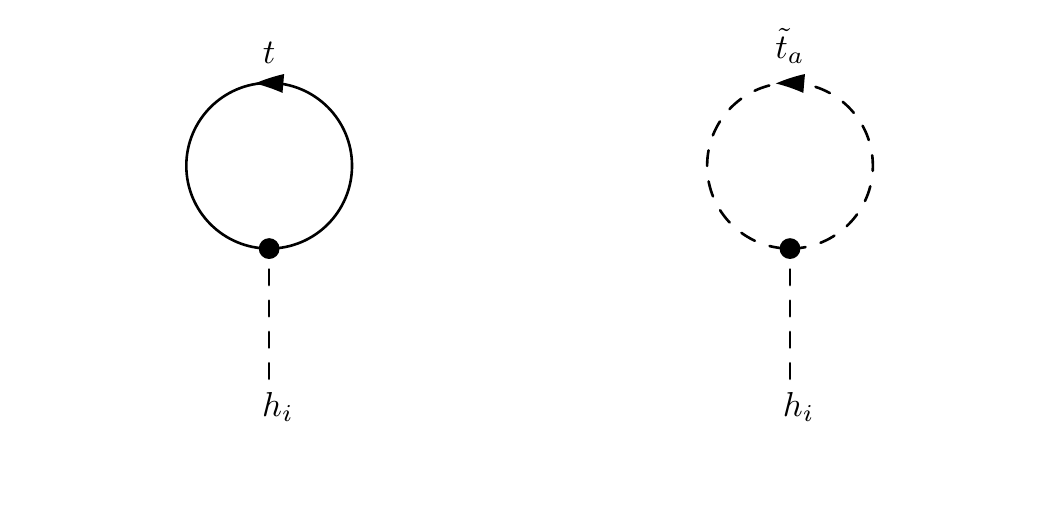}
\caption{One-loop tadpole diagrams of the neutral Higgs bosons
  $h_i$ ($i=1,...,5$). A summation over the index $a=1,2$ of the
  internal stop is implicit.}  
\label{fig:oneLoopTadpoles}
\end{figure} 
\begin{figure}[t]
\centering
\includegraphics[width=\linewidth, trim=0.31cm 0.1cm 0.75cm 0.6cm, clip]{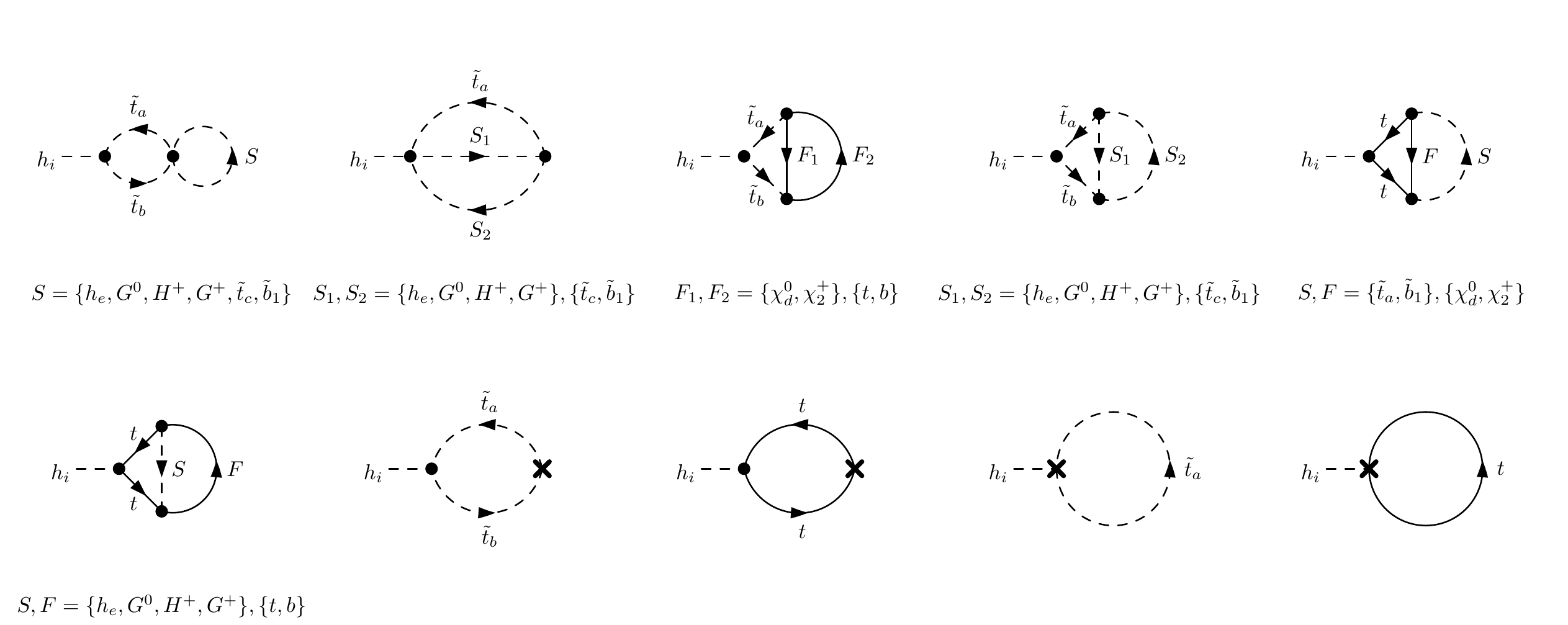}
\caption{Two-loop tadpole diagrams of the neutral Higgs bosons
  $h_i$ ($i=1,...,5$). A summation over all internal particles with indices $a,b,c=1,2$, 
$d=3,4,5$ and $e=1,...,5$ is implicit.}
\label{fig:twoLoopTadpoles}
\end{figure} 
The renormalization conditions for the tadpoles are chosen such
that the minimum of the Higgs potential does not change at two-loop order,
leading to ($\phi=h_d,h_u,h_s,a_d,a_s$)
\begin{align}
\delta ^{\text{\tiny{(1)}}} t_{\phi}  &= t_{\phi}^{\text{\tiny{(1)}}} \\ 
\delta ^{\text{\tiny{(2)}}} t_{\phi} &= t_{\phi}^{\text{\tiny{ (2)}}}+t_{\phi}^{\text{\tiny{ (2)}}}
                                       \wave_{\phi\phi} \;,  
\end{align}
where $t_{\phi}^{\text{\tiny{ (1)}}}$ and $t_{\phi}^{\text{\tiny{ (2)}}}$
represent the one- and two-loop tadpole contributions shown in
Fig.~\ref{fig:oneLoopTadpoles} and Fig.~\ref{fig:twoLoopTadpoles} for
$\mathcal{O} (\alpha _t )$ and $\order$, respectively. \s 

\noindent 
\underline{\it The charged Higgs boson mass counterterm}\\
\begin{figure}[t]
\centering
\includegraphics[width=0.8\linewidth, trim=0cm 0cm 0cm 0.8cm, clip]{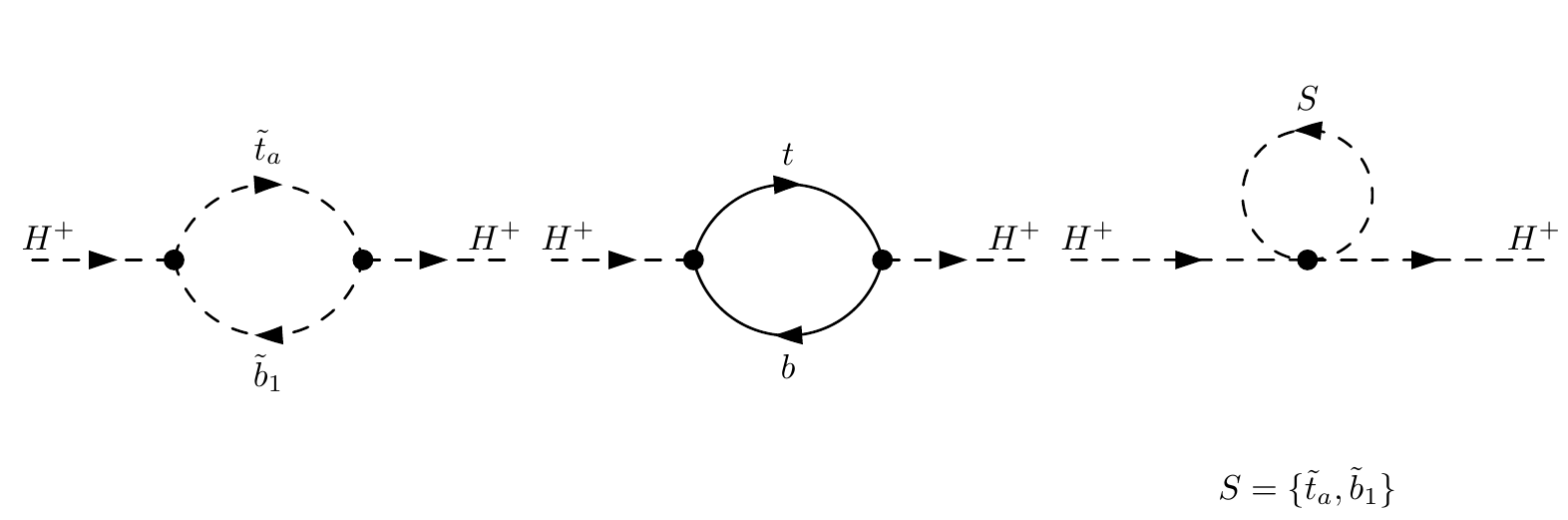}
\caption{Generic one-loop self-energies of the charged Higgs bosons
  contributing at $\mathcal{O} (\alpha _t)$ for the renormalization of
  $M_{H^\pm}^2$. A summation over the index $a=1,2$ of the
    internal stops is implicit.}  
\label{fig:oneLoopSelfEnergiesChargedHiggs}
\end{figure} 
If the charged Higgs boson mass is chosen as independent input, hence defined as OS parameter, we
renormalize it in the OS scheme accordingly. In the approximation of vanishing external momentum, the OS
counterterm of the charged Higgs mass at one-loop order is given by 
\begin{equation}
\deltaone\mhpm = \Sigma_{H^-H^-}^{\text{\tiny{(1)}}}(0) - \mhpm
\wave_{H^-H^-}
\label{eq:mhpmone}
\end{equation}
and the counterterm at two-loop level reads 
\beq
\deltatwo \mhpm &=& \Sigma_{H^-H^-}^{\text{\tiny{(2)}}}(0) -\fr{1}{4}\mhpm
(\wave_{H^-H^-})^2 - \wave_ {H^-H^-} \deltaone\mhpm- \wave_ {H^-G^-}
\deltaone M_{H^-G^-} \nonumber \\
&& - \mhpm\wavetwo_{H^-H^-} \;,
\label{eq:mhpmtwo}
\eeq
with
\bea 
\delta {\cal Z}^{(n)}_{H^-H^-}&=& \cos^2\!\beta \Delta^{(n)} Z_{H_u}+\sin^2\!\beta
\Delta^{(n)} Z_{H_d}\\ 
\wave_ {H^-G^-} &=& \cos\beta \sin\beta(-\deltaone
Z_{H_d} + \deltaone Z_{H_u})\\ 
\deltaone M_{H^-G^-} &=&\frac{-\cbeta^2\mhpm v \deltaone \,\tbeta
  +\cbeta \deltaone \,\thu-\deltaone \,\thd \sbeta}{v}+\frac{i \deltaone
  \,\tad}{\sbeta v}\;,
\eea
where
\beq
\Delta^{(1)} Z_i \equiv \delta^{(1)} Z_i \label{eq:del1zdef}
\eeq
and $\Delta^{(2)} Z_i$ ($i=H_u,H_d,S$) has been defined in Eq.~(\ref{eq:del2zdef}).
The $\mathcal{O} (\alpha _t )$ contributions to the unrenormalized
self-energy of the charged Higgs boson are depicted in
Fig.\,\ref{fig:oneLoopSelfEnergiesChargedHiggs}. The sum of all
diagrams yields the following analytic expression of the 
unrenormalized one-loop self-energy of the charged Higgs boson at $\mathcal{O}
(\alpha _t )$, 
\bea 
\Sigma_{H^\pm}^{\text{\tiny{(1)}}}(0)&=&\frac{3 m_t^2c_{\beta}^2}{8 \pi^2 s_{\beta}^2 v^2}
\bigg\{ A_0 (m_{\tilde Q_3}^2)-2 
A_0(m_t^2)+ |\mathcal{U}_{\tilde{t}_{12}}|^2
A_0(m_{\tilde{t}_1}^2) +
|\mathcal{U}_{\tilde{t}_{22}}|^2 A_0 (m_{\tilde{t}_2}^2)] \,
\nonumber \\
&+&\left| m_t |\mathcal{U}_{\tilde{t}_{11}}|+|A_t| 
  e^{i \phi_x} |\mathcal{U}_{\tilde{t}_{12}}|+\frac{|\lambda| \tbeta
    v_s |\mathcal{U}_{\tilde{t}_{12}}|}{\sqrt{2}}\right|^2
B_0(0,m_{\tilde Q_3}^2,m_{\tilde{t}_1}^2)
\nonumber \\
&+&\left| m_t |\mathcal{U}_{\tilde{t}_{21}}|+|A_t| 
  e^{i \phi_x} |\mathcal{U}_{\tilde{t}_{22}}|+\frac{|\lambda| \tbeta
    v_s |\mathcal{U}_{\tilde{t}_{22}}|}{\sqrt{2}}\right|^2
B_0(0,m_{\tilde Q_3}^2,m_{\tilde{t}_2}^2) \bigg\} \;,
\eea
with the one-loop scalar integrals $A_0$ and $B_0$ defined in
App.\,\ref{append:loopintegrals}. Note that since the mass of the charged Higgs
boson is calculated at vanishing external momentum the counterterm of
the mass of the charged Higgs boson also involves the field
renormalization constants. The two-loop $\order$ contributions to the
unrenormalized charged Higgs self-energies are depicted in
Fig.\,\ref{fig:twoLoopSelfEnergiesChargedHiggs} in
App.\,\ref{append:twoLoopSelfEnergiesChargedHiggs}. Due to its lengthy
structure, we do not display the analytic result of the corresponding
unrenormalized two-loop self-energy explicitly here. \s  
 
If $\ReAlambda$ is chosen as independent input parameter instead of
$M_{H^\pm}^2$, then the counterterms $\deltaone M_{H^\pm}^2$ and
$\deltatwo M_{H^\pm}^2$ of the charged Higgs boson mass at one- and
two-loop order, respectively, are calculated as functions of all other
counterterms by inserting the two-loop expansions of
Eqs.~(\ref{eq:rencond1}) to (\ref{eq:rencond10}) in the formula for
the charged Higgs boson mass, \eqref{eq:charalam}. The explicit
formulae of the counterterms are presented in
App.\,\ref{sec:dMHp2loop}. The loop-corrected charged Higgs mass 
$(M_{H^\pm}^2)^{(2)}$ is calculated iteratively by solving
\be 
p^2 - \mhpm +
\hat\Sigma_{H^-H^-}^{\text{\tiny{(1)}}}(p^2)
+ \hat\Sigma_{H^-H^-}^{\tiny{(2)},\alpha_t\alpha_s}(0) +
\hat\Sigma_{H^-H^-}^{\tiny{(2)},\alpha_t ^2}(0) = 0 
\ee
with the renormalized one-loop self-energy of the charged Higgs boson
\be 
\hat\Sigma_{H^-H^-}^{\text{\tiny{(1)}}}(p^2) =
\Sigma_{H^-H^-}^{\text{\tiny{(1)}}}(p^2) + \left(p^2 -  \mhpm\right) \wave_{H^-H^-} - \deltaone \mhpm
\ee 
evaluated at the scale of the squared tree-level charged Higgs mass
and the renormalized two-loop self-energies at
$\mathcal{O}(\alpha_t\alpha_s)$ and $\mathcal{O}(\alpha_t^2)$, respectively,
\beq 
\hat\Sigma_{H^-H^-}^{{\text{\tiny{(2)}}},\alpha_t\alpha_s / \alpha _t^2}(0) &=&\Sigma_{H^-H^-}^{{\text{\tiny{(2)}}},\alpha_t\alpha_s / \alpha _t^2}(0)-\fr{1}{4}\mhpm
(\wave_{H^-H^-})^2 - \wave_ {H^-H^-} \deltaone\mhpm \nonumber \\
&&- \wave_ {H^-G^-}
\deltaone M_{H^-G^-} - \mhpm\wavetwo_{H^-H^-}   -\deltatwo \mhpm \;, 
\eeq  
evaluated in the approximation of zero external momentum. \s

\begin{figure}[t]
\centering
\includegraphics[width=0.8\linewidth, trim=0cm 0cm 0cm 0.8cm, clip]{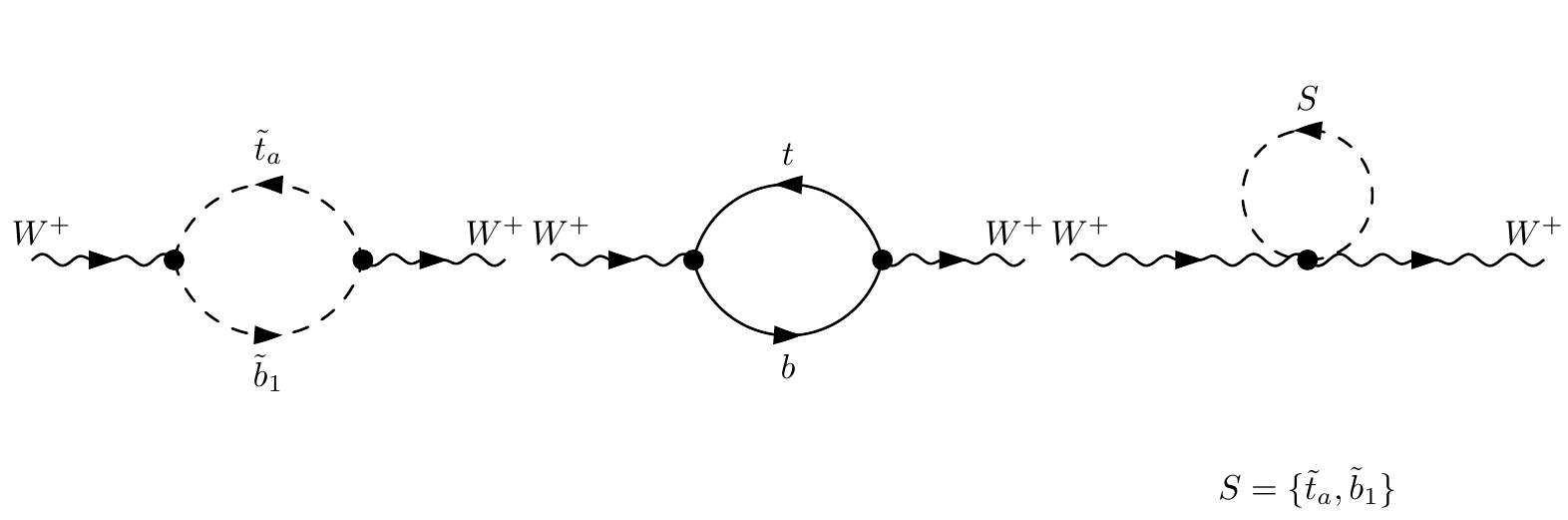}
\caption{Generic one-loop self-energies of the $W$ boson contributing
  at $\mathcal{O} (\alpha _t)$ to the renormalization of $v$. A
  summation over the index $a=1,2$ of the internal stops is implicit.}  
\label{fig:oneLoopSelfEnergiesWBoson}
\end{figure} 
\begin{figure}[t]
\centering
\includegraphics[width=0.8\linewidth, trim=0cm 0cm 0cm 0.8cm, clip]{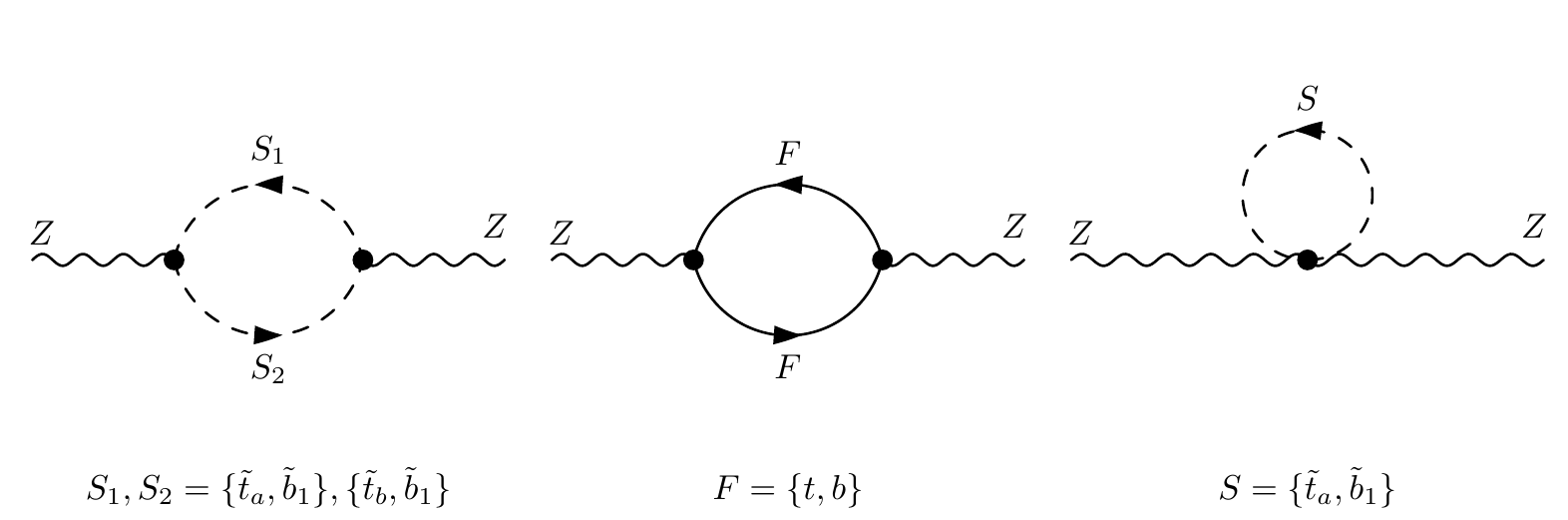}
\caption{Generic one-loop self-energies of the $Z$ boson contributing
  at $\mathcal{O} (\alpha _t)$ to the renormalization of $v$. A
  summation over the indices $a,b=1,2$ of the internal stops is implicit.}  
\label{fig:oneLoopSelfEnergiesZBoson}
\end{figure}
\noindent
\underline{\it The VEV counterterm}\\
The OS renormalized one-loop counterterm $\deltaone v/v$ of the VEV is
given by
\begin{equation}
\fr{\deltaone v}{v}=\fr{c_{\theta_W}^2}{2 s_{\theta_W}^2}\left(
  \fr{\deltaone M_Z^2}{M_Z^2} -\fr{\deltaone M_W^2}{M_W^2} \right) +
\fr{\deltaone M_W^2}{2M_W^2} \;, \label{eq:dvexpression}
\end{equation}
with the gauge bosons renormalized OS. Their OS counterterms in the
gaugeless approximation read
\begin{equation}
\fr{\deltaone M_W^2}{M_W^2}=\fr{\Sigma_{W}^{\scriptsize{ (1),T}}(0)}{M_W^2} \qquad
\text{and } \qquad \fr{\deltaone M_Z^2}{M_Z^2}=\fr{\Sigma_{Z}
  ^{(1),T}(0)}{M_Z^2}\,. 
\end{equation}
Here, $\Sigma^{(1),T}_{V}(0)$ ($V=W,Z$) is the transverse part of the
unrenormalized one-loop vector boson self-energy evaluated at vanishing
external momentum. The relevant diagrams at $\mathcal{O} (\alpha _t)$
are depicted in Figs.\,\ref{fig:oneLoopSelfEnergiesWBoson} and
\ref{fig:oneLoopSelfEnergiesZBoson}. \s

Note that while $\deltaone M_V^2$ and $M_V^2$ are separately zero in the
gaugeless limit, their ratio is non-zero for each gauge boson $V$ and
hence contributes to our $\order$ 
calculation. The explicit evaluation of the UV divergent part of
$\deltaone v/v$ 
shows that it is related to $\deltaone Z_{H_u}$ as
\beq
\left.\frac{\deltaone v}{v}\right|_{\text{div}} = \frac{s_\beta^2}{2}
\deltaone Z_{H_u} \;,
\eeq
as expected from Refs.~\cite{Sperling:2013eva,Sperling:2013xqa}. \s

\noindent
\underline{\it The $\tan\beta$ counterterm}\\
The ratio of the vacuum expectation values of the Higgs doublets,
$\tan\!\beta$, is renormalized in the $\DRb$ scheme with the counterterm given by 
\cite{Brignole:1992uf,Chankowski:1992ej,Chankowski:1992er,Dabelstein:1994hb,Dabelstein:1995js,Freitas:2002um} 
\begin{equation}
 \delta ^{(n)} \tan\beta=\frac{1}{2}\tan\!\beta\big(\delta ^{(n)}
 Z_{H_u}-\delta ^{(n)} Z_{H_d}\big)\big|_{\text{div}}=
 \frac{1}{2}\tan\!\beta\,\delta ^{(n)} Z_{H_u}\big|_{\text{div}}  \,,
\end{equation}
where the subscript $(n)$ again indicates the loop level. Note that
the last identity only holds at $\mathcal{O} (\alpha _t)$ and $\order$
in our approximation, but is not valid in general. \s

\noindent
\underline{\it The remaining $\overline{\mbox{DR}}$ counterterms}\\
The counterterms of the remaining $\DRb$ parameters $|\lambda|,
|\kappa|, v_s, \mbox{Re} A_\kappa, \varphi_\lambda,\varphi_\kappa,\varphi_u,$ and
  $\varphi_s$ have to cancel the left-over UV-divergent parts of the self-energies
of the neutral Higgs bosons. For the $\mathcal{O} (\alpha _t)$
one-loop counterterm of $|\lambda |$, we find 
\begin{equation}
\deltaone |\lambda| = - \frac{|\lambda|}{2} \left(\deltaone
  Z_{H_u} c_\beta^2 + 2 \left. \frac{\deltaone
      v}{v}\right|_{\text{div}} \right) = - \frac{|\lambda|}{2}
\deltaone Z_{H_u} ~,
\end{equation}
while the two-loop counterterm $\deltatwo |\lambda|$ as well as the
remaining one- and two-loop counterterms of $|\kappa|, v_s, \mbox{Re}
A_\kappa,$ $\varphi_u, \varphi_s, \varphi_\lambda$ and $\varphi_\kappa$ are not needed to
yield a UV-finite result of the neutral Higgs masses at $\order$. 

\subsection{Renormalization of the Quark/Squark
  Sector \label{ssect:TopStopCTs}}
The set of independent parameters to be renormalized in the third
generation quark/squark sector is given in Eq.\,(\ref{eq:stopparset}).   
The renormalization conditions of the corresponding counterterms 
\be 
 \delta m_t, \;\delta  m_{\tilde{Q}_3}, \;\delta m_{\tilde t_R} \quad
 \mbox{and} \quad \delta A_t \;,
\ee
are specified in the following. Since all counterterms of the
quark/squark sector necessary for the calculation of the $\order$
corrections to the neutral Higgs boson masses are of one-loop order,
we suppress the superscript $(1)$ of the counterterms in this
subsection. \s 

We proceed along the same lines as in our ${\cal
  O}(\alpha_t \alpha_s)$ calculation \cite{Muhlleitner:2014vsa} where
we have implemented in {\tt NMSSMCALC} both the OS and $\DRb$
renormalization scheme for the (s)quark sector.
We follow the SLHA
\cite{Skands:2003cj,Allanach:2008qq} where the input 
top quark mass is understood to be the pole mass whereas the soft
SUSY breaking masses and trilinear couplings are
$\overline{\mbox{DR}}$ parameters at the renormalization scale $\mu_R
= M_{\text{SUSY}}$. The translation between
the two schemes has to be done consistently both in the counterterm
part and at the level of the input parameters. Expanding the OS and
$\overline{\mbox{DR}}$ counterterms of the parameters $X=m_t,
m_{\tilde{Q}_3},m_{\tilde{t}_R}, A_t$ in terms of the dimensional
regulator $\varepsilon$, we have 
\bea 
\delta X^{\OS} &=& \fr{1}{\varepsilon} \delta X_{\text{pole}} +  \delta
X_{\text{fin}}  \label{eq:OScounterterm}\\
 \delta X^{\DRb} &=& \fr{1}{\varepsilon} \delta X_{\text{pole}}\,. 
\label{eq:DRbarcounterterm}
\eea
Note that in our definition of the parameters of the OS scheme we did
not take into account any terms proportional to $\varepsilon$, {\it
  i.e.}~$\varepsilon \delta X_\varepsilon$. These terms, that could also be
chosen to be included, {\it cf.}\,\cite{Degrassi:2014pfa, Borowka:2015ura},
would manifest themselves 
as additional finite contributions originating from the counterterm
inserted diagrams multiplying $1/\varepsilon$ terms from the one-loop
functions with the $\varepsilon$ parts of the
counterterms. We verified through explicit calculation that the contributions of finite
  terms arising through the inclusion of the $\mathcal{O}
  (\varepsilon)$ terms of the OS-defined counterterms cancel
  in the calculation of the $\mathcal O(\alpha_t^2)$ corrections. Therefore, we neglect the 
  $\mathcal{O} (\varepsilon )$ contributions from the counterterms and
  apply our thus defined OS scheme consistently throughout the
whole calculation. \s 

In case of $\DRb$ renormalization of the (s)quark sector the $\DRb$ top
quark mass $m_t^{\tiny{\overline{\mbox{DR}}}}$ has to be computed
from the corresponding top pole mass, as described in
App.\,\ref{append:mtrun}. If the OS scheme is chosen for the (s)quark
sector, then the translation of the parameters $m_{\tilde{Q}_3},
m_{\tilde{t}_R}$ and $A_t$ from the 
$\overline{\mbox{DR}}$ scheme to the OS scheme is performed by 
\beq
A_t^{\OS} &=& A_t^{\DRb} - \delta A_t^{\text{fin}} \label{eq:atfin}\\
(m_{\tilde Q_L}^2)^{\OS}  &=& (m_{\tilde Q_L}^2)^{\DRb} - \delta 
(m_{\tilde Q_L}^2)^{\text{fin}}  \label{eq:qlfin}\\
(m_{\tilde t_R}^2)^{\OS}  &=& (m_{\tilde t_R}^2)^{\DRb} - \delta
(m_{\tilde t_R}^2)^{\text{fin}} \label{eq:trfin} \;.
\eeq 
In the above equations, the finite counterterm parts have to be
computed with OS input parameters, which we achieve by applying an
iterative procedure. Note that we include both ${\cal O}(\alpha_s)$
and ${\cal O}(\alpha_t)$ corrections into these conversions in the numerical analysis. The OS conditions in the NMSSM (s)quark sector
are the same as the ones in the complex MSSM presented in
\cite{Heinemeyer:2007aq,Heinemeyer:2010mm}. We give here, for completeness, the
expressions of the counterterms:  
 \begin{figure}[t]
\centering
\includegraphics[width=0.33\linewidth, trim=0cm 0cm 0cm 1.8cm, clip]{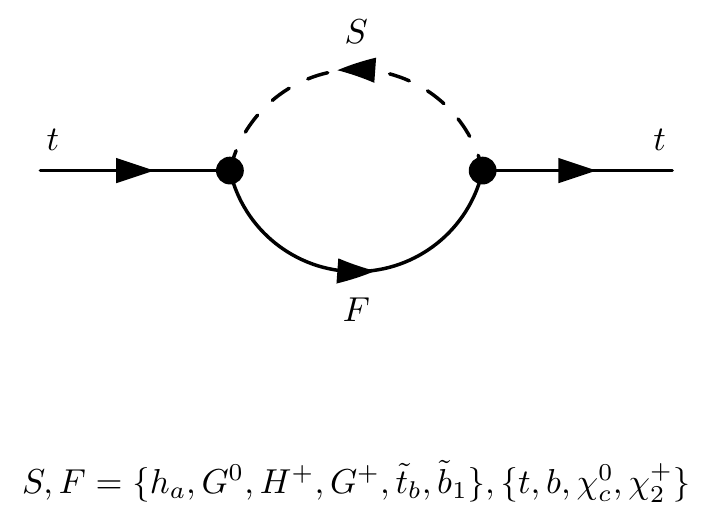}
\caption{Generic one-loop self-energies of the top quark contributing at $\mathcal{O}
  (\alpha _t)$ to the renormalization of $m_t$. A summation over all
  internal particles with indices $a=1,...,5$, $b=1,2$ and $c=3,4,5$ is implicit.}
\label{fig:oneLoopSelfEnergiesTop}
\end{figure}  
\begin{itemize}
\item Decomposing the unrenormalized top quark self-energy $\Sigma_t$ as 
\beq
\Sigma_t (p^2) = \slash{\hspace*{-0.2cm} p} P_L \Sigma^{VL}_t (p^2) +
\slash{\hspace*{-0.2cm} p} P_R 
\Sigma^{VR}_t (p^2) + P_L \Sigma^{SL}_t (p^2) + P_R \Sigma^{SR}_t (p^2) \;,
\eeq
in terms of the left- and right-handed projectors
$P_{L/R} = (1 \mp \gamma_5)/2$, the top mass counterterm reads
\begin{equation}
 \delta m_t=\frac{1}{2}\widetilde{\Re} \;\Big(m_t
 \Sigma_t^{VL}(m_t^2)+ m_t
 \Sigma_t^{VR}(m_t^2)+\Sigma_t^{SL}(m_t^2)+\Sigma_t^{SR}(m_t^2)\Big)\,.
\end{equation}
Here $\widetilde{\mbox{Re}}$ indicates that the real part is taken
only of the one-loop function, but not of the parameters.
The $\mathcal{O} (\alpha _t)$ contributions to the top mass
counterterm are depicted in Fig.\,\ref{fig:oneLoopSelfEnergiesTop}. 
 \begin{figure}[t]
\centering
\includegraphics[width=0.8\linewidth, trim=0cm 0cm 0cm 0.8cm, clip]{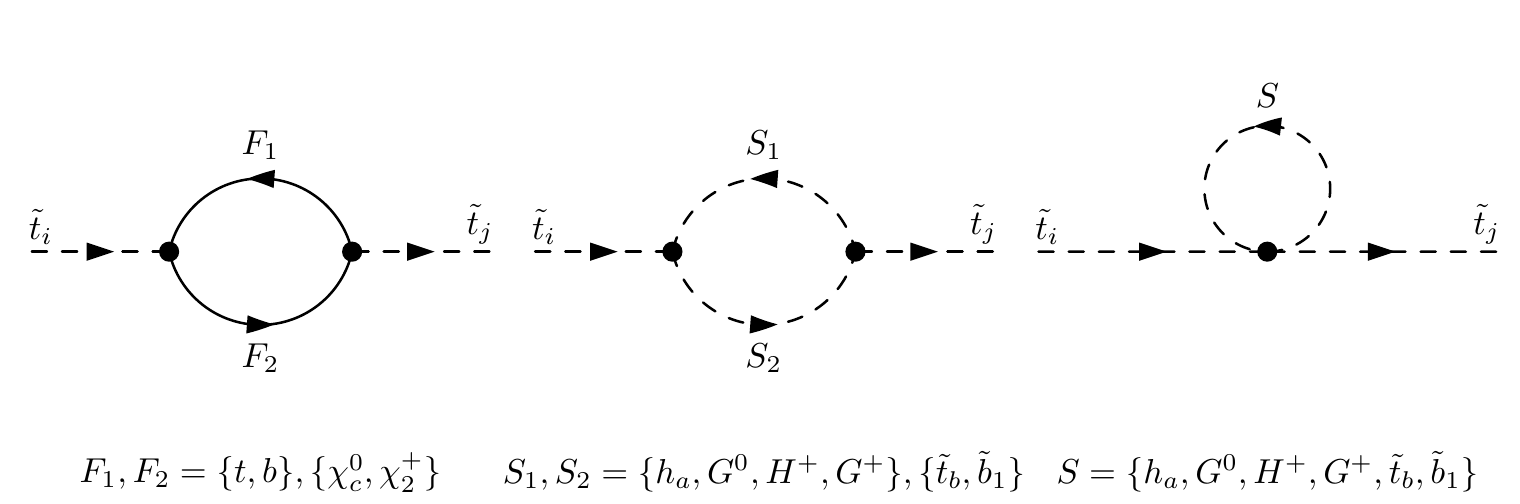}
\caption{Generic one-loop self-energies of the stops contributing at
  $\mathcal{O} (\alpha _t)$ to the renormalization of the stop
  sector. A summation over all internal particles with indices
  $a=1,...,5$, $b=1,2$ and $c=3,4,5$ is
  implicit.}  
\label{fig:oneLoopSelfEnergiesStops}
\end{figure} 
\item The counterterm of the trilinear stop coupling parameter reads
\beq
\de A_t &= &\fr{\expetam}{m_t}\bigg[ \mathcal{U}^{\ti t}_{11}\mathcal{U}^{\ti t*}_{12}\braket{\de m_{\ti t_1}^2 -
\de m_{\ti t_2}^2} + \mathcal{U}^{\ti t}_{11}\mathcal{U}^{\ti t*}_{22} (\de Y_{\ti t})^* +  \mathcal{U}^{\ti t}_{21}\mathcal{U}^{\ti t*}_{12} \de Y_{\ti t}  \crn
&-& \braket{A_t \expeta -\fr{\mueff^*}{\tan\beta}}\de m_t\bigg]
 - \fr{\expetam \mueff^* \de\tan\beta}{\tan^2\beta} +\fr{\expetam\de
   \mueff^*}{\tan\beta} \;,
\eeq 
where 
\bea 
\de m_{\ti t_1}^2& =&\Retilde\Sigma_{\ti t_1\ti t_1}(m_{\ti t_1}^2)\\
\de m_{\ti t_2}^2& =&\Retilde\Sigma_{\ti t_2\ti t_2}(m_{\ti t_2}^2)\\
\de Y_{\ti t} &=& \sbraket{\mathcal{U}^{\ti t} \de {\cal M}_{\ti t}
  \mathcal{U}^{\ti t\dagger}}_{12} = \sbraket{\mathcal{U}^{\ti t} \de
  {\cal M}_{\ti t} \mathcal{U}^{\ti t\dagger}}_{21}^* =
\fr{1}{2}\Retilde \braket{\Sigma_{\ti t_1^*\ti t_2^*}(m_{\ti t_1}^2)+
  \Sigma_{\ti t_1^*\ti t_2^*}(m_{\ti t_2}^2)}
\eea
with $\Sigma_{\ti t_i\ti t_j}$ $(i,j=1,2)$ denoting the unrenormalized self-energy
for the transition $\ti t_i\to \ti t_j$, whose contributions at $\mathcal{O}
(\alpha _t)$ are depicted in Fig.\,\ref{fig:oneLoopSelfEnergiesStops}. 
\item The counterterms for the soft SUSY breaking left-handed squark and
  right-handed stop mass parameters are given by
\bea
 \de \msq&=& \abs{\mathcal{U}^{\ti t}_{11}}^2 \de m_{\ti
   t_1}^2+\abs{\mathcal{U}^{\ti t}_{12}}^2 \de m_{\ti t_2}^2 +
 \mathcal{U}^{\ti t}_{21}\mathcal{U}^{\ti t *}_{11} \de
 Y_{\ti t} +  \mathcal{U}^{\ti
   t}_{11}\mathcal{U}^{\ti t *}_{21} \braket{\de
   Y_{\ti t}}^*-2 m_t \de m_t \\ 
 \de m_{\ti t_R}^2&=& \abs{\mathcal{U}^{\ti t}_{12}}^2 \de m_{\ti
   t_1}^2+\abs{\mathcal{U}^{\ti t}_{22}}^2 \de m_{\ti t_2}^2 +
 \mathcal{U}^{\ti t}_{22}\mathcal{U}^{\ti t*}_{12} \de
 Y_{\ti t} +  \mathcal{U}^{\ti
   t}_{12}\mathcal{U}^{\ti t*}_{22} \braket{\de
   Y_{\ti t}}^*-2 m_t \de m_t \,.
\eeq
\end{itemize}

\subsection{Tools and Checks \label{sec:toolschecks}}
We performed two independent calculations of the two-loop NMSSM Higgs
boson masses at $\order$ and cross-checked the results against each
other. In both cases, we used {\texttt{SARAH 4.14.0}}
\cite{Staub:2009bi, Staub:2010jh, Staub:2012pb, Staub:2013tta} to
generate a model file for the complex NMSSM that was used in
{\texttt{FeynArts 3.8}} \cite{Kublbeck:1990xc, Hahn:2000kx} to
generate all one- and two-loop Feynman diagrams necessary for the
calculation of the mass corrections. For the simplification of
expressions containing the Dirac matrices as well as for the calculation
of fermion traces, we used {\texttt{FeynCalc 8.2.0}} \cite{FeynCalc,
  SHTABOVENKO2016432}. The tensor reduction of the one- and two-loop
integrals was also performed in {\texttt{FeynCalc}}, where for the
two-loop case, we additionally used the {\texttt{TARCER}}
\cite{MERTIG1998265} extension of {\texttt{FeynCalc}}. \s

As additional cross-checks for the calculation of the fermion traces,
we used the {\texttt{Mathematica}} package {\texttt{FormTracer}}
\cite{CYROL2017346}. The fact that {\texttt{FormTracer}} treats
the Dirac matrix $\gamma _5$ in the \textit{Larin scheme}
\cite{LARIN1993113}, while {\texttt{FeynCalc 8.2.0}} treats $\gamma_5$
per default in the 'naive scheme' \cite{Jegerlehner2001}, allowed us
to compute the fermion traces in the 
framework of the two different $\gamma_5$ schemes. For the traces relevant for
our $\order$ computations, we explicitly verified that a change of the $\gamma_5$
scheme does not affect the calculation of the NMSSM Higgs boson masses
at $\order$.\s  

We have compared the results of the self-energies
  at $\mathcal O(\alpha_t^2)$ with the corresponding self-energies in
  {\tt FeynHiggs}~\cite{Hollik:2014bua,Hollik:2015ema,Heinemeyer:1998yj,Heinemeyer:1998np,
Degrassi:2002fi,Frank:2006yh,Hahn:2013ria,Bahl:2016brp,Bahl:2017aev}.
 Since the $\mathcal O(m_t^2 \alpha_t^2)$ corrections are equivalent
 in the MSSM and the NMSSM with $\mu$ in the MSSM identified as
 $\mu^\text{eff}$ in the NMSSM and with further contributions of
 $\mathcal O(\alpha_t^2)$ neglected, we found agreement after 
  adapting the counterterms for the weak mixing angle and the vacuum
  expectation value to the renormalization scheme applied in {\tt
    FeynHiggs}, see for example \cite{Drechsel:2016htw}, thus ensuring
  the same input values. In addition, we also compared to the on-shell
  and $\overline{\text{DR}} $ results for real parameters and the mass
  of the pseudoscalar Higgs boson $M_A$ as input of
  Ref.~\cite{Brignole:2001jy}, using the corresponding computer code,
  and found also complete agreement.

\section{Numerical Analysis \label{sec:analysis}}
\subsection{The Parameter Scan}
\begin{table*}
\begin{center}
{\small \begin{tabular}{l|cccccccccccccc} \toprule
& $M_1$ & $M_2$ & $A_t$ &
$A_b$ & $A_\tau$ & $m_{\tilde{Q}_3}$ & $m_{\tilde{t}_R}$ &  $m_{\tilde{b}_R}$ & $m_{\tilde{L}_3}$ 
& $m_{\tilde{\tau}_R}$ & $M_{H^\pm}$ & $A_\kappa$ & $\mu_{\text{eff}}$ \\ 
& \multicolumn{13}{|c}{in TeV} \TBstrut \\ \hline 
min & 0.4 & 0.4 & -2.0 & -2.0 & -2.0 & 0.4 & 0.4
& 2.0 & 0.4 & 0.4 & 0.2 & -2.0 & 0.2 \Tstrut \\
max & 1.0 & 1.0 & 2.0 & 2.0 & 2.0 & 3.0 & 3.0 & 3.0
& 3.0 & 3.0 & 1.0 & 2.0 & 0.3 \Bstrut \\ \bottomrule
\end{tabular}}
\caption{Scan ranges for the NMSSM scan, all parameters varied
  independently between the given minimum and maximum
  values. \label{tab:nmssmscan}} 
\end{center}
\end{table*}
For our numerical analysis we perform a scan in the NMSSM parameter
space in order to find scenarios that are compatible with the recent
experimental constraints. We proceed as described
in~\cite{Costa:2015llh,King:2014xwa,Azevedo:2018llq}, where also
further details can be found. We vary $\tan\beta$, $\lambda$ and
$\kappa$ in the ranges
\beq
1.5 \le \tan\beta \le 10 \;, \quad 10^{-4} \le \lambda \le 0.4 \;, \quad 0
\le \kappa \le 0.6 \;,
\eeq
thus not exceeding the rough constraint 
\beq
\lambda^2 + \kappa^2 < 0.7^2 \;,
\eeq
which ensures perturbativity.
Further parameter scan ranges are summarized in Tab.~\ref{tab:nmssmscan}. We set
\beq
M_3 = 1.85 \mbox{ TeV}
\eeq
and the mass parameters of the first and second generation sfermions
are chosen as
\begin{equation}   
m_{\tilde{u}_R,\tilde{c}_R} = 
m_{\tilde{d}_R,\tilde{s}_R} =
m_{\tilde{Q}_{1,2}}= m_{\tilde L_{1,2}} =m_{\tilde e_R,\tilde{\mu}_R}
= 3\;\mbox{TeV} \;. \label{eq:lightsquatmasses}
\end{equation}
The soft SUSY breaking trilinear couplings of the first two
generations are set equal to the corresponding values of the third
generation. Following the SLHA format~\cite{Skands:2003cj,Allanach:2008qq}, the 
soft SUSY breaking masses and trilinear couplings are understood as
$\DRb$ parameters at the scale 
\be 
\mu_R = M_{\text{SUSY}} = \sqrt{m_{\ti Q_3} m_{\ti t_R}} \;. 
\ee
The SM input parameters have been chosen
as~\cite{PhysRevD.98.030001,Dennerlhcnote} 
\begin{equation}
\begin{tabular}{lcllcl}
\quad $\alpha(M_Z)$ &=& 1/127.955, &\quad $\alpha^{\MSb}_s(M_Z)$ &=&
0.1181 \\
\quad $M_Z$ &=& 91.1876~GeV &\quad $M_W$ &=& 80.379~GeV  \\
\quad $m_t$ &=& 172.74~GeV &\quad $m^{\MSb}_b(m_b^{\MSb})$ &=& 4.18~GeV \\
\quad $m_c$ &=& 1.274~GeV &\quad $m_s$ &=& 95.0~MeV \\
\quad $m_u$ &=& 2.2~MeV &\quad $m_d$ &=& 4.7~MeV \\
\quad $m_\tau$ &=& 1.77682~GeV &\quad $m_\mu$ &=& 105.6584~MeV  \\
\quad $m_e$ &=& 510.9989~keV &\quad $G_F$ &=& $1.16637 \cdot 10^{-5}$~GeV$^{-2}$\,.
\end{tabular}
\end{equation}

The spectrum of the Higgs particles including the higher-order
corrections presented in this work is calculated with the new {\tt NMSSMCALC}
version which includes the higher-order corrections of ${\cal
  O}(\alpha_t^2)$ calculated in this paper. For the scan, $M_{H^\pm}$
has been used as input parameter, {\it
  cf.}~Eq.~(\ref{eq:inputset1}).\footnote{Note, however,
    that in {\tt NMSSMCALC} we also have the option to set $A_\lambda$
    as input parameter which is then renormalized in the
    $\overline{\mbox{DR}}$ scheme.}
 {\tt NMSSMCALC} also checks for the constraints from the electric
 dipole moments (EDMs) that become relevant for the CP-violating case
 \cite{King:2015oxa}. One of the neutral CP-even Higgs
 bosons is identified with the SM-like Higgs boson and its mass is
 required to lie in the range  
\beq
122 \mbox{ GeV } \le m_h \le 128 \mbox{ GeV} \;.
\eeq
Agreement with the Higgs exclusion limits from LEP, Tevatron and LHC
is checked by using {\tt HiggsBounds 5.3.2}
\cite{Bechtle:2008jh,Bechtle:2011sb,Bechtle:2013wla}, and with the
Higgs rates by using {\tt
  HiggsSignals 2.2.3}~\cite{Bechtle:2013xfa}. 
We demand the total $\chi^2$ computed by {\tt HiggsSignals} with our
given effective coupling factors to be compatible 
with the total  $\chi^2$ of the SM within 1$\sigma$.
The required input for {\tt HiggSignals} is computed with {\tt NMSSMCALC}. \s

We also take into account the most relevant LHC
exclusion bounds on the SUSY 
masses. These constrain the gluino mass and the lightest squark mass of
the second generation to lie above 1.8~TeV, see~\cite{Aad:2015iea}. The
stop and sbottom masses in general have to be above 800~GeV~\cite{Aaboud:2016lwz, 
  Aad:2015iea}, and the slepton masses above
400~GeV~\cite{Aad:2015iea}.\s 

We perform the scan in the limit of the CP-conserving NMSSM. For the numerical
analysis we start from a valid parameter point and subsequently turn
on various CP-violating phases. The thus obtained parameter points do not
necessarily fulfill the constraints from the EDMs any more but nevertheless, we
keep them for illustrative purposes. The strongest constraint
originates from the electron EDM~\cite{Inoue:2014nva}. We check the
EDMs of our parameter points against the experimental limit given by the ACME
collaboration~\cite{Andreev:2018ayy}.

\subsection{Results}
For our numerical analysis we have chosen two sample points among the 
parameter points compatible with all described constraints that we
obtained from our scan. They both feature a
SM-like Higgs boson with mass around 125~GeV at ${\cal O} (\alpha_t
\alpha_s + \alpha_t^2)$ when - in one case - the top/stop sector is
renormalized OS and - in the other case - the top/stop sector is
$\overline{\mbox{DR}}$ renormalized. We call the former point P1OS,
the latter P2DR. In the following we give the relevant input values for these
two points. Note that we deliberately chose parameter points with not too large
NMSSM-specific coupling values $\lambda$ and $\kappa$ as at two-loop order we
so far do not include the corrections proportional to these couplings.
This is left for future work. Since we include the 
complete set of one-loop corrections, however, our results for these
parameter points should not be affected significantly by the missing
corrections. This can also be inferred from the results given in \cite{Staub:2015aea}.\s

\paragraph{Parameter Point P1OS:}
Besides the SM values defined above, the parameter point is given by
the following soft SUSY breaking masses and trilinear couplings, 
\beq
&&  m_{\tilde{u}_R,\tilde{c}_R} = 
m_{\tilde{d}_R,\tilde{s}_R} =
m_{\tilde{Q}_{1,2}}= m_{\tilde L_{1,2}} =m_{\tilde e_R,\tilde{\mu}_R} = 3\;\mbox{TeV}\, , \;  
m_{\tilde{t}_R}=881\,\gev \,,\; \non \\ \non
&&  m_{\tilde{Q}_3}=1226\,\gev\,,\; m_{\tilde{b}_R}=2765\,\gev\,,\; 
m_{\tilde{L}_3}=1369\,\gev\,,\; m_{\tilde{\tau}_R}=2967\,\gev\,,
 \\ 
&& |A_{u,c,t}| = 1922\,\gev\, ,\; |A_{d,s,b}|=1885\,\gev\,,\; |A_{e,\mu,\tau}| = 1170\,\gev\,,\; \\ \non
&& |M_1| = 644\,\gev,\; |M_2|= 585\,\gev\,,\; |M_3|=1850\,\gev \;,
\label{eq:param1}
\eeq
with the CP phases given by
\beq
&&  \varphi_{A_{u,c,t}}=\varphi_{A_{d,s,b}}=\pi\,,\; 
\varphi_{A_{e,\mu,\tau}}=\varphi_{M_1}=\varphi_{M_2}=\varphi_{M_3}=0 
\;. \label{eq:param2}
\eeq
The remaining input parameters have been set to\footnote{The
  imaginary part of $A_\kappa$ is obtained from the tadpole
  condition.}
\beq
&& |\lambda| = 0.301 \;, \quad |\kappa| = 0.299 \; , \quad \mbox{Re}(A_\kappa) = -791\,\gev\;,\quad 
|\mu_{\text{eff}}| = 209\,\gev \;, \non \\ 
&&\varphi_{\lambda}=\varphi_{\kappa}=\varphi_{\mu_{\text{eff}}}=\varphi_u=0\;,
\quad  \tan\beta = 4.44 \;,\quad M_{H^\pm} = 898 \,\gev \;.
\eeq
As required by the SLHA, $\mu_{\text{eff}}$ is taken as input
parameter, from which $v_s$ and $\varphi_s$ are obtained through
Eq.~(\ref{eq:effectiveMuParameter}). The parameters $\lambda, \kappa,
A_\kappa, \mu_{\text{eff}}, \tan\beta$ as well as the soft 
SUSY breaking masses and trilinear couplings are understood as $\DRb$
parameters at the scale $\mu_R = M_{\text{SUSY}}$\footnote{For $\tan\beta$ this is
only the case if it is read in from the block EXTPAR as done in {\tt
  NMSSMCALC}. Otherwise it is the $\DRb$ parameter at the scale
$M_Z$.}. The charged Higgs mass, however, is 
an OS parameter. The SUSY scale $M_{\text{SUSY}}$ is set to be
\beq
M_{\text{SUSY}} = \sqrt{m_{\tilde Q_3}m_{\tilde t_R}} \;.
\eeq
In the following the subscript '$\text{eff}$' for $\mu$ is
dropped and we use the expressions OS and $\DRb$ in order to 
refer to the chosen renormalization conditions in the top/stop sector
only, while all the other renormalization conditions remain
unchanged. 

\paragraph{Parameter Point P2DR:}
Besides the SM values defined above, the parameter point is given by
the following soft SUSY breaking masses and trilinear couplings, 
\beq
&&  m_{\tilde{u}_R,\tilde{c}_R} = 
m_{\tilde{d}_R,\tilde{s}_R} =
m_{\tilde{Q}_{1,2}}= m_{\tilde L_{1,2}} =m_{\tilde e_R,\tilde{\mu}_R} = 3\;\mbox{TeV}\, , \;  
m_{\tilde{t}_R}=1247\,\gev \,,\; \non \\ \non
&&  m_{\tilde{Q}_3}=1353\,\gev\,,\; m_{\tilde{b}_R}=3\;\mbox{TeV}\,,\; 
m_{\tilde{L}_3}=3\;\mbox{TeV}\,,\; m_{\tilde{\tau}_R}=3\;\mbox{TeV}\,,
 \\ 
&& |A_{u,c,t}| = 2987\,\gev\, ,\; |A_{d,s,b}|=753\,\gev\,,\; |A_{e,\mu,\tau}| = 173\,\gev\,,\; \\ \non
&& |M_1| = 614\,\gev,\; |M_2|= 528\,\gev\,,\; |M_3|=1850\,\gev \,,\\ \non
&&  \varphi_{A_{u,c,t}}=\varphi_{A_{d,s,b}}=\varphi_{A_{e,\mu,\tau}}=0
= \varphi_{M_1}=\varphi_{M_2}=\varphi_{M_3}=0
 \;. \label{eq:param3}
\eeq
The remaining input parameters have been set to
\beq
&& |\lambda| = 0.096 \;, \quad |\kappa| = 0.372 \; , \quad \mbox{Re}(A_\kappa) = -61.8\,\gev\;,\quad 
|\mu_{\text{eff}}| = 237\,\gev \;, \non \\ 
&&\varphi_{\lambda}=\varphi_{\kappa}=\varphi_{\mu}=\varphi_u=0\;,
\quad \tan\beta = 9.97 \;,\quad M_{H^\pm} = 793 \,\gev \;.
\eeq
%
\subsection{Results and Analysis Parameter Point P1OS \label{subsec:p1os}} 
In Tab.~\ref{tab:massP1OS} we
summarize the values of the masses that we obtain for the chosen
P1OS at tree level, at one-loop level, at two-loop level including
only the ${\cal O}(\alpha_t \alpha_s)$ corrections and at two-loop
level including furthermore our newly calculated ${\cal
  O}(\alpha_t^2)$ corrections. In Tab.~\ref{tab:massP1DR} the results
are given for the $\DRb$ scheme in the top/stop sector. The tables
also contain the information on the main singlet/doublet and
scalar/pseudoscalar component of the respective mass eigenstate. The
tree-level stop masses obtained within the OS and $\DRb$ scheme are
given by
\beq
\begin{array}{llll}
\mbox{OS} &:& m_{\tilde{t}_1} = 811 \mbox{ GeV} \;, \qquad &
m_{\tilde{t}_2} = 1276 \mbox{ GeV} \;, \\
\DRb &:& m_{\tilde{t}_1} = 837 \mbox{ GeV} \;, \qquad & m_{\tilde{t}_2} = 1271      
\mbox{ GeV} \;.
\end{array}
\eeq
The $\DRb$ top quark mass in our scenario amounts to $m_t^{\DRb} =
141.8$~GeV, and has been computed as described in App.~\ref{append:mtrun}.\s 

\begin{table}[t]
\begin{center}
 \begin{tabular}{|l||c|c|c|c|c|}
\hline
 &${H_1}$&${H_2}$&${H_3}$&${H_4}$&${H_5}$\\ \hline \hline
tree-level &74.29&91.43&704.12&895.91&897.83\\ 
main component&$h_s$&$h_u$&$a_s$&$a$&$h_d$\\ \hline  
one-loop &86.58 & 135.0 & 700.03 & 895.83 & 897.83\\ 
main component&$h_s$&$h_u$&$a_s$&$a$&$h_d$\\ \hline  
two-loop ${\cal O}(\alpha_t \alpha_s)$ &86.16 & 118.11 & 700.04 & 895.83 & 897.76 \\  
main component&$h_s$&$h_u$&$a_s$&$a$&$h_d$\\ \hline
two-loop ${\cal O}(\alpha_t \alpha_s+ \alpha_t^2)$
 &86.35 & 125.05 & 700.04 & 895.83 & 897.79\\  
main component&$h_s$&$h_u$&$a_s$&$a$&$h_d$\\ \hline
\end{tabular}
\caption{P1OS: Mass values in GeV and main components of the neutral Higgs
  bosons at tree-level, one-loop, two-loop ${\cal O}(\alpha_t
  \alpha_s)$ and at two-loop ${\cal O}(\alpha_t \alpha_s +
  \alpha_t^2)$ obtained by using OS renormalization in the top/stop sector.}
\label{tab:massP1OS}
\end{center}
\end{table}
\begin{table}[t!]
\begin{center}
 \begin{tabular}{|l||c|c|c|c|c|}
\hline
 &${H_1}$&${H_2}$&${H_3}$&${H_4}$&${H_5}$\\ \hline \hline
tree-level &74.29&91.43&704.12&895.91&897.83\\
main component&$h_s$&$h_u$&$a_s$&$a$&$h_d$\\ \hline
one-loop &85.93 & 112.77 & 700.05 & 895.79 & 897.71 \\ 
main component&$h_s$&$h_u$&$a_s$&$a$&$h_d$\\ \hline
two-loop ${\cal O}(\alpha_t \alpha_s)$ &86.21 & 118.62 & 700.04 & 895.78 & 897.73 \\ 
main component&$h_s$&$h_u$&$a_s$&$a$&$h_d$\\ \hline
two-loop ${\cal O}(\alpha_t \alpha_s+ \alpha_t^2)$
 &86.22 & 119.1 & 700.04 & 895.78 & 897.73\\                 
main component&$h_s$&$h_u$&$a_s$&$a$&$h_d$\\ \hline
\end{tabular}
\caption{P1OS: Mass values in GeV and main components of the neutral Higgs
  bosons at tree-level, one-loop, two-loop ${\cal O}(\alpha_t
  \alpha_s)$ and at two-loop ${\cal O}(\alpha_t \alpha_s +
  \alpha_t^2)$ obtained using $\drbar$ renormalization in the top/stop sector.}
\label{tab:massP1DR}
\end{center}
\end{table}
The scenario features three heavy Higgs bosons with masses between 700
and 900~GeV. Additionally, we have a light Higgs boson with mass value
below 125~GeV. For the correct interpretation of the importance of the
loop corrections, the Higgs bosons with a similar admixture must be
compared and not the ones according to their mass
ordering. In the following plots we therefore label the Higgs bosons according to
their dominant admixture and not by their mass ordering. However, in the
scenario P1OS both orderings lead to the same result. 
The admixtures of the Higgs bosons are detailed in the Tables. While
the dominance is very pronounced at loop level, at tree level,
for $H_1$ and $H_2$ the $h_s$ and $h_u$ admixtures are
approximately the same, respectively. For $H_1$, however, the $h_s$
admixture is larger, and for $H_2$ the $h_u$ admixture is larger. 
In order to comply with the Higgs rate measurements of the LHC the SM-like Higgs
boson has to be $h_u$ dominated, as is the case for the 125~GeV Higgs
boson in our scenario (at ${\cal O}(\alpha_t \alpha_s + \alpha_t^2)$ with OS
renormalization in the top/stop sector). \s

Defining the absolute value of the relative change in the
  mass value when going from loop order $a$ to loop order $b$
  including the next level of corrections, as $|m^b-m^a|/m^a$, we see that 
the lightest singlet-like Higgs boson $H_1$ receives rather large
one-loop corrections of ${\cal O}(16\%)$. The two-loop corrections
are below the per-cent level. The $h_u$-dominated Higgs boson $H_2$ 
receives important one-loop corrections of ${\cal O}(48\%)$ in the OS scheme, 
respectively ${\cal O}(23\%)$ in the $\DRb$ scheme. The two-loop
${\cal O}(\alpha_t \alpha_s)$ corrections reduce the mass value by $12\%$
in the OS scheme and add 5\% in the $\DRb$ scheme so that the 
mass values in the two renormalization schemes move close to each other
with values of $\sim 118$~GeV. The two-loop ${\cal O}(\alpha_t
\alpha_s +\alpha_t^2)$ correction adds another 6\% to the ${\cal
  O}(\alpha_t \alpha_s)$ result in the OS scheme
while in the $\DRb$ scheme this loop correction barely alters the mass
value, so that at ${\cal O}(\alpha_t \alpha_s +\alpha_t^2)$ the mass
values in the two renormalization schemes move further away from each
other.  \s

\begin{figure}[t]
  \includegraphics[width=0.5\textwidth]{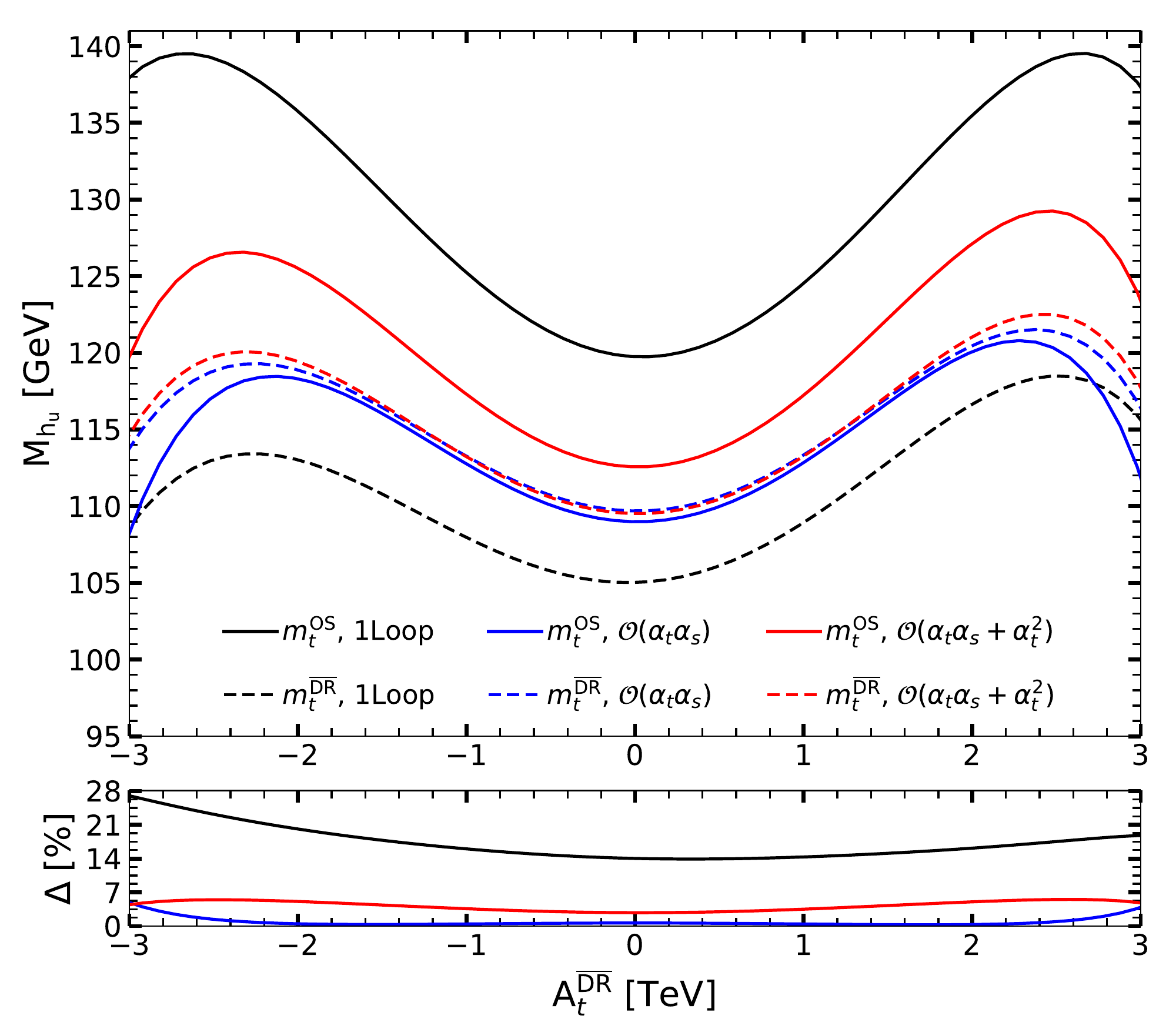}\includegraphics[width=0.5\textwidth]{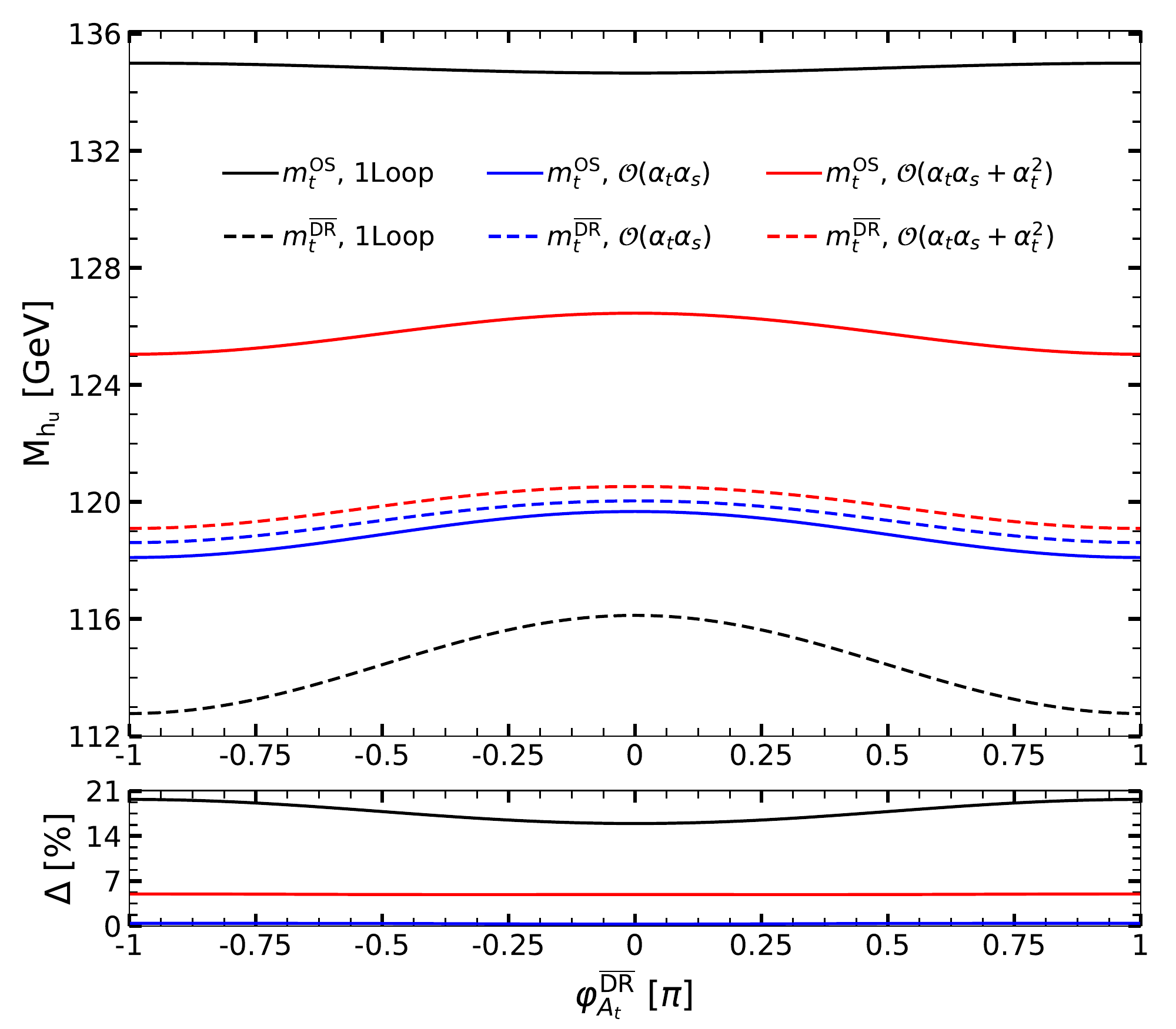}
\caption{Upper Panel: The mass of the $h_u$-like Higgs boson at
  one-loop level (black/two outer lines) and at two-loop level including the
  ${\cal O}(\alpha_t \alpha_s)$ corrections (blue/two lower middle lines) 
  and the ${\cal O}(\alpha_t \alpha_s+\alpha_t^2)$ corrections
  (red/two upper middle lines) as a function of $A_t^{\DRb}$ (left) and
  $\varphi_{A_t}^{\DRb}$ (right). Solid lines: OS, dashed lines:
  $\DRb$ renormalization in the top/stop sector. Lower Panel:
  Absolute value of the relative deviation of the result with OS renormalization in the top and stop sector with respect to the result using a $\DRb$ scheme --
  {\it i.e.} $\Delta=|M_{{h_u}}^{m_t(\drbar)}-M_{{h_u}}^{m_t({\tiny
      \mbox{OS}})}|/M_{{h_u}}^{m_t({\tiny \drbar})}$ -- in percent as a
  function of $A_t^{\DRb}$ (left) and $\varphi_{A_t}^{\DRb}$ at
  two-loop ${\cal O}(\alpha_t \alpha_s)$ (blue/lower line), at two-loop ${\cal
    O}(\alpha_t \alpha_s+\alpha_t^2)$  
  (red/middle line) and one-loop order (black/upper
  line).}
\label{fig:onelooptwoloop}
\end{figure}
In Fig.~\ref{fig:onelooptwoloop}, we display the one-loop
and two-loop corrected mass values $M_{h_u}$ of the $h_u$-like Higgs boson including
at two-loop level the ${\cal O} (\alpha_t \alpha_s)$ and the ${\cal O} (\alpha_t
\alpha_s + \alpha_t^2)$ corrections, as a function of the
$\DRb$ parameters $A_t$ (left) and $\varphi_{A_t}$ (right) for both OS
(full lines) and $\DRb$ (dashed lines) renormalization in the top/stop
sector. Starting from our initial parameter point P1OS, we vary $A_t$
in the left plot (keeping $\varphi_{A_t}$ unchanged) and for the right
plot we vary $\varphi_{A_t}$ while $A_t$ is fixed at its initial absolute
value. The point P1OS corresponds to the values at
$A_t^{\DRb} = -1922$~GeV (left) and vanishing CP-violating
  phase $\sin\varphi_{A_t}=0$, which corresponds to $\varphi_{A_t}=\pm \pi$
in the right plot. The two lower plots show the relative difference in
the masses (at the same loop order, which is
  indicated by the colour of the lines) obtained using the two 
renormalization schemes in the top/stop sector,
\beq
\Delta=\frac{|M_{h_u}^{m_t(\drbar)}-M_{h_u}^{m_t({\tiny
      \mbox{OS}})}|}{M_{h_u}^{m_t({\tiny \drbar})}} \;, \label{eq:renschemechange}
\eeq 
as a function of $A_t$ and $\varphi_{A_t}$, respectively. 
We show the corrections for the $h_u$-like Higgs boson mass, as it is affected
most by the ${\cal O} (\alpha_t\alpha_s)$ and ${\cal O}(\alpha_t^2)$ corrections. 
The plots confirm what the discussion of P1OS already revealed. For the $\DRb$
renormalization scheme the two-loop corrections of ${\cal
  O}(\alpha_t \alpha_s)$ do not change the one-loop result as much as
in the OS scheme. This behaviour can be understood as follows. The
$\DRb$ renormalization in the top/stop sector requires the conversion
of the input OS top-quark mass to the $\DRb$ mass at the scale $\mu_R
= M_{\text{SUSY}}$. This conversion is described in
App.~\ref{append:mtrun}. It includes the ${\cal O}(\alpha_s + \alpha_t
+ \alpha_s^2)$ contributions in the conversion of the
SM OS top-quark mass to the $\overline{\mbox{MS}}$ mass at $\mu_R =
M_Z$ and the ${\cal O}(\alpha_s+\alpha_t)+{\cal O}((\alpha_s
+\alpha_t)^2)$ corrections in the renormalization group equations
needed for the running from $\mu_R = M_Z$ to $\mu_R=
M_{\text{SUSY}}$. In this way the one-loop mass in the $\DRb$
renormalization scheme already includes higher-order corrections
beyond the one-loop level.
Furthermore, the ${\cal O}(\alpha_t \alpha_s +
\alpha_t^2)$ corrections barely change the ${\cal
  O}(\alpha_t \alpha_s)$ mass value for the $\DRb$ scheme, whereas the
OS renormalization leads to a further change of a few GeV, so that
the relative difference in the mass values at ${\cal O}(\alpha_t \alpha_s +
\alpha_t^2)$ with values of 5-6\% is larger compared to the relative
difference at ${\cal O}(\alpha_t \alpha_s)$ with values below 4-5\%
down to almost 0\%. As can be read off from the plots, the
effect on the Higgs mass values due to the change of the
renormalization scheme becomes more pronounced when moving from ${\cal
O} (\alpha_t \alpha_s)$ to ${\cal O}(\alpha_t \alpha_s +
\alpha_t^2)$, increasing thus the estimate of the remaining
theoretical uncertainty due to missing higher-order corrections from
the change of the renormalization scheme. At first sight, this might
be counter-intuitive. Both types of corrections are of two-loop order,
however. A reduction in the theoretical uncertainty cannot necessarily be
expected when further contributions at the same loop level are taken
into account. With the loop order included in our conversion of the
parameters, we estimate an uncertainty due to missing corrections of three-loop order ${\cal O} (\alpha_t^2
\alpha_s + \alpha_t \alpha_s^2)$ and of four-loop order ${\cal O} (\alpha_t^3
\alpha_s + \alpha_t^2 \alpha_s^2 + \alpha_s^3 \alpha_t)$ contributions 
when we include the ${\cal O}(\alpha_t \alpha_s)$ corrections. The
inclusion of the ${\cal O}(\alpha_t \alpha_s +
\alpha_t^2)$ corrections provides an estimate of the missing
contributions as before but additionally also of terms of the order ${\cal O}(\alpha_t^3)$ and ${\cal O}(\alpha_t^4)$. This shows that care has to be taken 
in the estimate of the uncertainty due to missing
higher-order corrections at a given loop-level when only parts of the
loop contributions at this given order are included. An estimate based
on the ${\cal O}(\alpha_t \alpha_s)$ corrections alone would be more
optimistic than after the inclusion of additionally the ${\cal O}(\alpha_t^2)$
corrections. \s

Note also that the one-loop corrections in the OS scheme are
almost symmetric 
with respect to the sign change of $A_t$ in contrast to the $\DRb$
scheme. This behaviour results from the threshold effect in the
conversion of the top OS to the $\DRb$ mass, which depends on the sign
of $A_t$. In contrast, the sign dependence in the conversion of the
$\DRb$ stop parameters to OS parameters for the OS scheme almost
cancels out.
The dependence of the loop-corrected mass values with varying $\varphi_{A_t}$ is on the one hand due to the genuine dependence on
the phase and on the other hand due to the dependence of the stop mass values
on the phase. The stronger dependence of the one-loop
$\DRb$ mass values compared to the higher-order and OS values again is
due to the necessary conversion from the OS to the $\DRb$ top mass
value in this renormalization scheme. Finally, we want to remark that the
non-zero $\sin\varphi_{A_t}$ may lead to scenarios that are not
compatible with the EDMs any more.\footnote{Actually, a
  check with {\tt NMSSMCALC} showed that the EDM constraints are not
  fulfilled anymore for a non-zero phase $|\varphi_{A_t}|\gsim
  0.08\pi$ (no additional CP violation from other phases).} We still
keep these scenarios in 
the plot for illustrative purposes. \s

\begin{figure}[t]
 \includegraphics[width=0.5\textwidth]{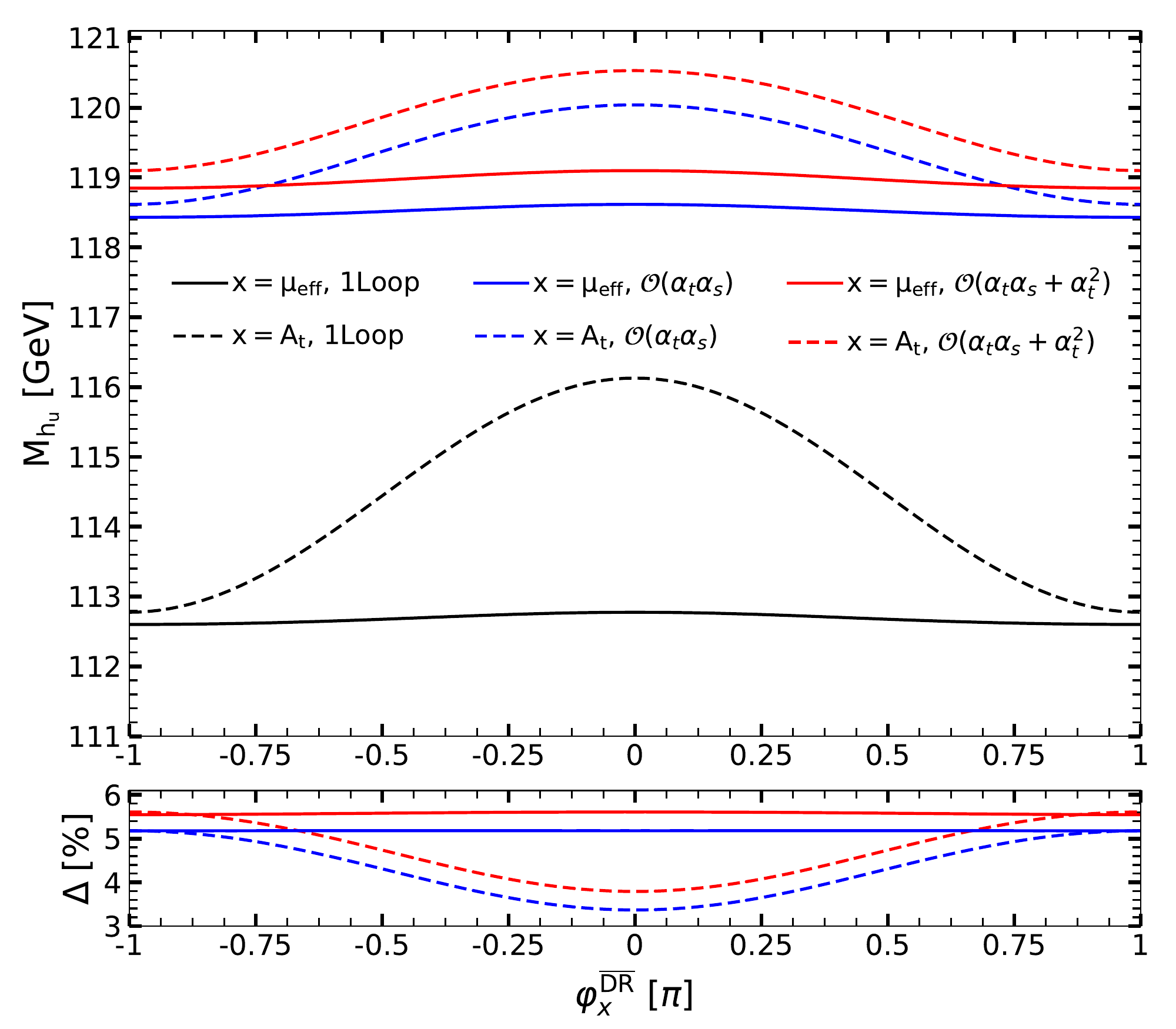} \includegraphics[width=0.5\textwidth]{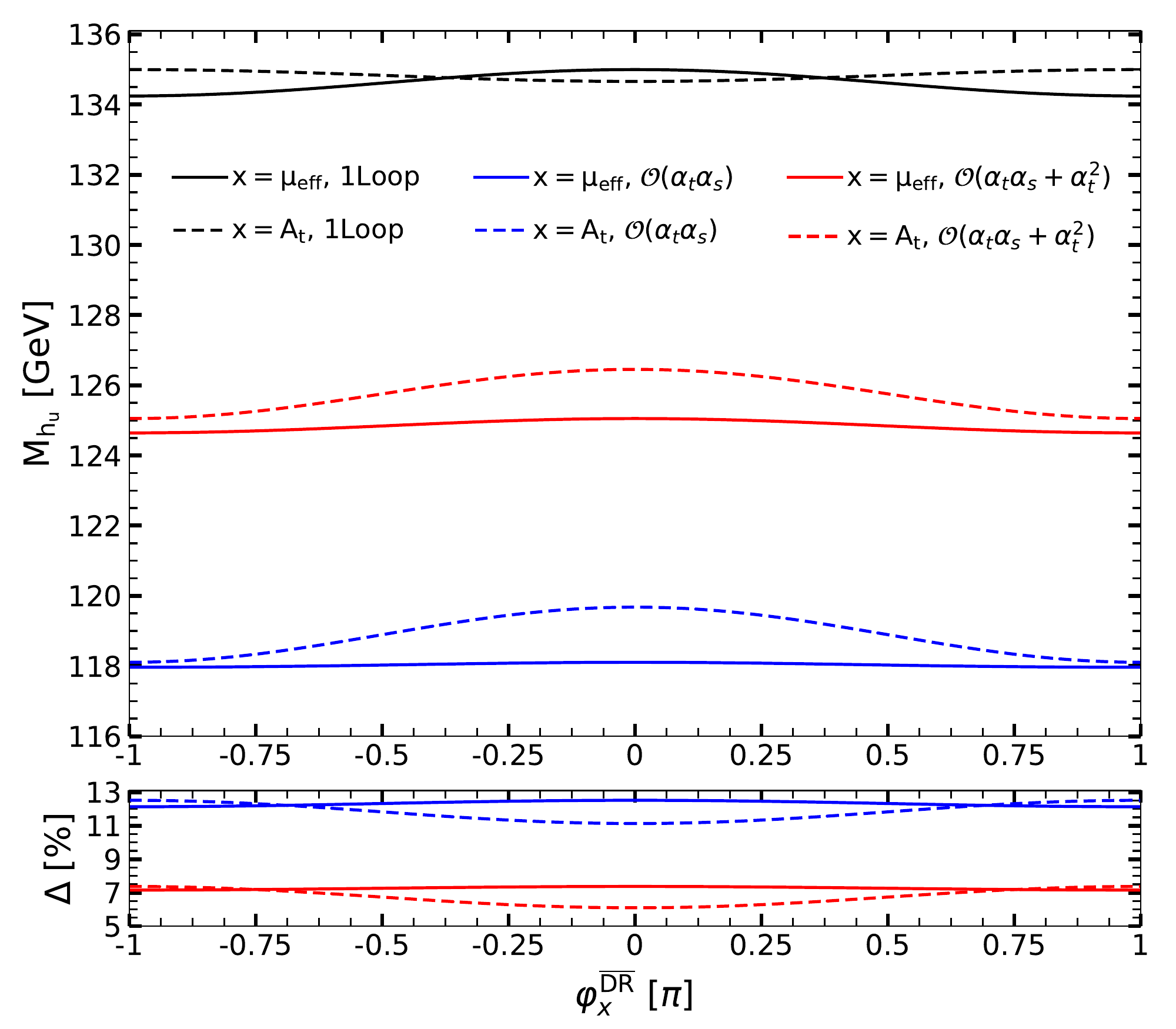}
\caption{Upper Panels: One-loop (black lines), two-loop ${\cal
    O}(\alpha_t \alpha_s)$ (blue lines) and two-loop ${\cal
    O}(\alpha_t \alpha_s + \alpha_t^2)$ (red lines) mass values
  $m_{h_u}$ of the SM-like Higgs boson as a function of the phases
  $\varphi_{\mu}$ (full lines) and 
  $\varphi_{A_t}$ (dashed lines). Lower Panels: Absolute value of the
  relative two-loop corrections to the mass of the SM-like Higgs boson
  with respect to the one-loop order -- {\it  
    i.e.}~$\Delta=|M_{{h_u}}^{(2,x)}-M_{{h_u}}^{(1)}|/M_{{h_u}}^{(1)}$
  -- in percent as a function of the phases $\varphi_{\mu}$ (full),
  and $\varphi_{A_t}$ (dashed) for $x={\cal O} (\alpha_t \alpha_s)$
  (blue lines) and $x={\cal O} (\alpha_t \alpha_s+\alpha_t^2)$ (red
  lines). Left: $\drbar$, right: OS scheme in the top/stop
  sector.}
\label{fig:phasevariation}
\end{figure}

Figure~\ref{fig:phasevariation} shows the one-loop and two-loop 
corrections of ${\cal O}(\alpha_t \alpha_s)$ and ${\cal O}(\alpha_t \alpha_s +
\alpha_t^2)$, respectively, to the $h_u$-like Higgs boson mass as a
function of the phases $\varphi_{A_t}$ and $\varphi_{\mu}$ where the latter denotes the phase of $\mu _\text{eff}$. The left
plot shows results obtained for the $\DRb$ renormalization scheme in
the top/stop sector, the right plot those for the OS scheme. Starting
from the above defined parameter point, we turn on separately one of
the two phases. For illustrative purposes we vary the phases also
beyond values already excluded by experiment.\footnote{The EDM constraints are not
  fulfilled any more for a non-zero phase $|\varphi_\mu| \gsim (9.5 \cdot
  10^{-10}) \pi$ (no additional CP violation from other phases).}  
For the plots, we have varied $\varphi_{\mu}$ in such a way that
the CP-violating phase, which appears already at tree level in the Higgs sector,
{\it i.e.}~$\varphi_y = \varphi_{\kappa} -\varphi_{\lambda} + 2\varphi_{s}-\varphi_{u}$, 
remains zero. For this, the phases $\varphi_{\lambda}$ and $\varphi_{s}$ were varied
at the same time, in particular we set $\varphi_\lambda= 2\varphi_s= 2/3
\varphi_\mu$. The phases $\varphi_\kappa$ and $\varphi_u$ were kept
zero ($|A_\kappa|$ is kept constant). We thus make sure that the main effect
in the plots on the dependence on the phases originates from the loop corrections.
\s

As can be inferred from the plots, for both schemes the shape of the
variation of the two-loop corrections with the phases is very
similar. The dependence on the phase $\varphi_{A_t}$ is
stronger than the one on $\varphi_\mu$. Overall, the influence of the
investigated complex phases 
on the loop corrections is quite small as we study purely
radiatively induced CP violation here. The effect of the phases on the mass
values is at most 3\% at one-loop and below the percent level at
two-loop order, and much smaller than the overall mass corrections at
each loop level. \s 

The conversion of the OS top-quark mass to the $\DRb$ mass in the
$\DRb$ renormalization scheme resums higher-order corrections into the 
fixed-order calculation, explaining the large difference between the
one-loop $\DRb$ and OS results, and the better convergence when going
from the one- to the two-loop level in the $\DRb$ scheme. Thus
Fig.~\ref{fig:phasevariation} shows that the
absolute relative corrections between the two loop levels, defined,
for a fixed renormalization scheme, as
\beq
\Delta=\frac{|M_{{h_u}}^{(2,x)}-M_{{h_u}}^{(1)}|}{M_{{h_u}}^{(1)}} \;,
\eeq
with the superscripts \{2, 1\} referring to the two- and one-loop level, respectively, and the superscript
$x$ to the two-loop ${\cal O}(\alpha_t \alpha_s)$ and 
${\cal O}(\alpha_t \alpha_s + \alpha_t^2)$ corrections, are larger in
the OS than in the $\DRb$ renormalization scheme. For the ${\cal
  O}(\alpha_t \alpha_s)$ corrections they amount to about 3-5\% in the
latter case, whereas the masses are reduced by 11-12\% in the former case. The ${\cal
  O}(\alpha_t \alpha_s + \alpha_t^2)$ corrections make up for a relative
correction between 3.7 and 5.5\% in the $\DRb$ scheme and between -6
and -7\% in the OS scheme with respect to the one-loop case. Overall, the two-loop corrections reduce the
one-loop masses in the OS scheme whereas they are positive in the
$\DRb$ scheme. As already discussed and commented on above, including the ${\cal
  O}(\alpha_t^2)$ corrections worsens the convergence of the
higher-order corrections. \s

\begin{figure}[t]
\begin{center}
  \includegraphics[width=1\textwidth]{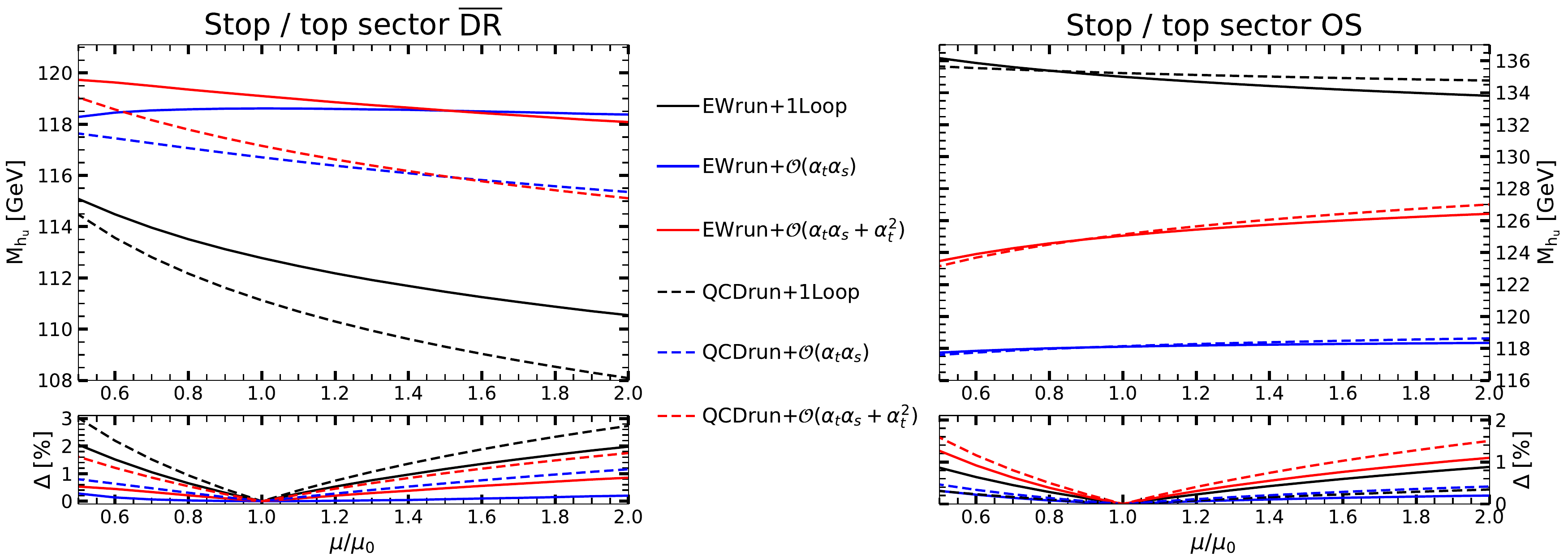}
\end{center}
\vspace*{-0.1cm}
\caption{Upper Panel: The mass of the $h_u$-like Higgs boson at
  one-loop level (black) and at two-loop level including the ${\cal
    O}(\alpha_t \alpha_s)$ corrections (blue) and the ${\cal
    O}(\alpha_t \alpha_s+\alpha_s^2)$ corrections (red) as a function
  of the renormalization scale $\mu_R = \mu$ varied between 1/2 and 2
  times the central value $\mu_R = \mu_0$, for EW running (full) and
  QCD running (dashed), see text for explanation. Lower Panel:
  Absolute value of the relative deviation of the result obtained at
  $\mu _R = \mu$ with respect to the result at $\mu_R = \mu_0$ -- {\it
    i.e.}
  $\Delta=|M_{{h_u}}^{\mu}-M_{{h_u}}^{\mu_0}|/M_{{h_u}}^{\mu_0}$
  -- in percent as a function of $\mu$ at one-loop (black), two-loop
  ${\cal O}(\alpha_t \alpha_s)$ (blue) and two-loop ${\cal O}(\alpha_t
  \alpha_s+\alpha_s^2)$ (red) level for EW running (full) and QCD
  running (dashed).}
\label{fig:renscalechange}
\end{figure}
An estimate of the theoretical uncertainty due to the missing higher-order
corrections can also be obtained from a change of the renormalization
scale $\mu$. In Fig.~\ref{fig:renscalechange} we depict the one- and two-loop
corrected mass values of the $h_u$-like Higgs boson, again including at
two-loop level the ${\cal O}(\alpha_t \alpha_s)$ and the ${\cal
  O}(\alpha_t \alpha_s+ \alpha_t^2)$ corrections, as a function of the
renormalization scale in terms of the default renormalization scale
$\mu_R \equiv \mu_0 = M_{\text{SUSY}}$. The renormalization scale is varied
between 1/2 and 2 times the value of the central scale $\mu_0$. The
lower panel shows the absolute value of the relative change of the
mass at the scale $\mu$ with respect to the value obtained for $\mu_0$,  
\beq
\Delta=\frac{|M_{{h_u}}^{\mu}-M_{{h_u}}^{\mu_0}|}{M_{{h_u}}^{\mu_0}} \;.
\eeq 
For this the loop order and type of conversion and RGE running is kept fix,
as indicated by the type and color of each line.
Note, that the consistent comparison of the results for different
renormalization scales also requires the conversion of the input
parameters to the new scale $\mu_R = \mu$ from the original set given at
$\mu_R =\mu_0$. We perform this conversion for the top quark mass by applying the
renormalization group equations and the procedure described in
App.~~\ref{append:mtrun}, including the ${\cal O}(\alpha_s +\alpha_t +
\alpha_s^2)$ contributions in the conversion of the
SM OS top-quark mass to the $\overline{\mbox{MS}}$ mass at $\mu_R =
M_Z$ and the ${\cal O}(\alpha_s+\alpha_t+ (\alpha_s
+\alpha_t)^2)$  corrections in the RGEs for the running values of
$m_t$ and $\alpha_s$ from $\mu_R = M_Z$ to $\mu_R=
  \mu$. The corresponding results are denoted by 'EWrun' in 
the plot. The relative change of the corrections obviously depends on
the running of the parameters that is applied. The plot also includes the
results 'QCDrun', where we include only the ${\cal O}(\alpha_s +
\alpha_s^2)$ contributions of the conversion of the top mass at the
scale $\mu_R = m_t$ and into RGEs, as done in
Ref.~\cite{Muhlleitner:2014vsa} where we computed the ${\cal
  O}(\alpha_t \alpha_s)$ corrections. The inclusion of the EW
contributions in the running is consistent with the ${\cal
  O}(\alpha_t^2)$ contributions in our two-loop fixed order masses,
leading (at all loop levels) to a 
better convergence of the higher-order results compared to those where
only QCD running is included. The remaining NMSSM $\DRb$ input
parameters are evaluated from the SLHA default input scale
$M_{\text{SUSY}}$ to the renormalization scale $\mu$ by applying the
RGEs given in App.~\ref{append:RGEs} and by using as input into the RGEs
  $\alpha_s$ and $y_t$ at 
the scale $\mu = M_{\text{SUSY}}$ as obtained by applying the
procedure 'EWrun' or 'QCDrun', respectively, in the curve with the
corresponding name. \s

For both schemes we observe that  the inclusion of the two-loop corrections of
${\cal O}(\alpha_t^2)$ worsens the scale dependence as compared to the two-loop
${\cal O}(\alpha_t \alpha_s)$ corrections alone. In the OS scheme this even
leads to a larger scale dependence than at one-loop order.\footnote{The
  one-loop scale dependence is, however, reduced in the OS scheme with
  respect to the $\DRb$ scheme due to the larger number of running
  parameters in the latter scheme.} In the $\DRb$ scheme 
we encounter a scale dependence that is not present in the OS scheme 
through the conversion of the OS to the $\DRb$ top-quark mass at
one-loop order, rendering the dependence on the scale $\mu$ larger
than in the OS scheme so that in the $\DRb$ scheme 
the black one-loop curve lies above the curves showing the two-loop scale
dependences. In the OS scheme, the running parameters are the $\DRb$ input
parameters whereas the top/stop parameters are treated OS in contrast to the
$\DRb$ scheme.\footnote{To be precise, in the OS scheme the values of the
running parameters are calculated in the same way as before for the $\DRb$
scheme. The soft SUSY breaking parameters $A_t$, $m_{\tilde{t}_R}$ and $m_{\ti Q_3}$,
however, are evolved to the scale $\mu$ and then converted to the corresponding
OS parameters. For the top quark mass the input OS value is used.} Additionally,
in both schemes $\alpha_s$ is running. The latter enters the one-loop result
only via the running of the other $\DRb$ parameters and directly only at
two-loop level. At ${\cal O}(\alpha_t \alpha_s)$ a cancellation between the
running parameters and the fixed order results leads to the observed flat
behaviour in the scale dependence. The inclusion of additional
${\cal O}(\alpha_t^2)$ terms, however, leads to a larger scale dependence
again. These observations show that care has to be taken with respect
to the estimate of the uncertainty due to missing higher-order
corrections based on a fixed order calculation that takes into account
only partial corrections at the given loop order and the conclusions
drawn from the scale dependence. Overall, as expected
and discussed above, however, the two-loop corrections are smaller than the
one-loop corrections and the renormalization scale dependence is
reduced. \s

\begin{figure}[t]
\begin{center}
  \includegraphics[width=0.5\textwidth]{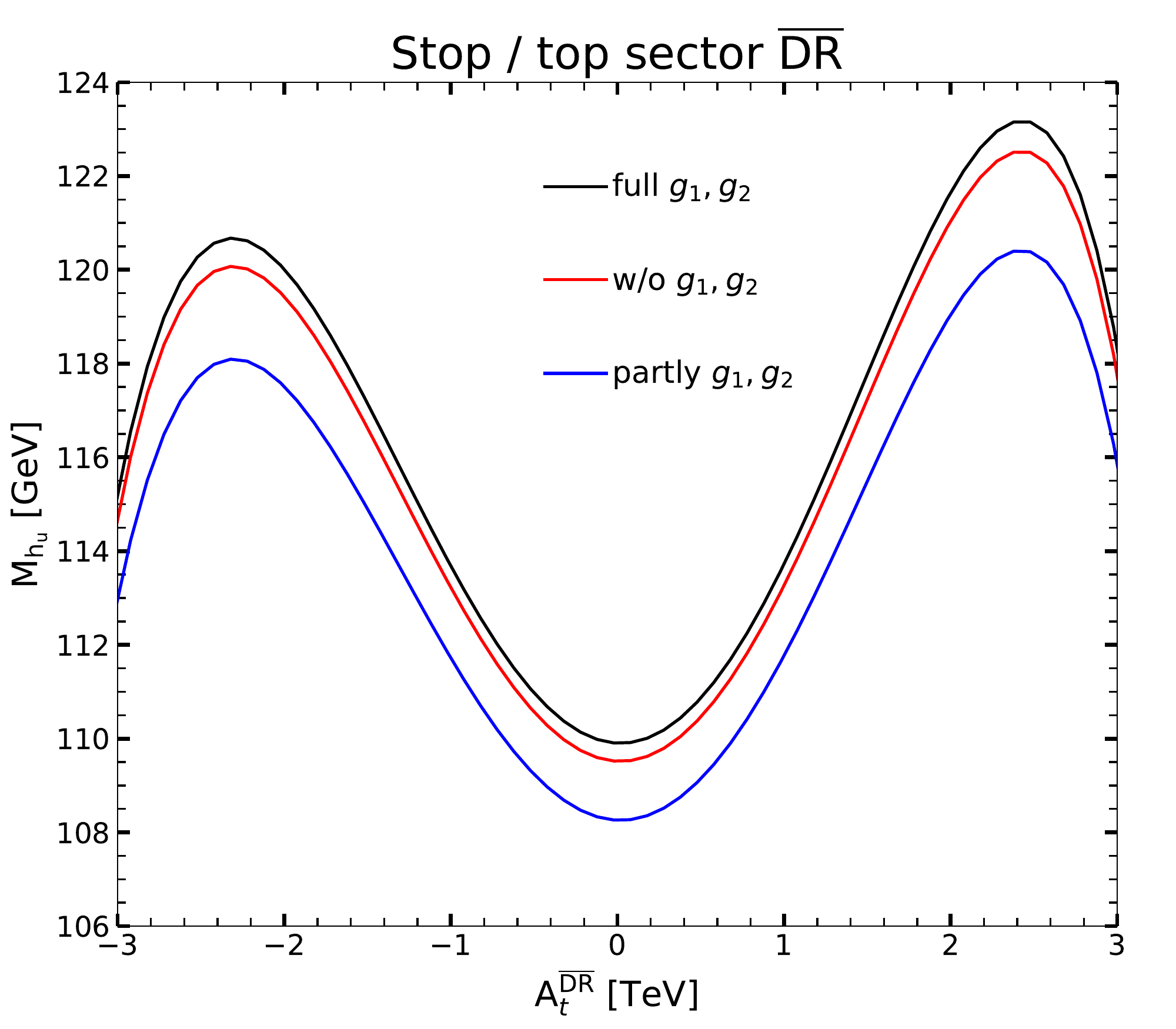}
\end{center}
\vspace*{-0.1cm}
\caption{The mass of the $h_u$-like Higgs boson at two-loop
  ${\cal O}(\alpha_t \alpha_s+\alpha_s^2)$ with $\overline{\mbox{DR}}$
  renormalization in the top/stop sector for full (black) non (red)
  or partial (blue) inclusion of the $g_1,g_2$ contributions in the
  computation of the running $\overline{\mbox{DR}}$ top mass from the
  OS input mass (see text for explanation), as a function of
  $A_t^{\overline{\text{DR}}}$.}
\label{fig:rgedependence}
\end{figure}
In order to further illustrate the interplay between the RGEs applied
on the running parameters and the fixed order results we show in Fig.~\ref{fig:rgedependence}
the two-loop ${\cal O}(\alpha_t \alpha_s + \alpha_t^2)$ corrections to
the SM-like Higgs boson mass with $\overline{\mbox{DR}}$
renormalization in the top/stop sector where the running $\overline{\mbox{DR}}$
top mass value is obtained as described in App.~\ref{append:mtrun}
and denoted by 'EWrun' in the previous plot (red line 'w/o
$g_1,g_2$'). The black line ('full $g_1,g_2$')
is obtained by including the $g_1$ and $g_2$ contributions in the
conversion and in the running, as described in
App.~\ref{append:mtrung1g2}. For the blue line ('partly $g_1,g_2$'),
we included the
$g_1$ and $g_2$ contributions in the RGEs of $m_t$, $\alpha_s$ and
$\lambda$ but not in the RGE for the VEV, and we did not include $g_1$
nor $g_2$ contributions in the conversion at the scale $M_Z$. As can
be inferred from the plot the results without the $g_1$, $g_2$
contributions - that have also not been included in our fixed-order
two-loop results - and those with their full inclusion are rather
close to each other, differing by less than 1~GeV. The partial
inclusion, however, drives the result away from the
  two previous ones by up to about 3~GeV. This is because here we do
not include the running of the VEV and therefore neglect the large
positive gauge contributions to the RGE of the VEV.
This once again shows that care has to be taken in
the conversion of the running parameters included in the higher-order
corrections and the estimate of the uncertainty due to the missing
higher-order contributions from the scale dependence. \s

\subsection{Results Parameter Point P2DR}
For completeness, we give also the results for a parameter point that
has been chosen such that the ${\cal O}(\alpha_t \alpha_s
+\alpha_t^2)$ corrections, now with $\DRb$ renormalization applied in
the top/stop sector, yield a SM-like Higgs boson with mass around
125~GeV. We called this point P2DR and presented its input values above.
Table~\ref{tab:massP2DR} summarizes the mass values that we obtain for
the $\DRb$ scheme in the top/stop sector at
tree level, at one-loop level and at two-loop level including
only the ${\cal O}(\alpha_t \alpha_s)$ and the ${\cal O}(\alpha_t \alpha_s
+\alpha_t^2)$ corrections, respectively. In Tab.~\ref{tab:massP2OS} the results
are given for the OS scheme in the top/stop sector. The tables
also contain the information on the main singlet/doublet and
scalar/pseudoscalar component of the respective mass eigenstate. The
tree-level stop masses obtained within the $\DRb$ and OS scheme are
given by 
\beq
\begin{array}{llll}
\DRb &:& m_{\tilde{t}_1} = 1121 \mbox{ GeV} \;, \qquad & m_{\tilde{t}_2} = 1473   
\mbox{ GeV} \;, \\
\mbox{OS} &:& m_{\tilde{t}_1} =  1100 \mbox{ GeV} \;, \qquad &
m_{\tilde{t}_2} = 1469 \mbox{ GeV} \;.
\end{array}
\eeq
The $\DRb$ top mass in our scenario amounts to $m_t^{\DRb} =
146.64$~GeV. \s              

\begin{table}[t]
\begin{center}
 \begin{tabular}{|l||c|c|c|c|c|}
\hline
 &${H_1}$&${H_2}$&${H_3}$&${H_4}$&${H_5}$\\ \hline \hline
tree-level &89.38&409.50&788.76&790.98&1828.56\\  
main component&$h_u$&$a_s$&$h_d$&$a$&$h_s$\\ \hline  
one-loop &120.86 & 407.68 & 788.64 & 791.01 & 1827.81\\   
main component&$h_u$&$a_s$&$h_d$&$a$&$h_s$\\ \hline  
two-loop ${\cal O}(\alpha_t \alpha_s)$ &124.58 & 407.69 & 788.65 & 791.0 & 1827.81 \\ 
main component&$h_u$&$a_s$&$h_d$&$a$&$h_s$\\ \hline
two-loop ${\cal O}(\alpha_t \alpha_s+ \alpha_t^2)$
 &125.67 & 407.69 & 788.65 & 791.0 & 1827.81\\  
main component&$h_u$&$a_s$&$h_d$&$a$&$h_s$\\ \hline
\end{tabular}
\caption{P2DR: Mass values in GeV and main components of the neutral Higgs
  bosons at tree-level, one-loop, two-loop ${\cal O}(\alpha_t
  \alpha_s)$ and at two-loop ${\cal O}(\alpha_t \alpha_s +
  \alpha_t^2)$ obtained by using $\DRb$ renormalization in the top/stop sector.}
\label{tab:massP2DR}
\end{center}
\end{table}
\begin{table}[t!]
\begin{center}
 \begin{tabular}{|l||c|c|c|c|c|}
\hline
 &${H_1}$&${H_2}$&${H_3}$&${H_4}$&${H_5}$\\ \hline \hline
tree-level &89.38&409.50&788.76&790.98&1828.56\\  
main component&$h_u$&$a_s$&$h_d$&$a$&$h_s$\\ \hline  
one-loop &142.91 & 407.74 & 788.62 & 790.9 & 1827.81\\ 
main component&$h_u$&$a_s$&$h_d$&$a$&$h_s$\\ \hline
two-loop ${\cal O}(\alpha_t \alpha_s)$ &123.92 & 407.71 & 788.57 & 790.91 & 1827.81\\ 
main component&$h_u$&$a_s$&$h_d$&$a$&$h_s$\\ \hline
two-loop ${\cal O}(\alpha_t \alpha_s+ \alpha_t^2)$
 &133.56 & 407.71 & 788.59 & 790.91 & 1827.81\\  
main component&$h_u$&$a_s$&$h_d$&$a$&$h_s$\\ \hline
\end{tabular}
\caption{P2DR: Mass values in GeV and main components of the neutral Higgs
  bosons at tree-level, one-loop, two-loop ${\cal O}(\alpha_t
  \alpha_s)$ and at two-loop ${\cal O}(\alpha_t \alpha_s +
  \alpha_t^2)$ obtained using OS renormalization in the top/stop
  sector.}
\label{tab:massP2OS}
\end{center}
\end{table}
In this scenario, the lightest Higgs boson corresponds to the SM-like
one with a mass around 125~GeV at ${\cal O}(\alpha_t \alpha_s +
\alpha_t^2)$ for $\DRb$ renormalization in the top/stop
sector. The remaining masses are around 400~GeV for the singlet-like
pseudoscalar, the masses of the CP-even and CP-odd MSSM-like Higgs
bosons are around 790~GeV, and the heaviest Higgs boson has a mass of
1828~GeV and is mostly CP-even singlet-like. The one-loop corrections to the
mass of the $h_u$-dominated SM-like Higgs boson $H_1$ are important
and lead to a relative increase of the mass value by 35\% in the $\DRb$
scheme, and by 59\% in the OS scheme. The two-loop 
${\cal O}(\alpha_t \alpha_s)$ corrections add a relative correction of
3\% in the $\DRb$ scheme and reduce the OS one-loop mass by
13\% in the OS scheme so that the mass values
in both renormalization schemes are close at this loop order. The ${\cal
  O}(\alpha_t \alpha_s +\alpha_t^2)$ lead to a relative increase of the mass value by 1\% in the $\DRb$ scheme, and the OS renormalization in the top/stop
sector increases the two-loop ${\cal O}(\alpha_t \alpha_s)$ mass by about 8\% when the
${\cal O}(\alpha_t^2)$ corrections are included as well. Overall, the
relative corrections are somewhat larger for P2DR than for
P1OS. Altogether, however, we observe the same behaviour, namely a
slightly worse convergence of the loop corrections after inclusion of
the ${\cal O}(\alpha_t^2)$ corrections. Since this is also a two-loop
correction, however, a better convergence cannot necessarily be
expected. Only the inclusion of all loop corrections at this given
order and finally also the three-loop corrections can be expected to
reduce the theoretical uncertainty on the mass values. 
We do not display the plots for P2DR corresponding to those shown for
P1OS. Their inspection shows the same qualitative behaviour as for P1OS. 

\subsection{Discussion of the Renormalization Schemes} 
\begin{table}[t!]
\centering
{\small \begin{tabular}{|l|cccccc|}
\hline
& TP1 & TP2 & TP3 & TP4 & TP5 & TP6 \\
\hline
 & \multicolumn{6}{|c|}{$h_1$}\\
\hline 
{\tt FlexibleSUSY }  & {\bf 123.55} & {\bf 122.84}       & {\it 91.11}      & {\bf 127.62}       & {\it 120.86}     &  {\bf 126.46}  \\ 
{\tt NMSSMCALC } 'QCDrun+${\cal O}(2,a)$' & {\bf 120.34}  & {\bf 118.57}       & {\it 90.88}      & {\bf 126.37}        & {\it 120.32}     &  {\bf 123.45}     \\
{\tt NMSSMCALC } 'EWrun+${\cal O}(2,b)$' & {\bf  123.58 }& {\bf  121.51 }& {\it  90.99 }& {\bf  127.38 }& {\it  120.82 }& {\bf  124.89 }\\  
{\tt NMSSMCALC } 'gaugerun+${\cal O}(2,b)$' &{\bf  124.31 }& {\bf  122.21 }& {\it  91.01 }& {\bf  127.69 }& {\it  120.92 }&  {\bf 125.32 } \\  
{\tt NMSSMTOOLS }    & {\bf 123.52}   & {\bf 121.83}       & {\it 90.78}      & {\bf 127.30}       & {\it 119.31}     &  {\bf 126.63}                                                                                       \\ 
{\tt SOFTSUSY }      & {\bf 123.84}   & {\bf 123.08}       & {\it 90.99}      & {\bf 127.52}        & {\it 120.81}     & {\bf 126.67}                                                                                       \\ 
{\tt SPHENO }        & {\bf 124.84}   & {\bf 124.74}       & {\it 89.54}      & {\bf 126.62}        & {\it 119.11}     & {\bf 131.29}                                                                                       \\ 
\hline
 & \multicolumn{6}{|c|}{$h_2$}\\
\hline 
{\tt FlexibleSUSY }  & {\it 1797.46}  & {\it 5951.36}      & {\bf 126.58}     & {\it 143.11}       & {\bf 125.08}      & {\it 700.80}                                                                                      \\ 
{\tt NMSSMCALC } 'QCDrun+${\cal O}(2,a)$' & {\it 1797.45}  & {\it 5951.36}      & {\bf 124.86}     & {\it 142.59}     & {\bf 123.14}      & {\it 701.02}     \\
{\tt NMSSMCALC } 'EWrun+${\cal O}(2,b)$' & {\it  1797.45 }& {\it  5951.36 }& {\bf  126.13 }& {\it  142.79 }& {\bf  124.16 }& {\it  701.06 } \\  
{\tt NMSSMCALC } 'gaugerun+${\cal O}(2,b)$' & {\it  1797.45 }& {\it  5951.36 }& {\bf  126.51 }& {\it  142.84 }& {\bf  124.51 }& {\it  701.06 }\\  
{\tt NMSSMTOOLS }    & {\it 1797.46}  & {\it 5951.36}      & {\bf 127.28}     & {\it 144.07}       & {\bf 126.95}      & {\it 700.46}                                                                                         \\ 
{\tt SOFTSUSY }      & {\it 1797.46}  & {\it 5951.36}      & {\bf 126.59}     & {\it 143.02}       & {\bf 125.12}      & {\it 701.01}                                                                                         \\ 
{\tt SPHENO }        & {\it 1798.01}  & {\it 5951.35}      & {\bf 126.77}     & {\it 144.04}       & {\bf 125.61}      & {\it 689.30}                                                                                         \\ 
\hline
 & \multicolumn{6}{|c|}{$h_3$}\\
\hline 
{\tt FlexibleSUSY }  & {2758.96}  & {6372.08}       & {652.95}     & {467.80}       &{627.28}     & {1369.53}                                                                                     \\ 
{\tt NMSSMCALC } 'QCDrun+${\cal O}(2,a)$' & {2756.70}   & {6371.48}      & {652.58}    & {467.48}       &{627.10}     & {1368.08} \\
{\tt NMSSMCALC } 'EWrun+${\cal O}(2,b)$' & { 2756.70 } & { 6368.58 } & { 652.70 } & { 467.73 } & { 627.16 } & { 1368.95 }  \\  
 {\tt NMSSMCALC } 'gaugerun+${\cal O}(2,b)$' &{ 2756.70 } & { 6368.47 } & { 652.67 } & { 467.72 } & { 627.15 } & { 1368.91 } \\ 
{\tt NMSSMTOOLS }    & {2758.51}  & {6345.72}      & {651.03}     & {466.38}        &{623.79}     & {1368.90}                                                                                         \\ 
{\tt SOFTSUSY }      & {2758.41}  & {6370.3}      & {652.78}     & {467.73}        & {627.14}     & {1369.19}                                                                                         \\ 
{\tt SPHENO }        & {2757.11}  & {6366.88}      & {651.21}     & {467.5}        & {624.02}     & {1363.02}                                                                                         \\ 
\hline
\hline
\end{tabular}}
\caption{Table adapted from Tab.~3 of \cite{Staub:2015aea} with the
  masses for the CP-even scalars (in GeV) for TP1--TP6 (defined in Tab.~2 of \cite{Staub:2015aea}) when using the spectrum generators ``out-of-the-box''. The values
  correspond to the two-loop results obtained by the different
  tools (the {\tt NMSSMTools} value corresponds to one using the option leading to the most
  precise calculation implemented in {\tt NMSSMTools}). For {\tt NMSSMCALC} the $\DRb$
  scheme in the top/stop sector is applied in three different variants
  as defined in the text. The masses for the SM-like scalar
  are written in bold fonts, those for the singlet-like scalar
  in italics.}
\label{tab:drbarcomparison}
\end{table}
After having presented our results, we want to finish with a discussion about the
renormalization schemes. Compared to our previous results of
\cite{Muhlleitner:2014vsa} we not only included additionally the
two-loop ${\cal O}(\alpha_t^2)$ corrections in our fixed order result to the
previously computed ${\cal O}(\alpha_t \alpha_s)$ corrections, we furthermore
changed the $\overline{\mbox{DR}}$ renormalization of the top/stop
sector. Thus we included also ${\cal O}(\alpha_t)$ contributions in
the conversion of the OS top quark mass to the $\DRb$ mass at
$\mu=M_Z$, and we included the resummation of $y_t$ in the
RGEs. Furthermore, we applied a running of the VEV. We found that in
particular the last three changes moved our results with 
$\DRb$ renormalization in the top/stop sector close to
those of the other codes applying solely $\DRb$ renormalization, {\tt
  FlexibleSUSY} \cite{Athron:2014yba}, {\tt NMSSMTools} 
\cite{Ellwanger:2004xm,Ellwanger:2005dv,Ellwanger:2006rn}, {\tt
  SOFTSUSY} \cite{Allanach:2001kg,Allanach:2013kza,Allanach:2014nba}
and {\tt SPHENO} \cite{Porod:2003um,Porod:2011nf}. The results of
these codes and {\tt NMSSMCALC} (applying the $\DRb$ scheme)
have been compared in 
\cite{Staub:2015aea}. In Table~\ref{tab:drbarcomparison} we show the
two-loop results for the masses of the CP-even scalars for six different test
points TP1,...,TP6 using all $\DRb$ spectrum generators 
``out-of-the-box''. This table is adapted from Tab.~3 of
\cite{Staub:2015aea}\footnote{For more details and the definition of the test
points, we refer the reader to \cite{Staub:2015aea}.} by adding
further results for {\tt NMSSMCALC}, for which we give three different
values. We give the result denoted by 'QCDrun+${\cal O}(2,a)$', with
'QCDrun' as defined in Subsec.~\ref{subsec:p1os} for the discussion of
Fig.~\ref{fig:renscalechange}, and '$2,a$' referring
to the two-loop correction of ${\cal O}(\alpha_t \alpha_s)$. It 
corresponds to our calculation of \cite{Muhlleitner:2014vsa} 
and reproduces the values given for {\tt NMSSMCALC} in
\cite{Staub:2015aea}\footnote{The small differences
    in the mass values of the $h_s$-like Higgs boson given here with respect to those of
    \cite{Staub:2015aea} are due to a correction in its counterterm.}. The numbers called 'EWrun+${\cal O}(2,b)$' are those for our newly computed fixed order
two-loop result of ${\cal O}(\alpha_t\alpha_s + \alpha_t^2)$ after
applying the changes in the $\DRb$ 
renormalization, described at the beginning of this subsection (and
in detail in App.~\ref{append:mtrun}), and
defined as 'EWrun' in Subsec.~\ref{subsec:p1os}. The third value,
with the label 'gaugerun+${\cal O}(2,b)$',
is obtained at ${\cal O}(\alpha_t \alpha_s + \alpha_t^2)$ after
including the gauge contributions in the conversion 
and running as described in App.~\ref{append:mtrung1g2}. 
The {\tt NMSSMCALC} values for the SM-like Higgs boson that were below
those of the other codes by up to slightly more than 3~GeV have moved
close to within 1~GeV mainly due to the adaption of our $\DRb$ renormalization
scheme. The exception is point TP6 which features a rather
large value for $\lambda$ with $\lambda=1.6$. We are using SM RGEs in
the running of $y_t$ and $\alpha_s$, while other codes ({\tt
  FlexibleSUSY}, {\tt SOFTSUSY} and {\tt SPHENO}) 
use NMSSM RGEs for scales $\mu\ge M_Z$, 
so that contributions which become
significant for large NMSSM-like couplings are not taken into
account in the same way. \s

\begin{figure}[t]
\begin{center}
  \includegraphics[width=0.5\textwidth]{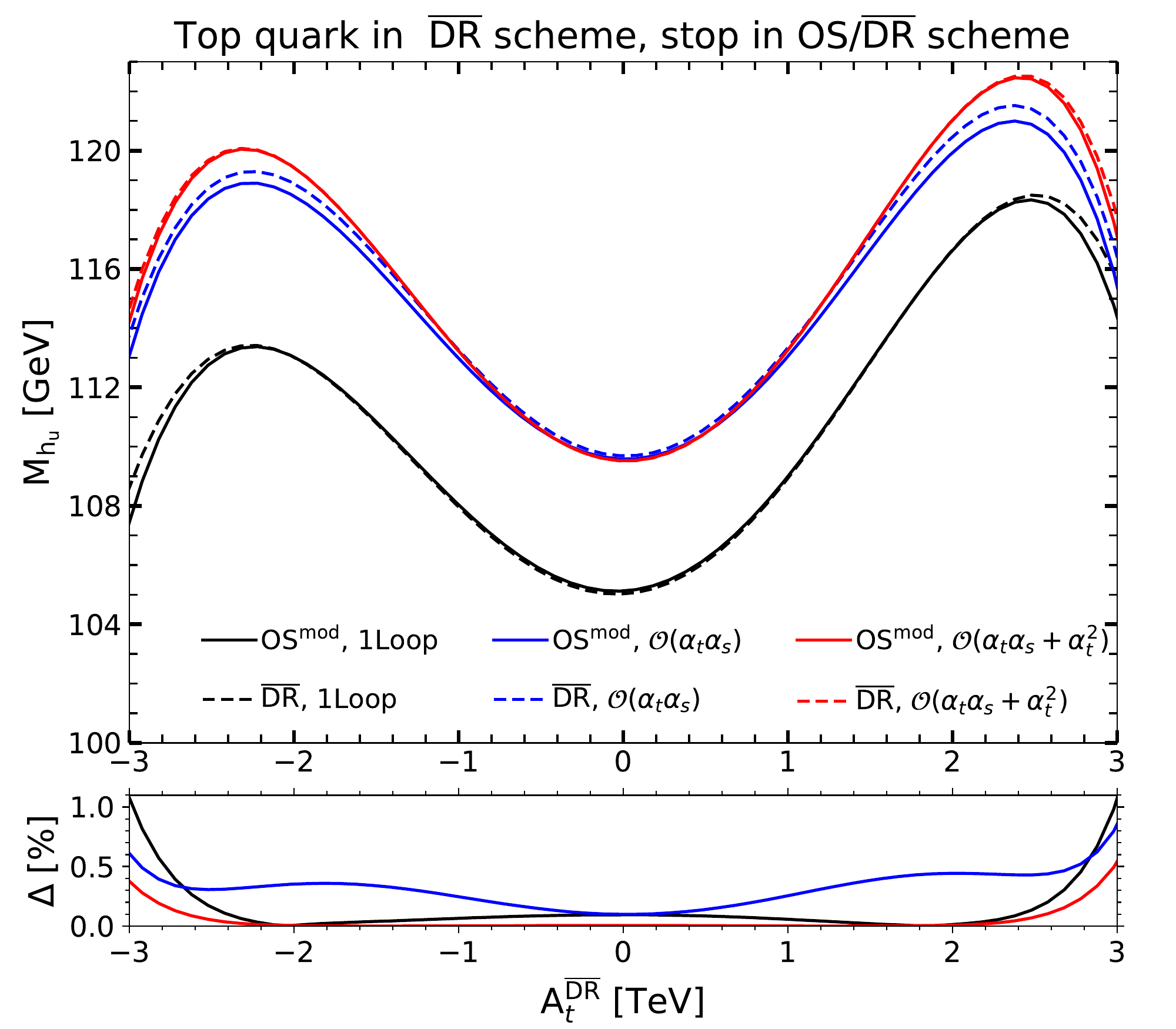}
\end{center}
\caption{Upper Panel: The mass of the $h_u$-like Higgs boson at
  one-loop level (black/two lower lines) and at two-loop level including the
  ${\cal O}(\alpha_t \alpha_s)$ corrections (blue) 
  and the ${\cal O}(\alpha_t \alpha_s+\alpha_t^2)$ corrections
  (red) as a function of $A_t^{\DRb}$. Solid lines: modified OS, dashed lines:
  $\DRb$ renormalization in the top/stop sector. Lower Panel:
  Absolute value of the relative deviation of the result with modified
  OS renormalization in the top and stop sector with respect to the
  result using a $\DRb$ scheme in percent at one-loop (black), 
  two-loop ${\cal O}(\alpha_t \alpha_s)$ (blue) and at two-loop ${\cal
    O}(\alpha_t \alpha_s+\alpha_t^2)$ (red) order.}
\label{fig:newscheme}
\end{figure}
Finally, we discuss our OS results where all top/stop parameters are
OS. The OS values of the stop parameters $A_t,m_{\tilde{Q}_L}$ and
$m_{\tilde{t}_R}$ are obtained from the default $\DRb$ input
parameters through the Eqs.~(\ref{eq:atfin})-(\ref{eq:trfin}) without the inclusion of any
resummation, whereas the top-quark mass is an OS input parameter
and hence does not require a conversion. The large difference between the $\DRb$ and OS
results originates from the inclusion of the $y_t$ running in the
$\DRb$ scheme. To verify this statement, we propose a third option for
the renormalization of the top/stop sector, calling it
OS$^{\text{mod}}$. In this scheme, we renormalize the three
stop parameters OS as before, but use the $\DRb$ top quark mass. The
results are shown in Fig.~\ref{fig:newscheme}. The figure has been
generated by starting from the point P1OS and varying $A_t$. The plot
shows the mass of the $h_u$ dominated SM-like Higgs boson for the
$\DRb$ scheme and the modified OS scheme
OS$^{\text{mod}}$, called for simplicity 'OS' in the plot,
at one-loop, two-loop ${\cal O}(\alpha_t \alpha_s)$ and
two-loop ${\cal O}(\alpha_t \alpha_s + \alpha_t^2)$. The lower plot
depicts for each loop order the relative change $\Delta$ in the mass when changing the
renormalization scheme from the $\DRb$ to the OS$^{\text{mod}}$
scheme, as defined in Eq.~(\ref{eq:renschemechange}) with OS replaced by
OS$^{\text{mod}}$. It is obvious that (at each loop order) the $\DRb$
and OS curves have moved much closer compared to
Fig.~\ref{fig:onelooptwoloop} with our original OS definition. The
relative error due to missing higher-order corrections based on the
change between the two renormalization schemes $\DRb$ and
OS$^{\text{mod}}$ has shrinked considerably and now amounts to less
than 1.1\% at one-loop level, less than about 0.9\% at ${\cal O}(\alpha_t
\alpha_s)$ and further improves at ${\cal O}(\alpha_t \alpha_s +
\alpha_t^2)$ with less than 0.5\%. We conclude our analysis with two
statements. First of all, care has to be taken when estimating the
remaining theoretical uncertainty. Depending on the applied
renormalization scheme, the conclusion drawn can be quite
different. Second, we found that the application of a $\DRb$ top-quark
mass and OS stop parameters moves the pure $\DRb$ result and the 'OS'
result much closer than the use of the OS top-quark mass, as is not
surprising. 

\section{Conclusions \label{sec:concl}}
We computed the fixed order ${\cal O}(\alpha_t^2)$ corrections to the
neutral Higgs bosons of the CP-violating NMSSM in the gaugeless limit
at vanishing external momentum, thus improving our previous results at
${\cal O}(\alpha_t \alpha_s)$. We applied a mixed
$\DRb$-OS renormalization scheme for the NMSSM input parameters. For
the top/stop sector 
which has to be renormalized at two-loop order, we apply either a $\DRb$ or
an OS definition. The two-loop corrections at ${\cal O}(\alpha_t
\alpha_s + \alpha_t^2)$ are found to amount to a few percent for the
SM-like Higgs boson mass. In order to discuss the remaining
theoretical uncertainty due to 
missing higher-order corrections we both vary the renormalization
scheme of the top/stop sector and the renormalization scale of the
$\DRb$ parameters. The discussion shows that care has to be taken when
drawing conclusions on the theory error. In particular, a modification
of the original OS definition of the top/stop sector to the inclusion
of a running top-quark instead of an OS mass considerably improves the convergence
between the $\DRb$ and the thus defined modified OS scheme
OS$^{\text{mod}}$. This calls for further
improvements in the fixed-order calculation including higher loop orders.

\section*{Acknowledgments}
We thank Philipp Basler for providing us with a set of NMSSM parameter
points. We are grateful to Pietro Slavich for sending us the $\al_t^2$
correction in the real MSSM for comparison. We thank Martin
  Gabelmann, Michael Spira and Florian Staub for useful discussions.
HR's work is partially funded by the Danish National Research
Foundation, grant number DNRF90. This research was supported in part by the Deutsche Forschungsgemeinschaft (DFG, German Research Foundation) under grant 396021762 - TRR 257. TND's work is funded by the Vietnam 
National Foundation for Science and Technology Development (NAFOSTED)
under grant number 103.01-2017.78. RG acknowledges support of the
'Berliner Chancengleichheitsprogramm'.

\newpage
\section*{Appendix}
\begin{appendix}
\section{Two-Loop Self-Energy Diagrams}
\subsection{Two-Loop Self-Energies of the Neutral Higgs Bosons}
\label{append:twoLoopSelfEnergiesNeutralHiggs}
Figure~\ref{fig:twoLoopSelfEnergiesNeutralHiggs} shows the two-loop 
self-energies of the neutral Higgs bosons, needed at ${\cal O}(\alpha_t^2)$.  
\begin{figure}[hb!]
\centering
\includegraphics[width=0.7\linewidth, trim=0cm 0.6cm 0cm 0.6cm,
clip]{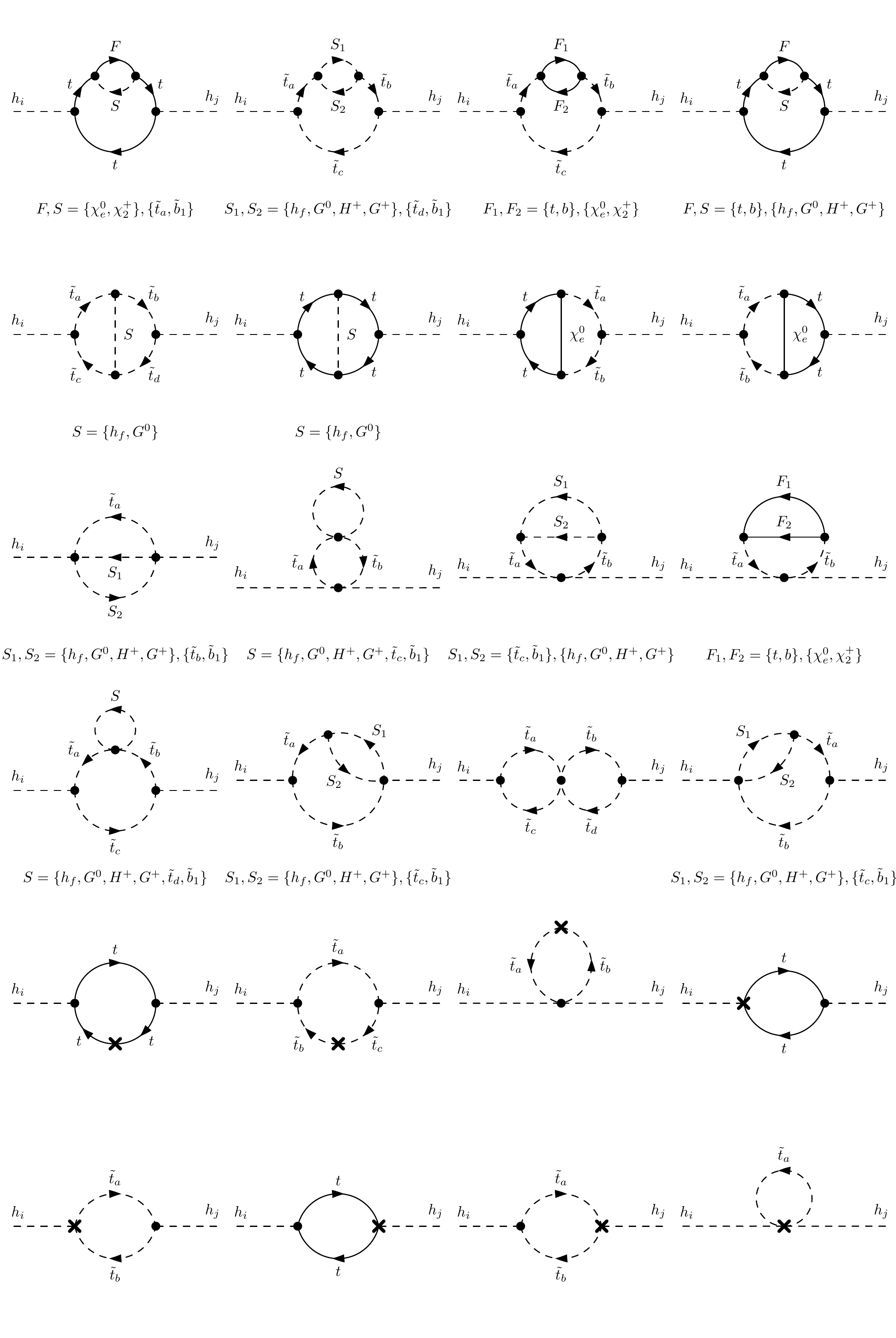}
\vspace*{-0.2cm}
\caption{Generic two-loop self-energies of the neutral Higgs bosons
  contributing to the two-loop corrections of the neutral Higgs boson
  masses at ${\cal O}(\alpha_t^2)$. 
  The placeholders $S_i$ and $F_i$ $(i=1,2)$ stand for the
  particle content of the diagrams specified below the respective
  diagrams. Diagrams with a cross denote the insertion of the
  respective one-loop counterterms of the vertices and masses. A
  summation over all internal particles with indices
  $a,b,c,d=1,2$, $e=3,4,5$ and $f=1,...,5$ 
  is implicit. Note that additional diagrams that differ from the ones
  shown only by the inversion of the fermion current are not explicitly
  shown.}  
\label{fig:twoLoopSelfEnergiesNeutralHiggs}
\end{figure} 

\subsection{Two-Loop Self-Energies of the Charged Higgs Boson}
\label{append:twoLoopSelfEnergiesChargedHiggs}
Figure~\ref{fig:twoLoopSelfEnergiesChargedHiggs} displays the generic
two-loop self-energies of the charged Higgs boson that contribute to
the computation of the two-loop counterterm of the charged Higgs boson mass.
\begin{figure}[h!]
\centering
\includegraphics[width=0.8\linewidth, trim=0cm 0cm 0cm 0.6cm, clip]{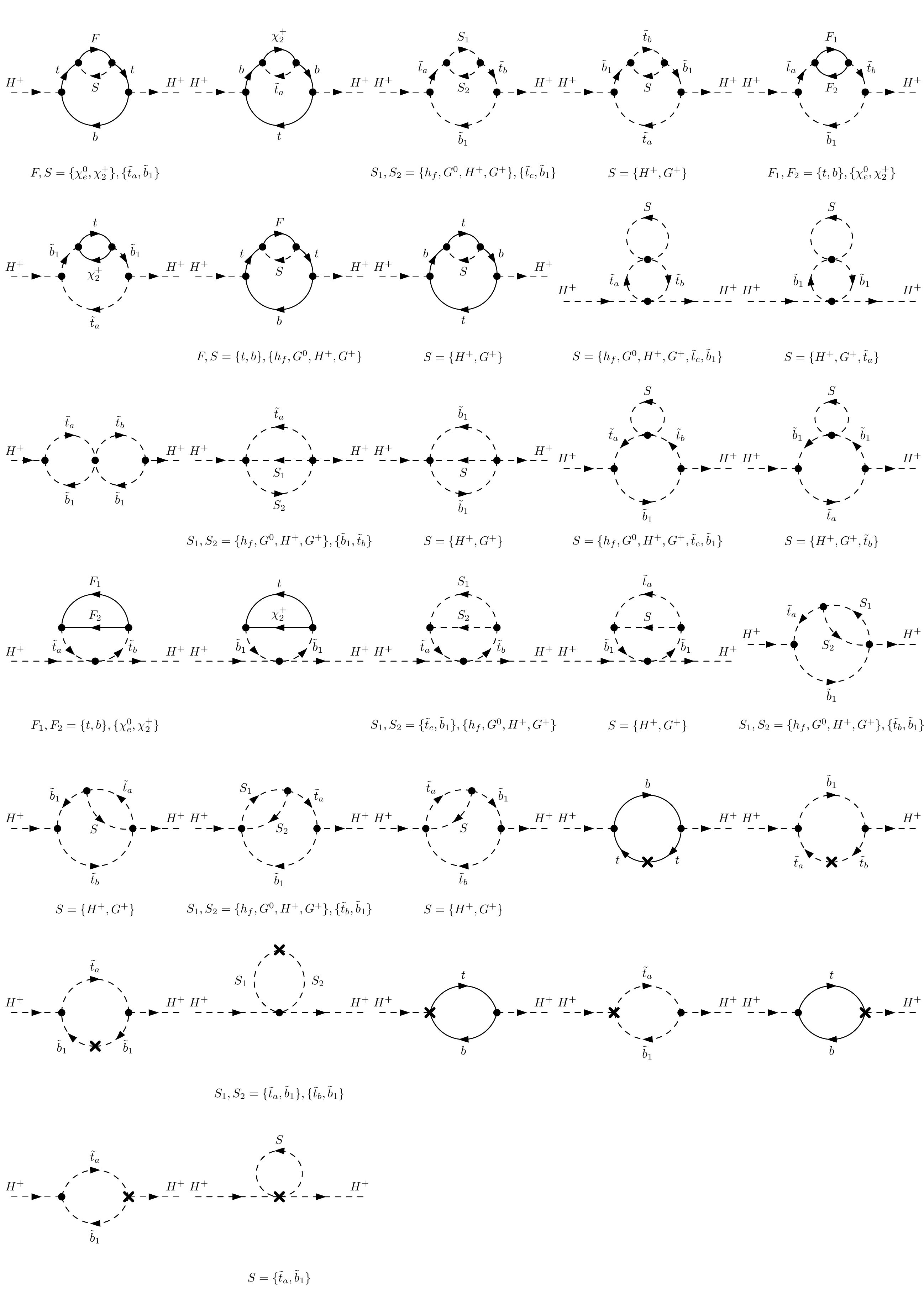}
\caption{Generic two-loop self-energies of the charged Higgs boson
  contributing to the computation of the two-loop counterterm
  of the charged Higgs boson mass. The placeholders $S_i$ and $F_i$ $(i=1,2)$ stand
  for the particle content of the diagrams specified below the
  respective diagrams. Diagrams with a cross denote the insertion of
  the respective one-loop counterterms of the vertices and masses. A
  summation over all internal particles with indices
  $a,b,c,d=1,2$, $e=3,4,5$ and $f=1,...,5$
  is implicit. Note that additional diagrams that differ from the ones
  shown only by the inversion of the fermion current are not explicitly
  shown.}  
\label{fig:twoLoopSelfEnergiesChargedHiggs}
\end{figure} 

\newpage

\section{Scalar One-Loop Integrals to $\mathcal{O} ( \varepsilon )$}
\label{append:loopintegrals}
In the following, we present the scalar one-loop one- and two-point
integrals expanded up to $\mathcal{O} ( \varepsilon )$ which are used
for our calculation of the two-loop corrected neutral Higgs boson masses. 
\subsection{Conventions}
For logarithms, we use the short-hand notation
\begin{equation}
	\overline{\ln } (m_i^2) \equiv \ln \left( \frac{m_i^2}{Q^2} \right) \;,
\end{equation}
where $m_i$ is an arbitrary parameter with mass dimension and $Q$ is
the modified renormalization scale, which is related to the
renormalization scale $\mu _R$ by 
\begin{equation}
	Q^2 \equiv 4\pi \mu _R^2 e^{-\gamma _E} ~,
\end{equation}
where $\gamma _E$ is the Euler-Mascheroni constant. With $\text{Li} _ 2$ we denote the dilogarithm, or Spence's function, defined for all $z\in \mathbb{C}$ as
	\begin{equation}
		\text{Li} _ 2 (z) \equiv - \int _0 ^z \text{d}x \frac{\ln (1-x)}{x} ~.
	\end{equation}
For taking several limits in the one-loop two-point integral,
the following specific values are useful, 
	\begin{equation}
		\text{Li} _ 2 ( -1 ) = - \frac{\pi ^2}{12} ~~, ~~~~ \text{Li} _ 2 (1) \equiv \zeta (2) = \frac{\pi ^2 }{6} ~,
	\end{equation}
	where $\zeta(2)$ denotes the Riemann zeta function $\zeta (s)$
        evaluated at $s=2$. \s
	
The solutions of the equation 
\begin{equation}
		\frac{m_2^2}{m_1^2} r^2 + \frac{p^2 - m_1^2 - m_2^2 + i \epsilon }{m_1 m_2} \frac{m_2}{m_1} r + 1 = 0 ~,
\end{equation}
needed later for the evaluation of the scalar two-point integral \cite{SCHARF1994523},
where $p^2$ is the squared four-momentum, are denoted by $r_1$ and
$r_2$. They can be cast into the form
\begin{equation}
r_{1/2} = \frac{-p^2 +m_1^2 + m_2^2 - i\epsilon \pm \sqrt{(p^2 - m_1^2 - m_2^2 + i\epsilon ) ^2 - 4m_1^2 m_2^2 }}{2m_2^2} ~,
\end{equation}
with $\epsilon > 0$ and $\epsilon \ll m_i^2$, $\epsilon \ll
p^2$. We choose $r_1$ to be the solution with the positive sign. 
The solutions satisfy the relations 
	\begin{equation}
		r_1 r_2 = y ~~, ~~~~ (1-r_1)(1-r_2) = x ~,
	\end{equation}	
	where $x$ and $y$ are dimensionless quantities defined as
	\begin{equation}
	x \equiv \frac{p^2}{m_2^2} ~~, ~~~~ y \equiv \frac{m_1^2}{m_2^2} ~.
	\end{equation}
	
\subsection{The Scalar One-Loop One-Point Integral $A_0$ at
  $\mathcal{O} (\varepsilon )$} 
The scalar one-loop one-point integral is defined as \cite{THOOFT1979365}
\begin{equation}
  A_0(m^2)=16\pi^2\mu_R^{4-D}\int
  \frac{d^Dq}{i(2\pi)^D}\frac{1}{(q^2-m^2)}  \label{eq:scalarOnePointIntegral} 
\end{equation}
in $D=4-2\varepsilon$ dimensions. The expansion of the integral up to
$\mathcal{O} (\varepsilon )$ reads
\begin{equation}
			A_0 (m^2) = \frac{m^2}{\varepsilon} + m^2 \left\{ 1 - \overline{\ln } (m^2) \right\} + m^2 \left\{ \frac{\zeta (2)}{2} + \frac{1}{2} \overline{\ln } ^2 (m^2) - \overline{\ln } (m^2) + 1 \right\} \varepsilon ~.
\end{equation}
For the special case $m^2 = 0$, the integral vanishes,
\begin{equation}
	A_0 (0) = 0 ~.
\end{equation}

\subsection{The Scalar One-Loop Two-Point Integral $B_0$ at $\mathcal{O} (\varepsilon )$}
The scalar one-loop two-point integral is defined as \cite{THOOFT1979365}
\begin{equation}
 B_0(p^2,m_1^2,m_2^2)=16\pi^2\mu_R^{4-D}\int
 \frac{d^Dq}{i(2\pi)^D}\frac{1}{(q^2-m_1^2)((q-p)^2-m_2^2)}
\label{eq:scalarTwoPointIntegral}
\end{equation}
in $D=4-2\varepsilon$ dimensions. The most general solution of $B_0$ up to $\mathcal{O} (\varepsilon)$ is given by \cite{SCHARF1994523}
	\begin{equation}
		\begin{split}
			B_0 &(p^2; m_1^2, m_2^2) = \frac{1}{\varepsilon} + \left\{ 2 -\frac{\overline{\ln } (m_1^2) + \overline{\ln} (m_2^2) }{2} - \frac{y-1}{2x} \ln (y) + \frac{r_1 - r_2}{2x} \big( \ln (r_1) - \ln (r_2) \big) \right\} \\
			&+\Bigg\{ \frac{\zeta (2)}{2} + 4 + \frac{\big( \overline{\ln} (m_1^2) + \overline{\ln} (m_2^2) \big) ^2}{8} + \frac{1}{8} \ln ^2 (y) - \frac{y-1}{x} \ln (y) \\
			&\hspace*{0.6cm}+\big( \overline{\ln} (m_1^2) + \overline{\ln} (m_2^2) \big) \left( -1 + \frac{y-1}{4x} \ln (y) - \frac{r_1 - r_2}{4x} ( \ln (r_1) - \ln (r_2) ) \right) \\
			&\hspace*{0.6cm}+\frac{r_1 - r_2}{2x} \bigg[ 2 \ln (r_1) - 2\ln (r_2) + \text{Li} _2 \left( \frac{-r_1 ( 1- r_2)}{r_2-r_1} \right) - \text{Li} _2 \left( \frac{-r_2 ( 1- r_1)}{r_1-r_2} \right) \\
			&\hspace*{2.4cm} + \ln \left( \frac{1-r_1}{r_2-r_1} \right) \ln \left( \frac{-r_1 (1-r_2)}{r_2-r_1} \right) - \ln \left( \frac{1-r_2}{r_1-r_2} \right) \ln \left( \frac{-r_2 (1-r_1)}{r_1-r_2} \right) \\
			&\hspace*{2.4cm} + \text{Li} _2 \left( \frac{1-r_1}{r_2-r_1} \right) - \text{Li} _2 \left( \frac{1-r_2}{r_1-r_2} \right) \bigg] \Bigg\} ~ \varepsilon ~~.
		\end{split}
		\end{equation}
This solution is valid for arbitrary values of $p^2$, $m_1^2$ and
$m_2^2$. However, for several limits of the parameters, the
computation becomes numerically unstable due to the appearance of
large logarithms or vanishing denominators. We therefore give
explicit expressions for several limiting cases:
\begingroup
\allowdisplaybreaks
\begin{align}
	B_0 (p^2; 0, m^2) &= \frac{1}{\varepsilon} + \left\{ 2 - \overline{\ln} (m^2) + \frac{m^2 - p^2}{p^2} \ln \left( \frac{-p^2 + m^2 -i\epsilon }{m^2} \right) \right\} \\
			&\hspace*{0.5cm}+ \Bigg\{ \frac{\zeta (2)}{2} + 4 + \frac{1}{2} \,\overline{\ln }^2 (m^2) - 2\,\overline{\ln } (m^2) + \frac{p^2 - m^2}{p^2}\,\overline{\ln} (m^2) \ln \left( \frac{-p^2 + m^2 -i \epsilon }{m^2 } \right) \nonumber \\
			&\hspace*{1.1cm}+ \frac{p^2-m^2}{p^2} \bigg[ \frac{1}{2} \ln ^2 \left( \frac{-p^2 + m^2 - i\epsilon }{m^2} \right) -2 \ln \left( \frac{-p^2 + m^2 -i\epsilon }{m^2} \right) \nonumber \\
			&\hspace*{3.1cm} - \text{Li} _2 \left( \frac{-p^2 -i\epsilon }{-p^2 + m^2 - i\epsilon } \right) \bigg] \Bigg\} ~ \varepsilon   \nonumber \\
	B_0 (p^2; 0, 0) &= \frac{1}{\varepsilon} + \bigg\{ 2 - \overline{\ln} (-p^2 -i\epsilon ) \bigg\}  \\
	 &\hspace*{0.5cm}+ \bigg\{ 4 - \frac{\zeta (2)}{2}+ \frac{\overline{\ln} ^2 (-p^2 -i\epsilon )}{2} - 2\,\overline{\ln} (-p^2 -i\epsilon ) \bigg\} ~ \varepsilon  \nonumber \\
	 B_0 (m^2; 0, m^2) &= \frac{1}{\varepsilon} + \bigg\{ 2 - \overline{\ln} (m^2) \bigg\} + \bigg\{ \frac{\zeta (2)}{2} + 4 + \frac{\overline{\ln }^2 (m^2)}{2} - 2\,\overline{\ln } (m^2) \bigg\} ~ \varepsilon 
\\
	 B_0 (0; m_1^2, m_2^2) &= \frac{1}{\varepsilon} + \left\{ 1 - \frac{m_1^2}{m_1^2 - m_2^2}\,\overline{\ln} (m_1^2) + \frac{m_2^2}{m_1^2 - m_2^2}\,\overline{\ln} (m_2^2) \right\} \\
			&\hspace*{0.5cm}+ \Bigg\{ \frac{\zeta (2)}{2} + 1 - \frac{y}{2(1-y)}\,\overline{\ln}^2 (m_1^2) + \frac{1}{2(1-y)}\,\overline{\ln} ^2 (m_2^2) + \frac{y}{1-y}\,\overline{\ln} (m_1^2) \nonumber \\
			&\hspace*{1.1cm}- \frac{1}{1-y}\,\overline{\ln} (m_2^2) \Bigg\} ~ \varepsilon  \nonumber \\
	B_0 (0; m^2, m^2) &= \frac{1}{\epsilon} - \overline{\ln} (m^2) + \bigg\{ \frac{\zeta (2)}{2} + \frac{1}{2}\,\overline{\ln} ^2 (m^2) \bigg\} ~ \varepsilon  \\
	B_0(0; 0, m^2 ) &= \frac{1}{\varepsilon} + \bigg\{ 1 - \overline{\ln} (m^2) \bigg\} + \bigg\{ \frac{\zeta (2)}{2} + 1 + \frac{\overline{\ln} ^2 (m^2)}{2} - \overline{\ln} (m^2) \bigg\} ~ \varepsilon  \\
	B_0 (0; 0,0) &\equiv \frac{1}{\varepsilon} - \frac{1}{\varepsilon _\text{IR}} ~.
\end{align}
\endgroup		
The parameter region of the last integral is a special case which
represents a pure divergence. It can be regularized in the UV regime
in the same way as the most general solution of the $B_0$
integral. However, for $p^2=m_i^2=0$, it additionally contains a
divergence in the infrared (IR) regime, which can be regularized in
dimensional regularization with the regulator $\varepsilon
_\text{IR}$. We want to emphasize that the results of our $\mathcal{O}
(\alpha _t^2)$ corrections to the neutral Higgs boson masses in the
gaugeless approximation do not depend on the IR regulator. 

\section{The Running $\DRb$ Top Mass}
\label{append:mtrun}
In this section, we present for the NMSSM the computation of the
$\DRb$ top quark mass at the SUSY scale from the given top quark pole
mass, as it is implemented in {\tt NMSSMCALC}. We follow the strategy
of \cite{Muhlleitner:2014vsa} where we assume that below the SUSY
scale the running top quark mass is described solely by the SM, and the NMSSM
contributions enter at the SUSY scale. The computation is done in the gaugeless approximation and performed in
four steps as follows: \s

\noindent
\underline{Step 1: Translation within the SM of the OS parameters to
  the $\MSb$ parameters at $\mu_R=M_Z$:}\\
We include in this translation the one-loop and two-loop QCD
corrections \cite{Chetyrkin:1999qi,Melnikov:2000qh} together with the one-loop EW corrections of $\calO(\al_t)$ in the
gaugeless limit,
 in particular
 \begin{align} 
m^{\MSb}_t(M_Z)=&\, m_t + \delta^{\al_s} m_t+ \delta^{\al_t} m_t
 \end{align}
 with
 \begin{align}
\delta^{\al_s} m_t &=\bigg[-\frac{\alpha_s(M_Z)}{\pi}\left(\frac{4}{3} + \ln \frac{M_Z^2}{m_t^2} \right)
 +\left( \frac{\alpha_s(M_Z)}{\pi}\right)^2 \bigg\{-\frac{3019}{288} - 2\zeta_2 - \frac 2 3\zeta_2\ln 2+\frac 16\zeta_3 \label{eq:dmtPoleMSbals} \\
&\hspace*{0.4cm} -\frac{445}{72} \ln \frac{M_Z^2}{m_t^2} - \frac{19}{24}\ln^2 \frac{M_Z^2}{m_t^2}  + n_f \left(\frac{71}{144} +\frac 1 3\zeta_2 + \frac{13}{36}\ln \frac{M_Z^2}{m_t^2}
  +\frac{1}{12}\ln^2 \frac{M_Z^2}{m_t^2}   \right)\bigg\}  \bigg]
  m_t \nonumber \\ 
\delta^{\al_t} m_t &= -\frac{ m_t^3}{16 \pi ^2 v^2} \Re \Big[B_1(m_t^2,m_t^2,M_H^2)+ B_1(m_t^2,m_t^2,0) + B_1(m_t^2,0,0) \label{eq:dmtPoleMSb} \\
&\hspace*{0.4cm}+B_0(m_t^2,m_t^2,0)- B_0(m_t^2,m_t^2,M_H^2) \Big] \;, \nonumber
\end{align}
where $n_f=5$. The one-loop
two-point functions are evaluated at the renormalization scale
$M_Z$. We denote the top quark pole mass by $m_t$ and the SM Higgs
mass by $M_H$, with $M_H=125.09\,\gev$ \cite{Aad:2015zhl}. 
Our $\calO(\al_t)$ correction in the gaugeless limit is in agreement
with the result presented in \cite{Denner:1991kt}. We use the following convention for the 
classical SM Higgs potential
\be V_H = m^2 \Phi^\dagger \Phi+ \lambda_{\text{SM}}(\Phi^\dagger \Phi)^2. \ee
The renormalized SM VEV is defined 
in such a way that it minimizes the loop-corrected Higgs
potential. The $\MSb$ running VEV at the scale $\mu_R=M_Z$
is related to the on-shell VEV via
\be  v(M_Z) = v + \delta v ,\ee
where $\delta v$ is the finite part of the on-shell VEV counterterm,
\be \delta v=\left[ \fr{c_{\theta_W}^2}{2 s_{\theta_W}^2}\left(
  \fr{\delta M_Z^2}{M_Z^2} -\fr{\delta M_W^2}{M_W^2} \right) +
\fr{\delta M_W^2}{2M_W^2} \right] v\;,
 \ee
with the finite part of the 
$W$ and $Z$ boson mass counterterms
\begin{align}
\fr{\delta M_Z^2}{M_Z^2}&= -\fr{M_H^2 -12  m_t^2 \ln(m_t^2/M_Z^2)}{32\pi^2  v^2},\crn
\fr{\delta M_W^2}{M_W^2} &=- \fr{M_H^2 - 12 m_t^2 \ln(m_t^2/M_Z^2) + 6 m_t^2}{32\pi^2 v^2 },\label{eq:MZMWcounterterm}
\end{align}
being computed at one-loop level in the gaugeless approximation. 
Additionally, we need
the running Higgs mass at the renormalization scale $M_Z$, which we obtain from 
\begin{equation}
m_H^{\MSb}(M_Z) = M_H\sqrt{1+\delta_H } \,,
\end{equation}
with
 \begin{align} \delta_H &= \frac{3}{32 v^2 \pi ^2 M_H^2
                          }\bigg[M_H^4 B_0(M_H^2,0,0)+3 M_H^4
                          B_0(M_H^2,M_H^2,M_H^2)-16 m_t^4
                          B_0(M_H^2,m_t^2,m_t^2) \nonumber \\
                          &\hspace*{0.4cm}+4 M_H^2 m_t^2 B_0(M_H^2,m_t^2,m_t^2)+M_H^2 A_0(M_H^2)-8 m_t^2  A_0(m_t^2)\bigg] \;, 
\end{align}
with the scalar one- and two-loop functions evaluated at the scale
$M_Z$. 

The running $\MSb$ top quark Yukawa coupling $y_t$ and the Higgs quartic
coupling $\lambda _\text{SM}$ at $M_Z$ are computed via the
relations\footnote{We computed $\delta m_t$, $\delta v$ and $\delta_H
  $ in the general $R_\xi$ gauge with the full one-loop EW correction
  and we confirmed that these formula give gauge-independent results
  for the running top quark Yukawa and quartic Higgs couplings, as
  expected.} 
\beq
y_t(M_Z) &=& \fr{\sqrt{2} m_t}{v}\braket{1+\fr{\delta m_t}{m_t} - \fr{\delta v}{v} }  \label{eq:ytatMZ}\\
\lambda_{\text{SM}}(M_Z) &=& \fr{M_H^2}{2 v^2 }\braket{1+ \delta_H - 2\fr{\delta v}{v} - \fr{t_H}{v M_H^2} }\,, \label{eq:lambdaatMZ}
\eeq 
where the tadpole at one-loop in the gaugeless approximation is given by
\be t_H = 3\fr{ M_H^4(1-\ln\fr{M_H^2}{M_Z^2}) -8 m_t^4(1 -\ln\fr{m_t^2}{M_Z^2} ) }{32 \pi^2 v}. \ee

The $\MSb$ running parameters $y_t(M_Z), \lambda_{\text{SM}}(M_Z)$ and $v(M_Z)$  together with $\alpha_s(M_Z)$ are input
parameters of the renormalization group equations (RGEs) for the running from
$M_Z$ to the SUSY scale, which is described in the second step. \s

\noindent
\underline{Step 2: RGE running to the SUSY scale $\mu_R=M_{\text{SUSY}}$:}\\
We  use the two-loop RGEs for $g_3,y_t,\lambda_{\text{SM}}$ where $g_3$ is the strong gauge coupling with $\alpha_s = g_3^2/(4 \pi)$ and $v$ in
the gaugeless limit that are relevant for the running of the top quark
mass to the SUSY scale, namely 
\cite{MACHACEK198383,MACHACEK1984221,MACHACEK198570,PhysRevLett.90.011601}
\begingroup
\allowdisplaybreaks
  \begin{align}
\frac{dg_3^2}{d\log\mu^2} &= \frac{g_3^4}{(4\pi)^2} \braket{-11 + \fr 23 n_f}  
+ \frac{g_3^4}{(4\pi)^4} \bigg[ g_3^2\left( -102+\frac{38}{3}n_f\right)-2 y_t^2 \bigg]\,, \\
\frac{d\lambda_\text{SM}}{d\log\mu^2}&= \frac{1}{(4\pi)^2} \bigg[\lambda_\text{SM}  \left(12 \lambda_\text{SM} +6 y_t^2-\frac{9 g_2^2}{2}-\frac{9 g_1^2}{10}\right) -3 y_t^4
+\frac{9 g_2^4}{16}+\frac{27 g_1^4}{400}+\frac{9 g_2^2 g_1^2}{40} \bigg]  
\\ &\hspace*{0.4cm}+\frac {1}{(4\pi)^4} \left[-156\lambda _\text{SM}^3 - 72 y_t^2\lambda _\text{SM}^2 - \frac 32 y_t^4 \lambda _\text{SM} + 15 y_t^6 + 40g_3^2 y_t^2\lambda _\text{SM} -16 g_3^2 y_t^4\right] \nonumber  \,,
\\
\frac{dy_t}{d\log\mu^2}&= \frac{y_t}{(4\pi)^2} \bigg[ \frac{9 y_t^2}{4}
-4 g_3^2-\frac{9 g_2^2}{8}-\frac{17 g_1^2}{40}\bigg]
 \\ &
\hspace*{0.4cm}+  \frac{y_t}{(4\pi)^4}  \bigg[3 \lambda _\text{SM}^2 - 6 y_t^2 \lambda _\text{SM} - 6 y_t^4 + 18 g_3^2 y_t^2 +\left( -\frac{202}{3}+n_f \frac{20}{9} \right)g_3^4\bigg] \nonumber \,,
\\
\frac{dv}{d\log\mu^2}&=- \frac{v}{(4\pi)^2}  \frac{3 y_t^2}{2}+
 \frac{v}{(4\pi)^4}  \bigg[ -10 g_3^2 y_t^2 + \fr{27}{8} y_t^4 - 3 \lambda^2 \bigg]\,,
 \end{align}
\endgroup
where $n_f =5$ for $\mu _R < m_t$ and $n_f=6$ for $\mu _R \ge
m_t$. For the solution of the coupled system of differential equations
we use the Runge-Kutta algorithm evaluated to fourth order
\cite{Runge1895,Kutta1901}. 
The $\overline{\mbox{MS}}$ top quark mass at the
SUSY scale $\mu _R = M_\text{SUSY}$ is then obtained from 
\begin{align}
	m_t^{\MSb} (M_{\text{SUSY}}) &= 2^{-1/2}  y_t (M_{\text{SUSY}}) v(M_{\text{SUSY}}) \,. \label{eq:HiggsMassAtSusyScale}
\end{align}

\noindent
\underline{Step 3: Conversion from $\overline{\text{MS}}$ to
  $\overline{\text{DR}}$:}\\
Within the SM, the $\overline{\text{DR}}$ top quark mass is computed
from the $\overline{\text{MS}}$ top quark mass at the SUSY scale by
using the two-loop
relation\cite{Avdeev:1997sz,Harlander:2006rj,Harlander:2007wh} 
 \begin{equation}
m_{t}^{\DRb, \text{SM}}(M_{\text{SUSY}})=m_{t}^{\MSb}(M_{\text{SUSY}})\left[1 -
  \frac{\alpha_s (M_{\text{SUSY}})}{3\pi} -
  \frac{\alpha_s^2 (M_{\text{SUSY}})}{144\pi^2}(73-3 n_f)
\right] \;.\label{eq:SMDRBmt}
\end{equation}
\noindent
\underline{Step 4: Adding the NMSSM contributions:}\\
In the final step, the NMSSM $\DRb$ top quark mass is calculated from the SM
$\DRb$ top quark mass by
 \begin{equation}
m_t^{\DRb,\text{NMSSM}} = m_t^{\DRb, \text{SM}}(M_{\text{SUSY}}) + dm_t^{\al_s} +
dm_t^{\al_t}\,,
\end{equation}
where $dm_t^{\alpha_s/\alpha_t}$ are the SUSY contributions relevant at
$\mu_R=M_{\text{SUSY}}$ at order ${\cal O}(\alpha_s)$ and ${\cal
  O}(\alpha_t)$, respectively,
\begin{align}
dm_t^{\al_s} &= \fr{\alpha_s(M_{\text{SUSY}})}{6\pi}\bigg\{-2m_t\text{Re}\Big[B_1(m_t^2,m_{\tilde{g}}^2,m_{\tilde t_1}^2)+ B_1(m_t^2,m_{\tilde{g}}^2,m_{\tilde t_2}^2) \Big] \label{eq:susyqcd_dmt} \\ \nonumber 
&\hspace*{0.4cm}+  2m_{\tilde{g}}\text{Re}\left[B_0(m_t^2,m_{\tilde{g}}^2,m_{\tilde t_1}^2) -B_0(m_t^2,m_{\tilde{g}}^2,m_{\tilde t_2}^2 ) \right] \\ \nonumber
&\hspace*{0.4cm}\times \left(e^{i(\varphi_{M_3}+ \varphi_u)} \mathcal{U}_{\tilde t_{22}} \mathcal{U}_{\tilde t_{21}}^*+e^{-i(\varphi_{M_3}+ \varphi_u)} \mathcal{U}_{\tilde t_{21}} \mathcal{U}_{\tilde t_{22}}^*\right) \bigg\} \\
 dm_t^{\al_t} &=-\fr{m_t^3 }{16 \pi^2 \sbeta^2v^2}\Re\bigg[\cbeta^2 B_1(m_t^2,0,\mhpm)  + \Re B_1(m_t^2, \abs{\mueff}^2, \msq)\label{eq:dmt} \\
&\hspace*{0.4cm}  +  2\cbeta^2 B_1(m_t^2,m_t^2, \mhpm) +  B_1(m_t^2, \abs{\mueff}^2,m_{\ti t_1}^2 )+B_1(m_t^2, \abs{\mueff}^2 ,m_{\ti t_2}^2 )\bigg] \nonumber \,,
\end{align}
where $m_t$ is the NMSSM $\DRb$ running mass at the SUSY
scale and $\varphi_{M_3}$ the phase of the gaugino parameter
$M_3$. Note that we keep only the heavy Higgs boson and Higgsino
contributions in $dm_t^{\al_t}$, since the light Higgs boson
contribution is identified with the SM Higgs contribution in \eqref{eq:dmtPoleMSb} and has  already been taken into account at scale $M_Z$.

\section{The Running $\DRb$ Top Mass with the Inclusion of the Gauge Couplings \label{append:mtrung1g2}}
Since we want to investigate the effect of the gauge contributions on the
loop-corrected Higgs masses, we specify in this section the
computation of the running $\overline{\mbox{DR}}$ top quark mass when
the gauge contributions are included. Due to their inclusion, gauge
dependences appear at several places 
in the formulae. We therefore performed the computation in the general $R_\xi$
gauge and kept the explicit dependence on the gauge parameter
$\xi$. In our numerical analysis, however, we use the t' Hooft-Feynman gauge
({\it i.e.}~$\xi=1$) to be consistent with the gauge that we use in
our one-loop calculation of the Higgs boson masses. In the following,
we present analogously to \sect{append:mtrun} the steps necessary to
obtain the running ${\overline{\mbox{DR}}}$ top mass at the scale
$M_{\text{SUSY}}$. Since the fourth step does not differ from the
gaugeless approximation, we solely describe here the changes in the
steps 1-3: \s

\noindent
\underline{Step 1: Translation within the SM of the OS parameters to
  the $\MSb$ parameters at $\mu_R=M_Z$:}\s 
Including the gauge contribution, the $\MSb$ top mass at $M_Z$ reads 
\be 
m^{\MSb}_t(M_Z)= m_t + \delta^{\al_s} m_t + \delta^{\al} m_t \;,
\ee
where the QCD contribution is presented in \eqref{eq:dmtPoleMSbals}
and the full one-loop EW correction is given by 
\allowdisplaybreaks
 \begin{align}
\delta^{\al} m_t =&\fr{1}{32\pi^2 v^2}\Bigg[m_t^3 (- 2B_1(m_t^2 ,m_t^2, M_H^2) + 2 B_0(m_t^2 ,m_t^2, M_H^2)+ B_0(m_t^2,0,M_W^2)) \\
&  -m_t \left(A_0\left(m_t^2\right)-2 A_0(M_W^2 
\xi_W)+A_0(M_W^2)-A_0(M_Z^2 \xi_Z)\right) \nonumber \\
&+M_W^2 \bigg(\left(m_t-\frac{2 M_W^2}{m_t}\right) B_0\left(m_t^2,0,M_W^2\right)-\frac{64}{9} m_t s_W^2 B_0\left(m_t^2,0,m_t^2\right) \nonumber \\
&+\frac{2 A_0(M_W^2)-\frac{32}{9} s_W^2 A_0\left(m_t^2\right)}{m_t}+m_t \left(\frac{32 s_W^2}{9}-2\right)\bigg)\nonumber \\
&+M_Z^2 \Bigg\{\left(\frac{1}{9} m_t \left(-64 s_W^4+48 s_W^2+9\right)-\frac{M_Z^2 \left(32 s_W^4-24 s_W^2+9\right)}{9 m_t}\right) B_0\left(m_t^2,m_t^2,M_Z^2\right) \nonumber \\
&-\frac{\left(32 s_W^4-24 s_W^2+9\right) \left(A_0\left(m_t^2\right)-A_0(M_Z^2)\right)}{9 m_t}+m_t \left(\frac{32 s_W^4}{9}-\frac{8 s_W^2}{3}-1\right)\Bigg\}\Bigg]\;,  \nonumber
\end{align}
where $\xi_{W,Z}$ is the gauge fixing parameter related to the $W$ and
$Z$ boson, respectively. 
For the running $\MSb$ VEV at the scale $\mu_R=M_Z$, we still use the relation
\be  
v(M_Z) = v + \delta v \;,
\ee
where $\delta v$ including the gauge contribution reads
\be 
\delta v=\left[\fr{c_W^2}{2 s_W^2}\left(
  \fr{\delta M_Z^2}{M_Z^2} -\fr{\delta M_W^2}{M_W^2} \right) +
\fr{\delta M_W^2}{2M_W^2} - \delta Z_e^{\alpha(M_Z)} \right] v\;.
\ee
Since  we use the fine structure constant at  the $Z$ boson mass
$M_Z$, $\al=\al(M_Z)$, as an input, the counterterm of the electric
coupling is defined as
\beq
\delta Z_e^{\alpha(M_Z)}&=&\delta Z_e^{\alpha(0)} - \fr{1}{2}\Delta\alpha(M_Z^2),\crn
\Delta\alpha(M_Z^2)&=&-\fr{\partial \Sigma_T^{AA}}{\partial 
  k^2}\bigg\vert_{k^2=0}-\fr{\Re\Sigma_T^{AA}(M_Z^2)}{M_Z^2} \;, 
\eeq
where the transverse part of the photon self-energy $\Sigma_T^{AA}$
includes only the light fermion contributions.
The analytical result for the counterterm of the electric coupling constant is given by
\begin{equation}
\delta Z_e^{\alpha(M_Z)} = -\frac{\alpha (M_Z)}{\pi} \left[ \frac{2}{9} \ln \left(
\frac{m_t^2}{M_Z^2} \right) - \frac{7}{8} \ln \left( \frac{M_W^2}{M_Z^2}
\right) - \frac{191}{108} \right] \;,
\label{eq:deltaze}
\end{equation}
where we have set the light fermion masses to zero. The finite parts
of the $W$ and $Z$ boson mass counterterms are 
\begingroup
\allowdisplaybreaks
\begin{align}
\delta M_W^2&=-\fr{\alpha(M_Z)}{4\pi}\Bigg[\frac{(2 c_{2W}+1) M_W^2 (32 c_{2W}+3 c_{4W}+31) B_0(M_W^2,M_W^2,M_Z^2)}{24 c_W^4 s_W^2}
\label{eq:dmw2f} \\
&\hspace*{0.4cm}-\frac{\left(M_H^4-4 M_H^2 M_W^2+12 M_W^4\right) B_0(M_W^2,M_H^2,M_W^2)}{12 M_W^2 s_W^2}-\frac{3 M_W^2 B_0(M_W^2,0,0)}{s_W^2}
\nonumber \\
&\hspace*{0.4cm}+\frac{\left(m_t^4+m_t^2 M_W^2-2 M_W^4\right) B_0\left(M_W^2,0,m_t^2\right)}{2 M_W^2 s_W^2}+4 M_W^2 B_0(M_W^2,0,M_W^2)
\nonumber \\
&\hspace*{0.4cm}-\frac{(8 c_{2W}+3 c_{4W}+4) A_0(M_Z^2)}{12 c_W^2 s_W^2}-\frac{A_0(M_W^2) (c_{2W} (M_H^2+12 M_W^2)+M_H^2+14 M_W^2)}{24 c_W^2 M_W^2 s_W^2}
\nonumber \\
&\hspace*{0.4cm}+\frac{(M_H^2-3 M_W^2) A_0(M_H^2)}{12 M_W^2 s_W^2}-\frac{\left(m_t^2-2 M_W^2\right) A_0\left(m_t^2\right)}{2 M_W^2 s_W^2}
-\frac{A_0(M_W^2 \xi_W)}{2 s_W^2}-\frac{A_0(M_Z^2 \xi_Z)}{4 s_W^2}
\nonumber \\
&\hspace*{0.4cm}-\frac{c_{2W} \left(-3 M_H^2+18 m_t^2-80 M_W^2\right)-3 M_H^2+18
        m_t^2-86 M_W^2}{36 c_W^2 s_W^2}\Bigg] \nonumber
\end{align}
\endgroup
and
\begin{align}
\delta M_Z^2&=-\fr{\alpha(M_Z)}{4\pi}\Bigg[\frac{M_W^2 (97 c_{2W}+35 c_{4W}+3 c_{6W}+63) B_0(M_Z^2,M_W^2,M_W^2)}{24 c_W^4 s_W^2}\label{eq:dmz2f} \\
&\hspace*{0.4cm}-\frac{M_W^2 (-20 c_{2W}+20 c_{4W}+63) B_0(M_Z^2,0,0)}{18 c_W^4 s_W^2}+\frac{(-4 c_{2W}+4 c_{4W}+9) A_0\left(m_t^2\right)}{9 c_W^2 s_W^2} \nonumber \\
&\hspace*{0.4cm}+\frac{1}{36 c_W^4 s_W^2}\Big(B_0\left(M_Z^2,m_t^2,m_t^2\right) \left(c_{2W} \left(13 m_t^2+8
      M_W^2\right)-4 c_{4W} \left(m_t^2+2 M_W^2\right) \right. \nonumber \\
&\hspace*{0.4cm} \left. -4 c_{6W} m_t^2+13 m_t^2-18 M_W^2\right) \Big) -\frac{(8 c_{2W}+3 c_{4W}+4) A_0(M_W^2)}{6 c_W^2 s_W^2} \nonumber \\
&\hspace*{0.4cm}-\frac{\left(12 M_Z^4-4 M_H^2 M_Z^2+M_H^4\right) B_0(M_Z^2,M_H^2,M_Z^2)}{12 M_W^2 s_W^2 }-\frac{A_0(M_Z^2) \left(M_H^2-\frac{2 M_W^2}{c_W^2}\right)}{12 M_W^2 s_W^2} \nonumber \\ 
&\hspace*{0.4cm}+\frac{A_0(M_H^2) \left(M_H^2-\frac{3 M_W^2}{c_W^2}\right)}{12 M_W^2 s_W^2} -\frac{A_0(M_W^2 \xi_W)}{2 c_W^2 s_W^2}-\frac{A_0(M_Z^2 \xi_Z)}{4 c_W^2 s_W^2} \nonumber \\ 
&\hspace*{0.4cm}-\frac{1}{72 c_W^4 s_W^2}\bigg(c_{2W} \left(-6 M_H^2+28 m_t^2-75 M_W^2\right)+c_{4W} \left(8 m_t^2-74 M_W^2\right) \nonumber \\
&\hspace*{0.4cm} +8 c_{6W} m_t^2-9 c_{6W} M_W^2-6 M_H^2+28 m_t^2-174 M_W^2\bigg)\Bigg].\nonumber 
\end{align}
For the running Higgs mass at the renormalization scale $M_Z$, we use
\begin{equation}
m_H^{\MSb}(M_Z) = M_H\sqrt{1+\delta_H } \,,
\end{equation}
with
 \begin{align} 
 \delta_H &= \frac{ 1}{16\pi^2 v^2}\Bigg[\frac{\left(2 M_H^4-8 M_H^2 M_W^2+24 M_W^4\right) B_0(M_H^2,M_W^2,M_W^2)}{2 M_H^2} - \frac{24 M_W^4+12 M_Z^4}{2 M_H^2}
\\ &\hspace*{0.4cm}
+\frac{\left(M_H^4-4 M_H^2 M_Z^2+12 M_Z^4\right) B_0(M_H^2,M_Z^2,M_Z^2)}{2 M_H^2}+\frac{\left(12 M_H^2 m_t^2-48 m_t^4\right) B_0\left(M_H^2,m_t^2,m_t^2\right)}{2 M_H^2}
\nonumber \\ &\hspace*{0.4cm}
+\frac{9}{2} M_H^2 B_0(M_H^2,M_H^2,M_H^2)-\frac{12 m_t^2 A_0\left(m_t^2\right)}{M_H^2}+\frac{(12 M_W^2-4 M_H^2) A_0(M_W^2)}{2 M_H^2}
\nonumber \\ &\hspace*{0.4cm}
+\frac{(6 M_Z^2-2 M_H^2) A_0(M_Z^2)}{2 M_H^2}
+\frac{3 A_0(M_H^2)}{2}+3 A_0(M_W^2 \xi_W)+\frac{3}{2} A_0(M_Z^2 \xi_Z)\Bigg] \;. \nonumber 
\end{align}
For the $\MSb$ top Yukawa coupling and SM coupling $\lambda$ at $M_Z$
we again use the Eqs.~(\ref{eq:ytatMZ}) and (\ref{eq:lambdaatMZ}) 
 with the  one-loop tadpole given by
 \begin{align} 
t_H &=\frac{1}{32 \pi ^2 v}\bigg[12  M_W^2 A_0(M_W^2)+6 M_Z^2 A_0(M_Z^2)-4 \left(2 M_W^4+M_Z^4\right)
\\ &\hspace*{0.4cm} +M_H^2 \left(3 A_0(M_H^2)+2 A_0(M_W^2 \xi_W)+A_0(M_Z^2
  \xi_Z)\right)-24 m_t^2 A_0\left(m_t^2\right)\bigg] \;. \nonumber
 \end{align} 

We also need the $\MSb$ SM gauge couplings $g_1,g_2$, with the
$U(1)_Y$ gauge coupling $g_1= \sqrt{5/3} g_Y$, where $g_2$ and $g_Y$
are the $SU(2)_L$ gauge coupling and the $U(1)_Y$ gauge coupling at
the GUT scale, respectively.  
We remind the reader that we have used the running fine structure
constant $\alpha(M_Z), M_Z, M_W$ as input parameters, so that the
gauge couplings $g_2$ and $g_Y$ read 
\be  
g_2= \fr{\sqrt{4\pi \al(M_Z)}}{s_W} \quad \mbox{and} \quad g_Y=  \fr{\sqrt{4\pi
    \al(M_Z)}}{c_W} \;,
\ee
where $c_W= M_W/M_Z$ and $s_W=\sqrt{1 - M_W^2/M_Z^2 }$ are OS
parameters. As result the running $\MSb$ couplings $g_2$ and $g_Y$ at
the scale $M_Z$ are given by\footnote{Note that the counterterms $\de s_W, \de
  c_W$ are gauge-independent quantities.} 
\bea 
g_2(M_Z) &= \fr{\sqrt{4\pi \al(M_Z)}}{s_W} \braket{1 +\delta Z_e^{\alpha(M_Z)} - \fr{\de s_W }{s_W} },\crn
g_Y(M_Z) &= \fr{\sqrt{4\pi \al(M_Z)}}{c_W} \braket{1 +\delta
  Z_e^{\alpha(M_Z)} - \fr{\de c_W }{c_W} } \;, 
\eea
where
 \bea 
\de s_W  = \fr{c_W^2}{2s_W}\braket{\fr{\delta M_Z^2}{M_Z^2}
  -\fr{\delta M_W^2}{M_W^2} } \quad \mbox{and} \quad \de c_W =
-\fr{s_W\de s_W}{c_W} \;, 
\eea
with the analytic expressions for $\de M_W$, $\de M_Z$ and $\delta Z_e^{\alpha(M_Z)}$ presented in \eqref{eq:dmw2f}, \eqref{eq:dmz2f} and \eqref{eq:deltaze},
respectively. \s

\noindent
\underline{Step 2: RGE running to the SUSY scale $\mu_R=M_{\text{SUSY}}$:}\\
Here we use the two-loop RGEs for $g_1,g_2,g_3,y_t,\lambda$ and $v$
\cite{MACHACEK198383,MACHACEK1984221,MACHACEK198570,PhysRevLett.90.011601}. The
computation of the RGE for the VEV is done with the help of the {\tt
  SARAH} package \cite{Staub:2010jh,Staub:2012pb,Staub:2013tta,Goodsell:2014bna,Goodsell:2014pla}.  The contributions from the light
fermions have been neglected here as well. We have
\begingroup
\allowdisplaybreaks
 \begin{align}
\frac{dg_1^2}{d\log\mu^2} &= \frac{g_1^4 (20 n_f+3)}{30 (4\pi)^2}+\frac{g_1^4 \left(g_1^2 (95 n_f+27)+5 \left(9 g_2^2 (n_f+3)+44 g_3^2 n_f-51 y_1^2\right)\right)}{150 (4\pi)^4} \\
\frac{dg_2^2}{d\log\mu^2} &=\frac{g_2^4 (4 n_f-43)}{6 (4\pi)^2}+ \frac{g_2^4 \left(3 g_1^2 (n_f+3)+5 \left(7 g_2^2 (7 n_f-37)+12 g_3^2 n_f-9 y_1^2\right)\right)}{30 (4\pi)^4} 
\\
\frac{dg_3^2}{d\log\mu^2} &=\frac{g_3^4 \left(\frac{2 n_f}{3}-11\right)}{(4\pi)^2}+ \frac{g_3^4 \left(11 g_1^2 n_f+15 \left(3 g_2^2 n_f-8 y_1^2\right)+40 g_3^2 (19 n_f-153)\right)}{60 (4\pi)^4}
\\
\frac{d\lambda_\text{SM}}{d\log\mu^2}&= \frac{1}{(4\pi)^2} \bigg[\lambda_\text{SM}  \left(12 \lambda_\text{SM} +6 y_t^2-\frac{9 g_2^2}{2}-\frac{9 g_1^2}{10}\right) -3 y_t^4
+\frac{9 g_2^4}{16}+\frac{27 g_1^4}{400}+\frac{9 g_2^2 g_1^2}{40} \bigg]  
 \\ &
\hspace*{0.4cm}+\frac {1}{(4\pi)^4} \bigg[-\frac{3 g_1^6 (160 n_f+177)}{4000}+g_1^4 \left(g_2^2 \left(-\frac{n_f}{5}-\frac{717}{800}\right)-\frac{171 y_1^2}{200}\right)
\nonumber \\ &
\hspace*{0.4cm}+\lambda_{\text{SM}} ^2 \left(\frac{54 g_1^2}{5}+54 g_2^2-72 y_1^2\right)+g_1^2 \left(g_2^4 \left(-\frac{n_f}{5}-\frac{97}{160}\right)+\frac{63 g_2^2 y_1^2}{20}-\frac{4 y_1^4}{5}\right)
\nonumber \\ &
\hspace*{0.4cm}+\lambda_{\text{SM}}  \left(\frac{1}{2} g_1^4
       \left(n_f+\frac{687}{200}\right)+g_1^2 \left(\frac{117
       g_2^2}{40}+\frac{17 y_1^2}{4}\right)+g_2^4 \left(\frac{5
       n_f}{2}-\frac{313}{16}\right)+\frac{45 g_2^2 y_1^2}{4} \right. \nonumber \\
&\hspace*{0.4cm} \left. +40 g_3^2 y_1^2-\frac{3 y_1^4}{2}\right)
+g_2^6 \left(\frac{497}{32}-n_f\right)-\frac{9 g_2^4 y_1^2}{8}-16 g_3^2 y_1^4-156 \lambda_{\text{SM}} ^3+15 y_1^6\bigg] \nonumber
\\
\frac{dy_t}{d\log\mu^2}&= \frac{y_t}{(4\pi)^2} \bigg[ \frac{9 y_t^2}{4}
-4 g_3^2-\frac{9 g_2^2}{8}-\frac{17 g_1^2}{40}\bigg]+
 \\ &
\hspace*{0.4cm}+  \frac{y_t}{(4\pi)^4}  \bigg[g_1^4 \left(\frac{29 n_f}{180}+\frac{9}{400}\right)+g_1^2 \left(-\frac{9 g_2^2}{40}+\frac{19 g_3^2}{30}+\frac{393 y_1^2}{160}\right)
+\frac{1}{8} g_2^4 (2 n_f-35)
\nonumber \\ &
\hspace*{0.4cm}+g_2^2 \left(\frac{9 g_3^2}{2}+\frac{225 y_1^2}{32}\right)+\frac{2}{9} g_3^4 (10 n_f-303)
+y_1^2 \left(18 g_3^2-6 \lambda_{\text{SM}} \right)+3 \lambda_{\text{SM}} ^2-6 y_1^4\bigg] \nonumber
\\
\frac{dv}{d\log\mu^2}&=\frac{v}{(4\pi)^2}  \left(\frac{3}{40} g_1^2 (\xi+3)+\frac{3}{8} g_2^2 (\xi+3)-\frac{3 y_1^2}{2}\right) \\ &
\hspace*{0.4cm}+
 \frac{v}{(4\pi)^4}  \bigg[\frac{3 g_1^4 (12 \xi (\xi+1)-431)}{1600}+g_1^2 \left(\frac{9}{160} g_2^2 (4 \xi (\xi+1)-3)+\frac{1}{80} (-36 \xi-85) y_1^2\right)
\nonumber \\ &
\hspace*{0.4cm}+\frac{1}{64} g_2^4 (108 \xi+271)-\frac{9}{16} g_2^2 (4 \xi+5) y_1^2-10 g_3^2 y_1^2-3 \lambda_{\text{SM}} ^2+\frac{27 y_1^4}{8}
 \bigg]\,. \nonumber
 \end{align}
\endgroup

\noindent
\underline{Step 3: Conversion from $\overline{\text{MS}}$ to $\overline{\text{DR}}$:} \\
With the inclusion of the electroweak gauge contributions the
conversion from the $\overline{\text{MS}}$ to the
  $\overline{\text{DR}}$ top quark mass is modified to
 \begin{align}
m_{t}^{\DRb, \text{SM}}(M_{\text{SUSY}})&=m_{t}^{\MSb}(M_{\text{SUSY}})\bigg[1 -
  \frac{\alpha_s (M_{\text{SUSY}})}{3\pi} -
  \frac{\alpha_s^2 (M_{\text{SUSY}})}{144\pi^2}(73-3 n_f) 
\label{eq:SMDRBmt1} \\&\hspace*{0.4cm} +\frac{\alpha(M_{\text{SUSY}}) (26 c_W^2+1)}{288 \pi  c_W^2 s_W^2}
\bigg] \;. \nonumber
 \end{align}
\section{NMSSM RGEs for Investigations of the Scale Dependence}\label{append:RGEs}
In order to estimate the remaining theoretical uncertainty due to
missing higher-order corrections, we investigate their scale
dependence. For a consistent investigation 
also the input parameters have to be evaluated at the respective
scales. According to the SLHA the SUSY parameters in the
input file are $\DRb$ parameters evaluated at the user-specified scale,
where the SUSY scale is the default choice. These parameters need to be
evolved to the different scales used to determine the scale
dependence. For this, the RGEs of the relevant parameters are needed. 
In order to match the approximations used for the computation of the
Higgs masses at order $\order$ and $\mathcal{O} (\alpha _t \alpha _s)$,
respectively, we use the RGEs in the gaugeless limit and set the light
Yukawa couplings to zero. We computed all RGEs with {\tt SARAH} and
cross-checked them with the results given in \cite{KING2012207, King:1995vk,
  Ellwanger:2009dp}, and with the results of \cite{Sperling:2013xqa} for the RGEs of
$\tan\beta$ and $v_s$. The RGEs are given by  
\begingroup
\allowdisplaybreaks
\begin{align}
\frac{dg_3}{d\log\mu^2}&= -\frac{ g_3^3}{16\pi^2} \frac{3 }{2} +\frac{g_3^3}{(16 \pi^2)^2}\bigg(7 g_3^2 - 2 y_t^2 \bigg)\\
\frac{dy_t}{d\log\mu^2}&= \frac{ y_t}{16\pi^2} \left( -\frac{8}{3}g_3^2  + 3 y_t^2 + \frac{1}{2}|\lambda|^2\right)\\ \nonumber&\hspace*{0.4cm}+\frac{y_t}{(16 \pi^2)^2}\left(-\frac{8}{9} g_3^4 + 8 g_3^2 y_t^2 - 11 y_t^4 - \frac{3}{2} y_t^2 |\lambda|^2 -
  |\kappa|^2 |\lambda|^2 - \frac{3}{2} |\lambda|^4 \right)\\
\frac{dv_u}{d\log\mu^2}&=\frac{v_u}{16\pi^2}\left( -\fr 32  y_t^2 -\fr12\abs{\lambda}^2\right)+\frac{v_u}{(16\pi^2)^2}\left(
 - 8 g_3^2 y_t^2 + \fr 92 y_t^4 + \abs{\kappa}^2\abs{\lambda}^2 + \fr 32 \abs{\lambda}^4\right)\\
\frac{dv_d}{d\log\mu^2}&=\frac{v_d}{16\pi^2}\left(-\fr12\abs{\lambda}^2\right)+\frac{v_d}{(16\pi^2)^2}\left(
\fr32 y_t^2 \abs{\lambda}^2 + \abs{\kappa}^2\abs{\lambda}^2 + \fr 32 \abs{\lambda}^4\right)\\
\frac{dv_s}{d\log\mu^2}&=\frac{v_s}{16\pi^2}\bigg(- |\kappa|^2 -  |\lambda|^2 \bigg)+\frac{v_s}{(16 \pi^2)^2}\bigg( 4 |\kappa|^4+ 3 y_t^2 |\lambda|^2 + 4 |\kappa|^2 |\lambda|^2 + 2 v_s |\lambda|^4\bigg)\\
\frac{d\tan\beta}{d\log\mu^2}&= -\frac{\tan\beta}{16\pi^2} \frac{3 }{2} y_t^2 +\frac{\tan\beta}{(16 \pi^2)^2}\left(-8 g_3^2 y_t^2 + \frac{9}{2} \tan\beta y_t^4 - \frac{3}{2} y_t^2 |\lambda|^2\right)\\
  \frac{d|\lambda|}{d\log\mu^2}&= \frac{| \lambda|}{16\pi^2} \left(\frac{3}{2} y_t^2 + |\kappa|^2 + 
 2 |\lambda|^2\right)\label{RGE:lambda} \\ \nonumber&\hspace*{0.4cm}+\frac{|\lambda|}{(16 \pi^2)^2}\left(8 g_3^2 y_t^2 - \frac{9}{2} y_t^4 - 
 4 |\kappa|^4 - \frac{9}{2}y_t^2 |\lambda|^2 - 6 |\kappa|^2 |\lambda|^2 - 
 5 |\lambda|^4\right)\\
  \frac{d|\kappa|}{d\log\mu^2}&= \frac{ |\kappa|}{16\pi^2} \bigg(3|\kappa|^2+3 |\lambda|^2\bigg)+\frac{|\kappa|}{(16 \pi^2)^2} \bigg(-12 |\kappa|^4 - 9 y_t^2 |\lambda|^2 - 12 |\kappa|^2 |\lambda|^2 - 6 |\lambda|^4 \bigg)\,.\label{RGE:kappa}
 \end{align}
\endgroup
Due to the supersymmetric non-renormalization theorem
\cite{PhysRevD.11.1521,GRISARU1979429} the RGEs of the superpotential
parameters are proportional to themselves. Hence, the phases
$\varphi_{\lambda}$ and $\varphi_{\kappa}$ do not need to be renormalized
separately and we can write the RGEs for $\lambda$ and $\kappa$ in
terms of their absolute values, see Eqs.~(\ref{RGE:lambda}) and
(\ref{RGE:kappa}). Instead, for the parameters
$A_{\lambda}$ and $A_{\kappa}$ we solve the RGEs for the real part and
imaginary part separately, 
\begingroup
\allowdisplaybreaks
 \begin{align}
  \frac{dA_{\lambda}}{d\log\mu^2}&=\frac{ 1}{16\pi^2} \bigg(3 A_t y_t^2 + 2 A_{\kappa} |\kappa|^2 + 
 4 A_{\lambda} |\lambda|^2 \bigg) \\
 &\hspace*{0.4cm}+\frac{ 1}{(16\pi^2)^2}\bigg( 16A_t g_3^2 y_t^2 - 16 g_3^2 M_3  y_t^2  - 18 A_t y_t^4 \nonumber  - 
 16 A_{\kappa} |\kappa|^4 - 9 A_t y_t^2 |\lambda|^2 - 
 9 A_{\lambda} y_t^2 |\lambda|^2 \\ 
&\hspace*{0.4cm} - 12 A_{\kappa} |\kappa|^2 |\lambda|^2 - 
 12 A_{\lambda} |\kappa|^2 |\lambda|^2 - 
 20 A_{\lambda} |\lambda|^4\bigg) \nonumber \\
  \frac{dA_{\kappa}}{d\log\mu^2}&=\frac{ 1}{16\pi^2} \bigg( 6 A_{\kappa} |\kappa|^2 + 
 6 A_{\lambda} |\lambda|^2 \bigg) \\
 &\hspace*{0.4cm}+\frac{ 1}{(16\pi^2)^2}\bigg( -48 A_{\kappa} |\kappa|^4 - 18 A_t y_t^2 |\lambda|^2 - 
 18 A_{\lambda} y_t^2 |\lambda|^2 - 24 A_{\kappa} |\kappa|^2 |\lambda|^2 - 
 24 A_{\lambda} |\kappa|^2 |\lambda|^2 \nonumber \\
 &\hspace*{0.4cm}-  24 A_{\lambda}|\lambda|^4\bigg)\,. \nonumber
 \end{align}
\endgroup
 For the soft SUSY breaking parameters entering the computation of the
 Higgs boson masses at one-loop level, we implemented the RGEs
 accordingly at one-loop order. They are given by 
\begingroup
\allowdisplaybreaks
\begin{align}
 \frac{dA_t}{d\log\mu^2}&=\frac{ 1}{16\pi^2}\left(\frac{16}{3} g_3^2 M_3 + 6 A_t y_t^2 + A_{\lambda} |\lambda|^2\right)\\
 \frac{dA_b}{d\log\mu^2}&=\frac{ 1}{16\pi^2}\left(\frac{16}{3} g_3^2 M_3 +  A_t y_t^2 + A_{\lambda} |\lambda|^2\right)\\
 \frac{dA_{\tau}}{d\log\mu^2}&=\frac{ 1}{16\pi^2} A_{\lambda} |\lambda|^2\\
    \frac{dm_{\tilde{Q}_3}^2}{d\log\mu^2}&=\frac{ 1}{16\pi^2}\left( m_{H_u}^2 y_t^2 + m_{\tilde{Q}_3}^2 y_t^2 + m_{\tilde{t}_R}^2 y_t^2 + y_t^2 |A_t|^2 - \frac{16}{3} g_3^2 |M_3|^2\right)\\
    \frac{dm_{\tilde{t}_R}^2}{d\log\mu^2}&=\frac{ 1}{16\pi^2}\left( 2 m_{H_u}^2 y_t^2 + 2 m_{\tilde{Q}_3}^2 y_t^2 + 2 m_{\tilde{t}_R}^2 y_t^2 + 2 y_t^2 |A_t|^2 - 
 \frac{16}{3} g_3^2 |M_3|^2\right)  \\
    \frac{dm_{\tilde{b}_R}^2}{d\log\mu^2}&=-\frac{ 1}{3\pi^2}  
  g_3^2 |M_3|^2 \\
   \frac{dM_1}{d\log\mu^2}&=\frac{ 1}{16\pi^2}\frac{33}{5} g_1^2 M_1\\
    \frac{dM_2}{d\log\mu^2}&=\frac{ 1}{16\pi^2} g_2^2 M_2\\  
   \frac{dM_3}{d\log\mu^2}&= - \frac{3}{16\pi^2} g_3
                            M_3 \,.
\end{align}
\endgroup
Like for $A_{\lambda}$ and $A_{\kappa}$, we implemented the RGEs
for the real and imaginary parts of $M_i$ ($i=1,2,3$) and $A_j$
($j=t,b,\tau$) separately. In addition, we need the RGEs for
$m_{H_u}^2$, $m_{H_d}^2$ and $m_{S}^2$ as they enter the RGEs of
$m_{\tilde{Q}_3}^2$ and $m_{\tilde{t}_R}^2$. They read
\begingroup
\allowdisplaybreaks
\begin{align}
   \frac{dm_{H_u}^2}{d\log\mu^2}&=\frac{ 1}{16\pi^2}\bigg(3 m_{H_u}^2 y_t^2 + 3 m_{\tilde{Q}_3}^2 y_t^2 + 3 m_{\tilde{t}_R}^2 y_t^2 + 3 y_t^2 |A_t|^2 + 
 m_{H_d}^2  |\lambda |^2 + m_{H_u}^2 |\lambda | ^2 \\  &\hspace*{0.4cm}+ 
 m_{S}^2 |\lambda|^2 + |A_{\lambda}|^2 |\lambda|^2 \bigg) \nonumber \\
   \frac{dm_{H_d}^2}{d\log\mu^2}&= \frac{ 1}{16\pi^2}\bigg(m_{H_d}^2 |\lambda|^2 + m_{H_u}^2 |\lambda|^2 + m_S^2 |\lambda|^2 +
  |A_{\lambda}|^2 |\lambda|^2 \bigg) \\
   \frac{dm_{S}^2}{d\log\mu^2}&= \frac{ 1}{16\pi^2}\bigg(6 m_S^2 |\kappa|^2 + 2 |A_{\kappa}|^2 |\kappa|^2 + 
 2 m_{H_d}^2 |\lambda|^2 + 2 m_{H_u}^2 |\lambda|^2 + 
 2 m_S^2 |\lambda|^2 \\
 &\hspace*{0.4cm}+ 2 |A_{\lambda}|^2 |\lambda|^2\bigg)\,. \nonumber
\end{align}
\endgroup
The RGEs are solved by means of the Runge-Kutta algorithm
evaluated to fourth order.
Note, that in the mixed $\DRb$-OS scheme we first convert all the
parameters to $\overline{\text{DR}}$ and subsequently solve the full
system of RGEs. 

\section{Tree-level Neutral Higgs Mass Matrix}\label{append:Hmass}
The matrix elements of the neutral Higgs boson mass matrix in terms of
the independent input parameters read 
\begin{align}
(\MH)_{h_dh_d} &=\frac{1}{2} \abslambda^2 \sbeta^2 v^2+\cbeta^2 M_Z^2 -M_W^2 \sbeta^2+\frac{\mhpm \sbeta^2}{\CDN^2}
+\frac{\CBN \thd (\cbeta \CBN+2 \sbeta \SBN)}{\CDN^2 v} \\
&\hspace*{0.4cm}-\frac{\CBN^2 \sbeta \thu}{\CDN^2 v} \nonumber
\displaybreak[0] \\[5mm]
(\MH)_{h_dh_u} &=\frac{1}{2} \abslambda^2 s_{2\beta} v^2 + \frac{M_W^2 s_{2\beta}}{2}-\frac{M_Z^2 s_{2\beta}}{2}-\frac{\mhpm s_{2\beta}}{2 \CDN^2}
+\frac{\cbeta \CBN^2 \thu}{\CDN^2 v}+\frac{\sbeta \SBN^2 \thd}{\CDN^2 v}
\displaybreak[0] \\[5mm]
(\MH)_{h_dh_s} &=-\frac{1}{2} \abskappa \abslambda \cphiy \sbeta v \vs-\frac{\abslambda^2 \cbeta v \left(\sbeta^2 v^2-2 \vs^2\right)}{2 \vs}-\frac{\cbeta \mhpm \sbeta^2 v}{\CDN^2 \vs}
\displaybreak[0]\\[2mm]&
\hspace*{0.4cm}+\frac{\cbeta M_W^2 \sbeta^2 v}{\vs} +\frac{\CBN^2 s_{2\beta} \thu}{2 \CDN^2 \vs}+\frac{\sbeta^2 \SBN^2 \thd}{\CDN^2 \vs}
\displaybreak[0] \nonumber \\[5mm]
(\MH)_{h_da} &=\frac{\CBN \tad}{\sbeta v}
\displaybreak[0] \\[5mm]
(\MH)_{h_da_s} &=\frac{3}{2} \abskappa \abslambda \sbeta \sphiy v \vs
+\frac{\tad}{\vs}
\displaybreak[0] \\[5mm]
(\MH)_{h_uh_u} &=\frac{1}{2} \abslambda^2 \cbeta^2 v^2+\frac{\cbeta^2 \mhpm}{\CDN^2}-\cbeta^2 M_W^2+M_Z^2 \sbeta^2+\frac{\SBN \thu (2 \cbeta \CBN+\sbeta \SBN)}{\CDN^2 v}
\displaybreak[0]\\[2mm]&
\hspace*{0.4cm}-\frac{\cbeta \SBN^2 \thd}{\CDN^2 v}
\nonumber\displaybreak[0]\\[5mm]
(\MH)_{h_uh_s} &=-\frac{1}{2} \abskappa \abslambda \cbeta \cphiy v \vs-\frac{\abslambda^2 \sbeta v \left(\cbeta^2 v^2-2 \vs^2\right)}{2 \vs}-\frac{\cbeta^2 \mhpm \sbeta v}{\CDN^2 \vs}
\displaybreak[0]\\[2mm]&
\hspace*{0.4cm}+\frac{\cbeta^2 M_W^2 \sbeta v}{\vs}+\frac{\cbeta^2 \CBN^2 \thu}{\CDN^2 \vs}+\frac{s_{2\beta} \SBN^2 \thd}{2 \CDN^2 \vs}
\nonumber\displaybreak[0]\\[5mm]
(\MH)_{h_ua} &=
\frac{\SBN \tad}{\sbeta v}
\displaybreak[0] \\[5mm]
(\MH)_{h_ua_s} &=\frac{3}{2} \abskappa \abslambda \cbeta \sphiy v \vs
+\frac{\cbeta \tad}{\sbeta \vs}
\displaybreak[0] \\[5mm]
(\MH)_{h_sh_s} &=2 \abskappa^2 \vs^2-\frac{\abskappa \abslambda s_{2\beta} v^2 (\cosks \cphiy+3\sinks \sphiy)}{4 \cosks}+\frac{\abskappa \ReAkappa \vs}{\sqrt{2} \cosks}+\frac{\mhpm s_{2\beta}^2 v^2}{4 \CDN^2 \vs^2}
\displaybreak[0]\\[2mm]&
\hspace*{0.4cm}+\frac{\abslambda^2 s_{2\beta}^2 v^4}{8 \vs^2}-\frac{M_W^2 s_{2\beta}^2 v^2}{4 \vs^2}-\frac{\cbeta^2 \CBN^2 \sbeta \thu v}{\CDN^2 \vs^2}-\frac{\cbeta \sbeta^2 \SBN^2 \thd v}{\CDN^2 \vs^2}+\frac{\ths}{\vs}
\nonumber\displaybreak[0]\\[2mm]&
\hspace*{0.4cm}+\frac{\sinks (\tas \vs-\cbeta \tad v)}{\cosks \vs^2}
\nonumber\displaybreak[0]\\[5mm]
(\MH)_{h_sa} &=-\frac{1}{2} \abskappa \abslambda \CDN \sphiy v \vs
+\frac{\CDN \tad}{\sbeta \vs}
\displaybreak[0] \\[5mm]
(\MH)_{h_sa_s} &=-\abskappa \abslambda s_{2\beta} \sphiy v^2
+\frac{2 \tas \vs-2 \cbeta \tad v}{\vs^2}
\displaybreak[0] \\[5mm]
(\MH)_{aa} &=\frac{1}{2} \abslambda^2 \CDN^2 v^2-\CDN^2 M_W^2+\mhpm
\displaybreak[0] \\[5mm]
(\MH)_{aa_s} &=-\frac{3}{2} \abskappa \abslambda \CDN \cphiy v \vs+\frac{\abslambda^2 \CDN s_{2\beta} v^3}{4 \vs}+\frac{\mhpm s_{2\beta} v}{2 \CDN \vs}-\frac{\CDN M_W^2 s_{2\beta} v}{2 \vs}
\displaybreak[0]\\[2mm]&
\hspace*{0.4cm}-\frac{\cbeta \CBN^2 \thu}{\CDN \vs}-\frac{\sbeta \SBN^2 \thd}{\CDN \vs}
\nonumber\displaybreak[0]\\[5mm]
(\MH)_{a_sa_s} &=\frac{3 \abskappa \abslambda s_{2\beta} v^2 (\cosks \cphiy+3 \sinks \sphiy)}{4 \cosks}-\frac{3 \abskappa \ReAkappa \vs}{\sqrt{2} \cosks}+\frac{\abslambda^2 s_{2\beta}^2 v^4}{8 \vs^2}-\frac{M_W^2 s_{2\beta}^2 v^2}{4 \vs^2}
\displaybreak[0]\\[2mm]&
\hspace*{0.4cm}+\frac{\ths}{\vs}-\frac{\cbeta^2 \CBN^2 \sbeta \thu v}{\CDN^2 \vs^2}-\frac{\cbeta \sbeta^2 \SBN^2 \thd v}{\CDN^2 \vs^2}+\frac{3 \sinks (\cbeta \tad v-\tas \vs)}{\cosks \vs^2}+\frac{\mhpm s_{2\beta}^2 v^2}{4 \CDN^2 \vs^2}
\nonumber\displaybreak[0]\,.
\end{align}
The gaugeless approximation is obtained by setting $M_W=M_Z=0$ but
keeping $v$ fixed. 

\section{Neutral Higgs Counterterm Mass Matrix}
\label{sec:dMH2loop}

\newcommand{\dphiksn}{(\delta^{(n)}\varphi_z)}
\newcommand{\dreakappan}{\delta^{(n)} \ReAkappa }
\newcommand{\dvsn}{\delta^{(n)} \vs}
\newcommand{\dsphiyn}{\delta^{(n)} \sphiy}
\newcommand{\dcphiyn}{\delta^{(n)} \cphiy}
\newcommand{\dtanBn}{\delta^{(n)} \tbeta}
\newcommand{\dabslambdan}{\delta^{(n)} \abslambda}
\newcommand{\dabskappan}{\delta^{(n)} \abskappa}
\newcommand{\dvvn}{\delta^{(n)} v}
\newcommand{\dmhpmn}{\delta^{(n)} \mhpm}
\newcommand{\dthun}{\delta^{(n)} \thu}
\newcommand{\dthdn}{\delta^{(n)} \thd}
\newcommand{\dthsn}{\delta^{(n)} \ths}
\newcommand{\dtadn}{\delta^{(n)} \tad}
\newcommand{\dtasn}{\delta^{(n)} \tas}
\newcommand{\dphiyn}{\delta^{(n)} \varphi_y}

In the following we present the explicit analytic form of the
counterterm mass matrix defined in Eq.~(\refeq{eq:CTmassH})  in the
basis $(h_d,h_u,h_s,a,a_s)$. At one-loop level, the matrix elements of
the Higgs counterterm mass matrix are given by
\be  
(\deltaone \MH)_{ij} = (\Deltaone \MH)_{ij} \;, \quad i,j=
h_d,h_u,h_s,a,a_s \;, 
\ee
where  the $(\Deltaone \MH)_{ij}$ are obtained from 
Eqs.~(\refeq{eq:setdeMH1})-(\refeq{eq:setdeMH2}) by setting $n=1$. At
two-loop level, the matrix elements of the Higgs counterterm mass 
matrix not only contain counterterms at two-loop order but also the product of two
one-loop counterterms,   
\be
(\deltatwo \MH)_{ij} = (\Deltatwo \MH)_{ij} +(\Delta^{\tol \tol}
\MH)_{ij} \;, 
\ee 
where the $(\Deltatwo \MH)_{ij}$ are given in
Eqs.~(\ref{eq:setdeMH1})-(\ref{eq:setdeMH2}) with $n=2$ and
$(\Delta^{\tol \tol} \MH)_{ij}$ in
Eqs.~(\ref{eq:setdeMH21})-(\ref{eq:setdeMH22}). In
  the following formulae for the counterterm mass matrix elements, we already
  applied the gaugeless approximation.

\begin{align}
 (\Delta^{(n)} \MH)_{h_dh_d}&= v |\lambda |^2 \sbeta^2 \delta^{(n)} v\,+\sbeta^2\delta^{(n)} M_{H^\pm}^2+\frac{\delta^{(n)}\thd 
   \left(1-\sbeta^4\right)}{v \cbeta }-\frac{\delta^{(n)} \thu \sbeta\cbeta^2}{v} \label{eq:setdeMH1} \displaybreak[0]\\[2mm]
\quad&\hspace*{0.4cm}+ \cbeta^3 \sbeta\braket{\abslambda^2 v^2 + 2\mhpm} \delta^{(n)} \tbeta +  \sbeta^2 v^2\abslambda \delta^{(n)}\abslambda\displaybreak[0] \nonumber\\[5mm]
 (\Delta^{(n)}\MH)_{h_dh_u}&=  v |\lambda |^2 \sbeta \cbeta\delta^{(n)} v\,-\sbeta \cbeta\delta^{(n)} \mhpm +\frac{ \sbeta^3\delta^{(n)} \thd}{v}+\frac{ \cbeta^3 \delta^{(n)} \thu}{v} \displaybreak[0]\\[2mm]
\quad&\hspace*{0.4cm}+\sbeta\cbeta v^2\abslambda\delta^{(n)}\abslambda +\fr 12 \cbeta^2(\cbeta^2-\sbeta^2)(\abslambda^2 v^2 -2 \mhpm)\delta^{(n)}\tbeta
\displaybreak[0] \nonumber \\[5mm]
 (\Delta^{(n)}\MH)_{h_dh_s}&=\frac{\cbeta^3 \dthun \sbeta}{\vs}+\frac{\dthdn \sbeta^4}{\vs}-\frac{\
\cbeta \dmhpmn \sbeta^2 v}{\vs} \displaybreak[0]\\[2mm]
&\hspace*{0.4cm} +\dvvn
\left(-\frac{\cbeta \mhpm \sbeta^2}{\vs}-\frac{1}{2} \abskappa \
\abslambda \cphiy \sbeta \vs-\frac{\abslambda^2
\cbeta \left(3 \sbeta^2 v^2-2 \vs^2\right)}{2 \vs}\right)
\nonumber \displaybreak[0]\\[2mm]
& \hspace*{0.4cm}
-\frac{1}{2} \abslambda \cphiy \dabskappan \sbeta v \vs+\frac{1}{2} \
\abskappa \abslambda \dphiyn
\sbeta \sphiy v \vs
\nonumber \displaybreak[0]\\[2mm]
&+\dabslambdan \left(-\frac{1}{2} \abskappa \cphiy \
\sbeta v \vs+\abslambda
\cbeta \left(-\frac{\sbeta^2 v^3}{\vs}+2 v \vs\right)\right)
\nonumber \displaybreak[0]\\[2mm]
&\hspace*{0.4cm}
+\dvsn \
\left(-\frac{1}{2} \abskappa \abslambda \cphiy
\sbeta v+\frac{\cbeta \mhpm \sbeta^2 v}{\vs^2}+\frac{\abslambda^2 \
\cbeta v \left(\sbeta^2 v^2+2 \vs^2\right)}{2
\vs^2}\right)
\nonumber \displaybreak[0]\\[2mm]
&\hspace*{0.4cm} 
+\dtanBn \bigg[\frac{\cbeta^2 \mhpm \sbeta \left(-2 \
\cbeta^2+\sbeta^2\right) v}{\vs}-\frac{1}{2}
\abskappa \abslambda \cbeta^3 \cphiy v \vs
\nonumber \displaybreak[0]\\[2mm]
&\hspace*{0.4cm}
-\frac{\abslambda^2 \
\cbeta^2 \sbeta v \left(\left(2 \cbeta^2-\sbeta^2\right)
v^2+2 \vs^2\right)}{2 \vs}\bigg]
\nonumber \displaybreak[0]\\[5mm] 
 (\Delta^{(n)}\MH)_{h_da}&= \fr{ \delta^{(n)} \tad}{\tbeta v}\displaybreak[0]\\[5mm] 
 (\Delta^{(n)}\MH)_{h_da_s}&=\frac{\dtadn}{\vs}+\frac{3}{2} \abskappa \abslambda \dvvn \sbeta \
\sphiy \vs +\frac{3}{2} \abskappa \abslambda \dvsn \sbeta \sphiy v  \displaybreak[0]\\[2mm]
&\hspace*{0.4cm}  
+\frac{3}{2} \abskappa \abslambda \cbeta^3 \
\dtanBn \sphiy v \vs+\frac{3}{2}
\abslambda \dabskappan \sbeta \sphiy v \vs+\frac{3}{2} \abskappa \
\dabslambdan \sbeta \sphiy v
\vs \nonumber \displaybreak[0]\\[2mm]
&\hspace*{0.4cm}
+\frac{3}{2} \
\abskappa \abslambda \cphiy
\dphiyn \sbeta v \vs \nonumber \displaybreak[0]\\[5mm]
 (\Delta^{(n)}\MH)_{h_uh_u}&=\cbeta^2 \dmhpmn-\frac{\cbeta \dthdn \sbeta^2}{v}+\frac{\dthun \
\left(2 \cbeta^2 \sbeta+\sbeta^3\right)}{v}
\displaybreak[0]\\[2mm]
&\hspace*{0.4cm} 
+\abslambda \cbeta^2 \dabslambdan v^2+\dtanBn \left(-2 \cbeta^3 \mhpm \
\sbeta-\abslambda^2 \cbeta^3 \sbeta
v^2\right) +\abslambda^2
\cbeta^2 \dvvn v   \nonumber \displaybreak[0]\\[5mm]
 (\Delta^{(n)}\MH)_{h_uh_s}&=\frac{\cbeta^4 \dthun}{\vs}+\frac{\cbeta \dthdn \sbeta^3}{\vs}-\frac{\
\cbeta^2 \dmhpmn \sbeta v}{\vs}\displaybreak[0]\\[2mm] &
\hspace*{0.4cm}+\dvvn
\left[-\frac{\cbeta^2 \mhpm \sbeta}{\vs}-\frac{1}{2} \abskappa \
\abslambda \cbeta \cphiy \vs-\frac{\abslambda^2
\sbeta \left(3 \cbeta^2 v^2-2 \vs^2\right)}{2 \vs}\right]
\nonumber\displaybreak[0]\\[2mm]
&\hspace*{0.4cm}
-\frac{1}{2} \abslambda \cbeta \cphiy \dabskappan v \vs+\frac{1}{2} \
\abskappa \abslambda \cbeta \dphiyn
\sphiy v \vs
\nonumber\displaybreak[0]\\[2mm]
&\hspace*{0.4cm}
+\dabslambdan \bigg[-\frac{1}{2} \abskappa \cbeta \cphiy \
v \vs+\abslambda \left(-\frac{\cbeta^2
\sbeta v^3}{\vs}+2 \sbeta v \vs\right)\bigg]\nonumber\displaybreak[0]\\[2mm]
&\hspace*{0.4cm}
+\dtanBn \bigg[-\frac{\
\cbeta^3 \mhpm \left(\cbeta^2-2 \sbeta^2\right)
v}{\vs}+\frac{1}{2} \abskappa \abslambda \cbeta^2 \cphiy \sbeta v \
\vs \nonumber\displaybreak[0]\\[2mm]
&\hspace*{0.4cm}
 -\frac{\abslambda^2 \cbeta^3
v \left(\left(\cbeta^2-2 \sbeta^2\right) v^2-2 \vs^2\right)}{2 \vs}\
\bigg]\nonumber\displaybreak[0]\\[2mm]
&\hspace*{0.4cm}
+\dvsn \left(-\frac{1}{2} \abskappa \abslambda
\cbeta \cphiy v+\frac{\cbeta^2 \mhpm \sbeta \
v}{\vs^2}+\frac{\abslambda^2 \sbeta v \left(\cbeta^2 v^2+2 \vs^2\right)}{2
\vs^2}\right)
\nonumber\displaybreak[0]\\[5mm]
 (\Delta^{(n)}\MH)_{h_ua}&= \frac{ \dtadn}{v} \displaybreak[0]\\[5mm]
 (\Delta^{(n)}\MH)_{h_ua_s}&= \frac{3}{2} \abskappa \abslambda \
\cbeta \dvvn \sphiy \vs  +\frac{3}{2} \abskappa \abslambda \cbeta \dvsn \sphiy v+\frac{3}{2} \abslambda \cbeta \dabskappan \sphiy v \vs \displaybreak[0]\\[2mm]
&\hspace*{0.4cm}
+\frac{\cbeta \dtadn}{\sbeta \vs}+\frac{3}{2} \
\abskappa \abslambda \cbeta \cphiy
\dphiyn v \vs+\
\frac{3}{2} \abskappa \cbeta
\dabslambdan \sphiy v \vs\nonumber\displaybreak[0]\\[2mm]
&\hspace*{0.4cm}
-\frac{3}{2} \abskappa \abslambda \cbeta^2 \
\dtanBn \sbeta \sphiy v \vs \nonumber\displaybreak[0]\\[5mm]
 (\Delta^{(n)}\MH)_{h_sh_s}&=\dvvn \left[-\frac{1}{2} \abskappa \abslambda s_{2\beta} (\cphiy+3 \
\sphiy \tanks) v+\frac{\mhpm
s_ {2\beta}^2 v}{2 \vs^2}+\frac{\abslambda^2 s_ {2\beta}^2 v^3}{2 \
\vs^2}\right]
\displaybreak[0]\\[2mm]
&\hspace*{0.4cm}
-\frac{\cbeta^4 \dthun \sbeta v}{\vs^2}-\frac{\cbeta
\dthdn \sbeta^4 v}{\vs^2}-\frac{\cbeta \dtadn \tanks v}{\vs^2}+\frac{\
\dmhpmn s_ {2\beta}^2 v^2}{4 \vs^2}
\nonumber\displaybreak[0]\\[2mm]
&\hspace*{0.4cm}
+\frac{\dthsn}{\vs}+\frac{\dtasn
\tanks}{\vs}  +  \frac{1}{4} \abskappa \abslambda \dphiyn s_{2\beta} (\sphiy-3 \cphiy \
\tanks) v^2
\nonumber\displaybreak[0]\\[2mm]
&\hspace*{0.4cm}
-\frac{3}{4} \abskappa
\abslambda \dphiksn s_ {2\beta} \sphiy \left(1+\tanks^2\right)v^2 \
\nonumber\displaybreak[0]\\[2mm]
&\hspace*{0.4cm}
+\dabslambdan \left(-\frac{1}{4} \abskappa
s_{2\beta} (\cphiy+3 \sphiy \tanks) v^2+\frac{\abslambda s_{2\beta}^2 v^4}{4 \vs^2}\right)
\nonumber\displaybreak[0]\\[2mm]
&\hspace*{0.4cm}
+\dtanBn \bigg[-\frac{1}{2}
\abskappa \abslambda c_ {2\beta} \cbeta^2 (\cphiy+3 \sphiy \tanks) \
v^2+\frac{2 c_ {2\beta} \cbeta^3 \mhpm
\sbeta v^2}{\vs^2}
\nonumber\displaybreak[0]\\[2mm]
&\hspace*{0.4cm}
+\frac{\abslambda^2 c_ {2\beta} \cbeta^3 \sbeta \
v^4}{\vs^2}\bigg]+\frac{\abskappa (\dreakappan+\dphiksn
\ReAkappa \tanks) \vs}{\sqrt{2} \cosks}
\nonumber\displaybreak[0]\\[2mm]
&\hspace*{0.4cm}
+\dvsn \left[\frac{\abskappa \
\ReAkappa}{\sqrt{2} \cosks}-\frac{\mhpm
s_ {2\beta}^2 v^2}{2 \vs^3}-\frac{\abslambda^2 s_ {2\beta}^2 v^4}{4 \
\vs^3}+4 \abskappa^2 \vs\right]
\nonumber\displaybreak[0]\\[2mm]
&\hspace*{0.4cm}
+\dabskappan
\left[-\frac{1}{4} \abslambda s_{2\beta} (\cphiy+3 \sphiy \tanks) \
v^2+\frac{\ReAkappa \vs}{\sqrt{2} \cosks}+4
\abskappa \vs^2\right]  \nonumber\displaybreak[0]\\[5mm]
 (\Delta^{(n)}\MH)_{h_sa}&= 
-\frac{1}{2} \
\abskappa \abslambda \cphiy \dphiyn
v \vs   -\frac{1}{2} \abslambda \dabskappan \sphiy v \vs-\frac{1}{2} \
\abskappa \dabslambdan \sphiy
v \vs
\displaybreak[0]\\[2mm]
&\hspace*{0.4cm}+ \frac{\dtadn}{\sbeta \vs}-\frac{1}{2} \abskappa \abslambda \dvvn \
\sphiy \vs -\frac{1}{2} \abskappa \abslambda \dvsn \sphiy v \nonumber\displaybreak[0]\\[5mm]
 (\Delta^{(n)}\MH)_{h_sa_s}&= -2 \abskappa \abslambda \dvvn s_ {2\beta} \sphiy v-\frac{2 \cbeta \
\dtadn v}{\vs^2}+\frac{2 \dtasn}{\vs}  -\abskappa
\dabslambdan s_ {2\beta} \sphiy v^2
\displaybreak[0]\\[2mm]
&\hspace*{0.4cm} +2 \abskappa \abslambda \cbeta^2 \
\dtanBn \left(\sbeta^2 - \cbeta^2 \right)
\sphiy v^2
-\abslambda \
\dabskappan s_ {2\beta} \sphiy v^2 \nonumber \displaybreak[0]\\[2mm]
&\hspace*{0.4cm}  -\abskappa \abslambda \cphiy \dphiyn s_{2\beta} v^2 
\nonumber\displaybreak[0]\\[5mm]
 (\Delta^{(n)}\MH)_{aa}&= \dmhpmn+\abslambda^2 \dvvn v +\abslambda \dabslambdan v^2 \displaybreak[0]\\[5mm]
 (\Delta^{(n)}\MH)_{aa_s}&=-\frac{\cbeta^3 \dthun}{\vs}-\frac{\dthdn \
\sbeta^3}{\vs}+\frac{\dmhpmn s_{2\beta} v}{2 \vs}
\displaybreak[0]\\[2mm]
&\hspace*{0.4cm}
+\dvvn
\left[\frac{\mhpm s_{2\beta}}{2 \vs}+\frac{3 \abslambda^2 s_{2\beta} \
v^2}{4 \vs}-\frac{3}{2} \abskappa \abslambda
\cphiy \vs\right]
\nonumber\displaybreak[0]\\[2mm]
&\hspace*{0.4cm}
   +\dvsn \left[-\frac{3}{2} \abskappa \abslambda \cphiy v-\frac{\mhpm \
s_{2\beta} v}{2 \vs^2}-\frac{\abslambda^2
s_{2\beta} v^3}{4 \vs^2}\right]
\nonumber\displaybreak[0]\\[2mm]
&\hspace*{0.4cm}
+\dtanBn \left[\frac{\cbeta^2 \mhpm \
\left(\cbeta^4-\sbeta^4\right) v}{\vs}+\frac{\abslambda^2
c_ {2\beta} \cbeta^2 v^3}{2 \vs}\right]
\nonumber\displaybreak[0]\\[2mm]
&\hspace*{0.4cm}
-\frac{3}{2} \abslambda \
\cphiy \dabskappan v \vs+\frac{3}{2} \abskappa
\abslambda \dphiyn \sphiy v \vs\nonumber\displaybreak[0]\\[2mm]
&\hspace*{0.4cm}+\dabslambdan \left[\frac{\abslambda \
s_{2\beta} v^3}{2 \vs}-\frac{3}{2}
\abskappa \cphiy v \vs\right] \nonumber\displaybreak[0]\\[5mm]
 (\Delta^{(n)}\MH)_{a_sa_s}&= \dvvn \left[\frac{3}{2} \abskappa \abslambda s_{2\beta} (\cphiy+3 \
\sphiy \tanks) v+\frac{\mhpm s_ {2\beta}^2
v}{2 \vs^2}+\frac{\abslambda^2 s_ {2\beta}^2 v^3}{2 \vs^2}\right] \label{eq:setdeMH2}
\displaybreak[0]\\[2mm]
&\hspace*{0.4cm}
-\frac{\cbeta^4 \dthun \sbeta v}{\vs^2}-\frac{\cbeta
\dthdn \sbeta^4 v}{\vs^2}+\frac{3 \cbeta \dtadn \tanks \
v}{\vs^2}
\nonumber\displaybreak[0]\\[2mm]
&\hspace*{0.4cm}
+\frac{\dmhpmn s_ {2\beta}^2 v^2}{4 \
\vs^2}+\frac{\dthsn}{\vs}-\frac{3
\dtasn \tanks}{\vs}  
\nonumber\displaybreak[0]\\[2mm]
&\hspace*{0.4cm}
-\frac{3}{4} \abskappa \abslambda \dphiyn s_{2\beta} (\sphiy-3 \cphiy \
\tanks) v^2
\nonumber\displaybreak[0]\\[2mm]
&\hspace*{0.4cm}
+\frac{9}{4} \abskappa
\abslambda \dphiksn s_ {2\beta} \sphiy \left(1+\tanks^2\right) \
v^2
\nonumber\displaybreak[0]\\[2mm]
&\hspace*{0.4cm}
+\dvsn \left(-\frac{3 \abskappa \ReAkappa}{\sqrt{2}
\cosks}-\frac{\mhpm s_ {2\beta}^2 v^2}{2 \vs^3}-\frac{\abslambda^2 s_{2\beta}^2 v^4}{4 \vs^3}\right)
\nonumber\displaybreak[0]\\[2mm]
&\hspace*{0.4cm}
+\dabslambdan
\left[\frac{3}{4} \abskappa s_{2\beta} (\cphiy+3 \sphiy \tanks) \
v^2+\frac{\abslambda s_ {2\beta}^2 v^4}{4 \vs^2}\right]
\nonumber\displaybreak[0]\\[2mm]
&\hspace*{0.4cm}
+\dtanBn
\left[\frac{3}{2} \abskappa \abslambda c_ {2\beta} \cbeta^2 (\cphiy+3 \
\sphiy \tanks) v^2+\frac{2 c_ {2\beta}
\cbeta^3 \mhpm \sbeta v^2}{\vs^2} \right.
\nonumber\displaybreak[0]\\[2mm]
&\hspace*{0.4cm} \left. +\frac{\abslambda^2 c_ {2\beta} \
\cbeta^3 \sbeta v^4}{\vs^2}\right]
-\frac{3 \abskappa
(\dreakappan+\dphiksn \ReAkappa \tanks) \vs}{\sqrt{2} \
\cosks}
\nonumber\displaybreak[0]\\[2mm]
&\hspace*{0.4cm}
+\dabskappan \left[\frac{3}{4} \abslambda
s_{2\beta} (\cphiy+3 \sphiy \tanks) v^2-\frac{3 \ReAkappa \
\vs}{\sqrt{2} \cosks}\right]
\nonumber\displaybreak[0]\,.
\end{align} 

\renewcommand{\dphiksn}{\deltaone\varphi_z}
\renewcommand{\dreakappan}{\deltaone \ReAkappa }
\renewcommand{\dvsn}{\deltaone \vs}
\renewcommand{\dsphiyn}{\deltaone \sphiy}
\renewcommand{\dcphiyn}{\deltaone \cphiy}
\renewcommand{\dtanBn}{\deltaone \tbeta}
\renewcommand{\dabslambdan}{\deltaone \abslambda}
\renewcommand{\dabskappan}{\deltaone \abskappa}
\renewcommand{\dvvn}{\deltaone v}
\renewcommand{\dmhpmn}{\deltaone \mhpm}
\renewcommand{\dthun}{\deltaone \thu}
\renewcommand{\dthdn}{\deltaone \thd}
\renewcommand{\dthsn}{\deltaone \ths}
\renewcommand{\dtadn}{\deltaone \tad}
\renewcommand{\dtasn}{\deltaone \tas}
\renewcommand{\dphiyn}{\deltaone \varphi_y}

In the following, we present the matrix elements of the Higgs 
counterterm mass matrix containing the product of two one-loop
counterterms which contribute at two-loop order.  All one-loop
counterterms $\dvsn,\dabskappan,$ $ \dreakappan, \dphiyn,\dphiksn$ are
zero.  

\begin{align}
 (\Delta^{\tol \tol} &\MH)_{h_dh_d}=\abslambda^2 (\dvvn)^2 \sbeta^2-\frac{2 \dthdn \dvvn \left(\cbeta^3+2 \cbeta \sbeta^2\right)}{v^2}+\frac{2 \cbeta^2 \dthun \dvvn \sbeta}{v^2}
\label{eq:setdeMH21}\displaybreak[0]\\[2mm]&\hspace*{0.4cm}
+4 \abslambda \dabslambdan \dvvn \sbeta^2 v+(\dabslambdan)^2 \sbeta^2 v^2+\cbeta^6 (\dtanBn)^2 \left(\abslambda^2 v^2+2 \mhpm\right)
\nonumber\displaybreak[0]\\[2mm]&\hspace*{0.4cm}
+4 \cbeta^3 \dtanBn \sbeta \left(\abslambda \dabslambdan v^2+\dmhpmn\right)+4 \abslambda^2 \cbeta^3 \dtanBn \dvvn \sbeta v
\nonumber\displaybreak[0]\\[2mm]&\hspace*{0.4cm}
-\frac{2 \cbeta^5 \dtanBn \dthun}{v}+\frac{2 \cbeta^4 \dtanBn \dthdn \sbeta}{v}\displaybreak[0]\nonumber\\[5mm]
(\Delta^{\tol \tol} &\MH)_{h_dh_u}=\frac{1}{2} \abslambda^2 (\dvvn)^2 s_{2\beta}-\frac{2 \cbeta^3 \dthun \dvvn}{v^2}-\frac{2 \dthdn \dvvn \sbeta^3}{v^2}+2 \abslambda \dabslambdan \dvvn s_{2\beta} v
\displaybreak[0]\\[2mm]&\hspace*{0.4cm}
+2 \abslambda^2 c_{2\beta} \cbeta^2 \dtanBn \dvvn v-\frac{2 \cbeta^4 \dtanBn \dthun \sbeta}{v}+\frac{2 \cbeta^3 \dtanBn \dthdn \sbeta^2}{v}
\nonumber\displaybreak[0]\\[2mm]&\hspace*{0.4cm}
+\cbeta^5 (\dtanBn)^2 \sbeta \left(2 \mhpm-\abslambda^2 v^2\right)-2 c_{2\beta} \cbeta^2 \dtanBn \left(\dmhpmn-\abslambda \dabslambdan v^2\right)
\nonumber\displaybreak[0]\\[2mm]&\hspace*{0.4cm}
+\frac{1}{2} (\dabslambdan)^2 s_{2\beta} v^2\nonumber\displaybreak[0]\\[5mm]
(\Delta^{\tol \tol} &\MH)_{h_dh_s}=-\frac{3 \abslambda^2 \cbeta \dvvn^2 \sbeta^2 v}{\vs}-\frac{2 \cbeta \dmhpmn \dvvn \sbeta^2}{\vs}+\cbeta \dabslambdan^2 \left(2 v \vs-\frac{\sbeta^2 v^3}{\vs}\right)
\displaybreak[0]\\[2mm]&\hspace*{0.4cm}
+\dabslambdan \dvvn \left(\abslambda \cbeta \left(4 \vs-\frac{6 \sbeta^2 v^2}{\vs}\right)-\abskappa \cphiy \sbeta \vs\right)+\frac{4 \cbeta^3 \dtanBn \dthdn \sbeta^3}{\vs}
\nonumber\displaybreak[0]\\[2mm]&\hspace*{0.4cm}
+\frac{\cbeta^2 \dtanBn v \left(2 \sbeta \left(\sbeta^2-2 \cbeta^2\right) \left(\abslambda \dabslambdan v^2+\dmhpmn\right)\right)}{\vs}
\nonumber\displaybreak[0]\\[2mm]&\hspace*{0.4cm}
-\frac{\dabslambdan \vs^2 (\abskappa \cbeta \cphiy+4 \abslambda
                                  \sbeta) +\cbeta^5 \dtanBn^2 v \left(\cbeta^2-2 \sbeta^2\right) \left(\abslambda^2 v^2+2 \mhpm\right)}{\vs}
\nonumber\displaybreak[0]\\[2mm]&\hspace*{0.4cm}
+\dtanBn \dvvn \Bigg[-\abskappa \abslambda \cbeta^3 \cphiy \vs-\frac{\abslambda^2 \cbeta^2 \sbeta \left(6 \cbeta^2 v^2-3 \sbeta^2 v^2+2 \vs^2\right)}{\vs}
\nonumber\displaybreak[0]\\[2mm]&\hspace*{0.4cm}
+\frac{2 \cbeta^2 \mhpm \sbeta \left(\sbeta^2-2 \cbeta^2\right)}{\vs}\Bigg]+\frac{2 c_{2\beta} \cbeta^4 \dtanBn \dthun}{\vs}\nonumber\displaybreak[0]\\[5mm]
(\Delta^{\tol \tol} &\MH)_{h_da}=-\frac{2 \cbeta \dtadn \dvvn}{\sbeta v^2}
-\frac{2 \cbeta^4 \dtadn \dtanBn}{\sbeta^2 v}\displaybreak[0]\\[5mm]
(\Delta^{\tol \tol} &\MH)_{h_da_s}=-\frac{2 \cbeta \dtadn \dvvn}{\sbeta v^2}
-\frac{2 \cbeta^4 \dtadn \dtanBn}{\sbeta^2 v}
+3 \abskappa \dabslambdan \dvvn \sbeta \sphiy \vs
\displaybreak[0]\\[2mm]&\hspace*{0.4cm}
+3 \abskappa \cbeta^3 \dabslambdan \dtanBn \sphiy v \vs
+3 \abskappa \abslambda \cbeta^3 \dtanBn \dvvn \sphiy \vs
\nonumber\displaybreak[0]\\[5mm]
(\Delta^{\tol \tol} &\MH)_{h_uh_u}=-\frac{2 \cbeta \dtadn \dvvn}{\sbeta v^2}
-\frac{2 \cbeta^4 \dtadn \dtanBn}{\sbeta^2 v}
+\frac{2 \cbeta \dthdn \dvvn \sbeta^2}{v^2}
\displaybreak[0]\\[2mm]&\hspace*{0.4cm}
+3 \abskappa \cbeta^3 \dabslambdan \dtanBn \sphiy v \vs+4 \abslambda \cbeta^2 \dabslambdan \dvvn v+\cbeta^2 \dabslambdan^2 v^2
\nonumber\displaybreak[0]\\[2mm]&\hspace*{0.4cm}
+3 \abskappa \abslambda \cbeta^3 \dtanBn \dvvn \sphiy \vs\abslambda^2 \cbeta^2 \dvvn^2-\frac{2 \dthun \dvvn \left(2 \cbeta^2 \sbeta+\sbeta^3\right)}{v^2}
\nonumber\displaybreak[0]\\[2mm]&\hspace*{0.4cm}
+\cbeta^4 \dtanBn^2 \sbeta^2 \left(\abslambda^2 v^2+2 \mhpm\right)-4 \cbeta^3 \dtanBn \sbeta \left(\abslambda \dabslambdan v^2+\dmhpmn\right)
\nonumber\displaybreak[0]\\[2mm]&\hspace*{0.4cm}
-4 \abslambda^2 \cbeta^3 \dtanBn \dvvn \sbeta v-\frac{2 \cbeta^3 \dtanBn \dthun \sbeta^2}{v}+\frac{2 \cbeta^2 \dtanBn \dthdn \sbeta^3}{v}
\nonumber\displaybreak[0]\\[2mm]&\hspace*{0.4cm}
+3 \abskappa \dabslambdan \dvvn \sbeta \sphiy \vs
\nonumber\displaybreak[0]\\[5mm]
(\Delta^{\tol \tol} &\MH)_{h_uh_s}=-\frac{3 \abslambda^2 \cbeta^2 \dvvn^2 \sbeta v}{\vs}-\frac{2 \cbeta^2 \dmhpmn \dvvn \sbeta}{\vs}-\frac{4 \cbeta^5 \dtanBn \dthun \sbeta}{\vs}
\displaybreak[0]\\[2mm]&\hspace*{0.4cm}
+\dabslambdan \dvvn \left[\abslambda \left(4 \sbeta \vs-\frac{6 \cbeta^2 \sbeta v^2}{\vs}\right)-\abskappa \cbeta \cphiy \vs\right] +\frac{2 c_{2\beta} \cbeta^2 \dtanBn \dthdn \sbeta^2}{\vs}
\nonumber\displaybreak[0]\\[2mm]&\hspace*{0.4cm}
+\frac{\cbeta^2 \dtanBn v \left(\dabslambdan \vs^2 (\abskappa \cphiy \sbeta+4 \abslambda \cbeta)-2 \cbeta \left(\cbeta^2-2 \sbeta^2\right) \left(\abslambda \dabslambdan v^2+\dmhpmn\right)\right)}{\vs}
\nonumber\displaybreak[0]\\[2mm]&\hspace*{0.4cm}
+\frac{\cbeta^4 \dtanBn^2 \sbeta v \left(2 \cbeta^2-\sbeta^2\right) \left(\abslambda^2 v^2+2 \mhpm\right)}{\vs}+(\dabslambdan)^2 \left[2 \sbeta v \vs-\frac{\cbeta^2 \sbeta v^3}{\vs}\right]
\nonumber\displaybreak[0]\\[2mm]&\hspace*{0.4cm}
+\dtanBn \dvvn \Bigg[\abskappa \abslambda \cbeta^2 \cphiy \sbeta \vs-\frac{\abslambda^2 \cbeta^3 \left(3 v^2 \left(\cbeta^2-2 \sbeta^2\right)-2 \vs^2\right)}{\vs}
\nonumber\displaybreak[0]\\[2mm]&\hspace*{0.4cm}
-\frac{2 \cbeta^3 \mhpm \left(\cbeta^2-2 \sbeta^2\right)}{\vs}\Bigg]\nonumber\displaybreak[0]\\[5mm]
(\Delta^{\tol \tol} &\MH)_{h_ua}=-\frac{2 \dtadn \dvvn}{v^2}
-\frac{2 \cbeta^3 \dtadn \dtanBn}{\sbeta v}\displaybreak[0]\\[5mm]
(\Delta^{\tol \tol} &\MH)_{h_ua_s}= 3 \abskappa \cbeta \dabslambdan \dvvn \sphiy \vs
+\cbeta^2 \dtanBn \left(-3 \abskappa \dabslambdan \sbeta \sphiy v \vs-\frac{2 \dtadn}{\sbeta^2 \vs}\right)\displaybreak[0]\\[2mm]&\hspace*{0.4cm}
-3 \abskappa \abslambda \cbeta^2 \dtanBn \dvvn \sbeta \sphiy \vs\nonumber\displaybreak[0]\\[5mm]
(\Delta^{\tol \tol} &\MH)_{h_sh_s}=\dvvn^2 \left(-\frac{1}{2} \abskappa \abslambda s_{2\beta} (\cphiy+3 \sphiy \tanks)+\frac{3 \abslambda^2 s_{2\beta}^2 v^2}{2 \vs^2}+\frac{\mhpm s_{2\beta}^2}{2 \vs^2}\right)
\displaybreak[0]\\[2mm]&\hspace*{0.4cm}
-\frac{2 \cbeta^4 \dthun \dvvn \sbeta}{\vs^2}
-\frac{2 \cbeta \dtadn \dvvn \tanks}{\vs^2}-\frac{2 \cbeta \dthdn \dvvn \sbeta^4}{\vs^2}+\frac{\dabslambdan^2 s_{2\beta}^2 v^4}{4 \vs^2}
\nonumber\displaybreak[0]\\[2mm]&\hspace*{0.4cm}
+\frac{\dmhpmn \dvvn s_{2\beta}^2 v}{\vs^2}+\dabslambdan \dvvn \left(\frac{2 \abslambda s_{2\beta}^2 v^3}{\vs^2}-\abskappa s_{2\beta} v (\cphiy+3 \sphiy \tanks)\right)
\nonumber\displaybreak[0]\\[2mm]&\hspace*{0.4cm}
+\frac{\cbeta^4 \dtanBn^2 v^2\left( \frac{1}{2} \abskappa \abslambda s_{2\beta} \vs^2 (\cphiy+3 \sphiy \tanks)+\left(\abslambda^2 v^2+2 \mhpm\right) \left(\cbeta^4-s_{2\beta}^2+\sbeta^4\right) \right) }{\vs^2} 
\nonumber\displaybreak[0]\\[2mm]&\hspace*{0.4cm}
+\dtanBn \dvvn \bigg[-2 \abskappa \abslambda c_{2\beta} \cbeta^2 v (\cphiy+3 \sphiy \tanks)+\frac{8 \abslambda^2 c_{2\beta} \cbeta^3 \sbeta v^3}{\vs^2}
+\frac{8 c_{2\beta} \cbeta^3 \mhpm \sbeta v}{\vs^2}\bigg]
\nonumber\displaybreak[0]\\[2mm]&\hspace*{0.4cm}
+\frac{\cbeta^2 \dtanBn v }{\vs^2}\bigg[c_{2\beta} v \left(4 \cbeta \sbeta \left(\abslambda \dabslambdan v^2+\dmhpmn\right)-\abskappa \dabslambdan \vs^2 
(\cphiy+3 \sphiy \tanks)\right)
\nonumber\displaybreak[0]\\[2mm]&\hspace*{0.4cm}
+2 \dtadn \sbeta \tanks\bigg] +\frac{2 \cbeta^2 \dtanBn \dthdn \sbeta^3 v \left(\sbeta^2-2 \cbeta^2\right)}{\vs^2}-\frac{2 \cbeta^5 \dtanBn \dthun v \left(\cbeta^2-2 \sbeta^2\right)}{\vs^2}
\nonumber\displaybreak[0]\\[5mm]
(\Delta^{\tol \tol} &\MH)_{h_sa}=-\abskappa \dabslambdan \dvvn \sphiy \vs
-\frac{2 \cbeta^3 \dtadn \dtanBn}{\sbeta^2 \vs}\displaybreak[0]\\[5mm]
(\Delta^{\tol \tol} &\MH)_{h_sa_s}=4 \abskappa \abslambda \cbeta^5 \dtanBn^2 \sbeta \sphiy v^2
-4 \abskappa \dabslambdan \dvvn s_{2\beta} \sphiy v -\frac{4 \cbeta \dtadn \dvvn}{\vs^2}
\displaybreak[0]\\[2mm]&\hspace*{0.4cm}
+\frac{4 \cbeta^2 \dtanBn v \left(\abskappa \dabslambdan \sphiy v \vs^2 \left(\sbeta^2-\cbeta^2\right)+\dtadn \sbeta\right)}{\vs^2}-2 \abskappa \abslambda \dvvn^2 s_{2\beta} \sphiy
\nonumber\displaybreak[0]\\[2mm]&\hspace*{0.4cm}
+8 \abskappa \abslambda \cbeta^2 \dtanBn \dvvn \sphiy v \left(\sbeta^2-\cbeta^2\right)\nonumber\displaybreak[0]\\[5mm]
(\Delta^{\tol \tol} &\MH)_{aa}=-2 \abskappa \abslambda \dvvn^2 s_{2\beta} \sphiy +\dabslambdan^2 v^2+4 \abskappa \abslambda \cbeta^5 \dtanBn^2 \sbeta \sphiy v^2
\displaybreak[0]\\[2mm]&\hspace*{0.4cm}
-\frac{4 \cbeta \dtadn \dvvn}{\vs^2}
+\frac{4 \cbeta^2 \dtanBn v \left(\abskappa \dabslambdan \sphiy v \vs^2 \left(\sbeta^2-\cbeta^2\right)+\dtadn \sbeta\right)}{\vs^2}
\nonumber\displaybreak[0]\\[2mm]&\hspace*{0.4cm}
+8 \abskappa \abslambda \cbeta^2 \dtanBn \dvvn \sphiy v \left(\sbeta^2-\cbeta^2\right)\abslambda^2 \dvvn^2
+4 \abslambda \dabslambdan \dvvn v 
\nonumber\displaybreak[0]\\[2mm]&\hspace*{0.4cm}
-4 \abskappa \dabslambdan \dvvn s_{2\beta} \sphiy v
\nonumber\displaybreak[0]\\[5mm]
(\Delta^{\tol \tol} &\MH)_{aa_s}=\frac{3 \abslambda^2 \dvvn^2 s_{2\beta} v}{2 \vs}+\frac{\dmhpmn \dvvn s_{2\beta}}{\vs}+\frac{2 \cbeta^4 \dtanBn \dthun \sbeta}{\vs}
\displaybreak[0]\\[2mm]&\hspace*{0.4cm}
+\dabslambdan \dvvn \left(\frac{3 \abslambda s_{2\beta} v^2}{\vs}-3 \abskappa \cphiy \vs\right)+\frac{\dabslambdan^2 s_{2\beta} v^3}{2 \vs}-\frac{2 \cbeta^3 \dtanBn \dthdn \sbeta^2}{\vs}
\nonumber\displaybreak[0]\\[2mm]&\hspace*{0.4cm}
+\frac{2 \cbeta^2 \dtanBn v \left(\cbeta^4-\sbeta^4\right) \left(\abslambda \dabslambdan v^2+\dmhpmn\right)}{\vs}-\frac{\cbeta^5 \dtanBn^2 \sbeta v \left(\abslambda^2 v^2+2 \mhpm\right)}{\vs}
\nonumber\displaybreak[0]\\[2mm]&\hspace*{0.4cm}
+\dtanBn \dvvn \left(\frac{3 \abslambda^2 c_{2\beta} \cbeta^2 v^2}{\vs}+\frac{2 \cbeta^2 \mhpm \left(\cbeta^4-\sbeta^4\right)}{\vs}\right)
\nonumber\displaybreak[0]\\[5mm]
(\Delta^{\tol \tol} &\MH)_{a_sa_s}=(\dvvn)^2 \left[\frac{3}{2} \abskappa \abslambda s_{2\beta} (\cphiy+3 \sphiy \tanks)+\frac{3 \abslambda^2 s_{2\beta}^2 v^2}{2 \vs^2}+\frac{\mhpm s_{2\beta}^2}{2 \vs^2}\right]\label{eq:setdeMH22}
\displaybreak[0]\\[2mm]&\hspace*{0.4cm}
-\frac{2 \cbeta^4 \dthun \dvvn \sbeta}{\vs^2}+\frac{6 \cbeta \dtadn \dvvn \tanks}{\vs^2}-\frac{2 \cbeta \dthdn \dvvn \sbeta^4}{\vs^2}+\frac{\dmhpmn \dvvn s_{2\beta}^2 v}{\vs^2}
\nonumber\displaybreak[0]\\[2mm]&\hspace*{0.4cm}
+\dabslambdan \dvvn \left[3 \abskappa s_{2\beta} v (\cphiy+3 \sphiy \tanks)+\frac{2 \abslambda s_{2\beta}^2 v^3}{\vs^2}\right]+\frac{\dabslambdan^2 s_{2\beta}^2 v^4}{4 \vs^2}
\nonumber\displaybreak[0]\\[2mm]&\hspace*{0.4cm}
+\dtanBn \dvvn \left[6 \abskappa \abslambda c_{2\beta} \cbeta^2 v (\cphiy+3 \sphiy \tanks)+\frac{8 \abslambda^2 c_{2\beta} \cbeta^3 \sbeta v^3}{\vs^2}+\frac{8 c_{2\beta} \cbeta^3 \mhpm \sbeta v}{\vs^2}\right]
\nonumber\displaybreak[0]\\[2mm]&\hspace*{0.4cm}
+\frac{2 \cbeta^2 \dtanBn \dthdn \sbeta^3 v \left(\sbeta^2-2 \cbeta^2\right)}{\vs^2}-\frac{2 \cbeta^5 \dtanBn \dthun v \left(\cbeta^2-2 \sbeta^2\right)}{\vs^2}
\nonumber\displaybreak[0]\\[2mm]&\hspace*{0.4cm}
+\frac{\cbeta^4 \dtanBn^2 v^2 }{\vs^2}\bigg[\left(\abslambda^2 v^2+2 \mhpm\right) \left(\cbeta^4-s_{2\beta}^2+\sbeta^4\right)-\frac{3}{2} \abskappa \abslambda s_{2\beta} \vs^2 (\cphiy+3 \sphiy \tanks)\bigg]
\nonumber\displaybreak[0]\\[2mm]&\hspace*{0.4cm}
+\frac{\cbeta^2 \dtanBn v}{\vs^2} \bigg[c_{2\beta} v \bigg(3 \abskappa \dabslambdan \vs^2 (\cphiy+3 \sphiy \tanks)+4 \cbeta \sbeta \left(\abslambda \dabslambdan v^2+\dmhpmn\right)\bigg)
\nonumber\displaybreak[0]\\[2mm]&\hspace*{0.4cm}
-6 \dtadn \sbeta \tanks\bigg]
\nonumber\displaybreak[0]\,.
\end{align}

\section{Charged Higgs Boson Mass Counterterm}
\label{sec:dMHp2loop}
In the following we present the explicit analytic form of the
counterterm of the charged Higgs boson mass, defined in
Eqs.~(\ref{eq:mhpmone}) and (\ref{eq:mhpmtwo}), which
is needed if $\ReAlambda$ is chosen as 
independent input parameter. At one-loop level, the counterterm of
the charged Higgs boson mass is given by
\be  
\deltaone M_{H^\pm} ^2  = \Deltaone M_{H^\pm} ^2 \,,
\ee
where the $\Deltaone M_{H^\pm} ^2$ is obtained from
\eqref{eq:chargedHiggsGenuineCT} by setting $n=1$. At
two-loop level, the charged Higgs mass counterterm not only contains
counterterms at two-loop order but also the product of two one-loop
counterterms,   
\be
\deltatwo M_{H^\pm} ^2 = \Deltatwo M_{H^\pm} ^2 + \Delta^{\tol \tol}
M_{H^\pm} ^2 \,, 
\ee 
where the $\Deltatwo M_{H^\pm} ^2$ are obtained from
\eqref{eq:chargedHiggsGenuineCT} for $n=2$ and $\Delta^{\tol \tol}
M_{H^\pm} ^2$ is given in \eqref{eq:chargedHiggsProductCT}. In
  the following formulae for the mass counterterm, we already applied
  the gaugeless approximation and present only the terms that are
  relevant for the calculation at order $\order$. The counterterm
  contribution $\Delta^{(n)} M_{H^\pm}^2$ reads
\begingroup
\allowdisplaybreaks
\begin{align}
 \Delta^{(n)} M_{H^\pm}^2 &=\left[\frac{  \vs \left(\abskappa \vs c_\phiom+\sqrt{2} \ReAlambda \right)}{s_{2\beta} c_{\phiom-\phiy}} -\abslambda v^2 \right]\delta^{(n)}\abslambda +\frac{\sbeta^2}{\cbeta v}\delta^{(n)} \thd  -\abslambda^2 v\delta^{(n)} v  \label{eq:chargedHiggsGenuineCT} \\
&\hspace*{0.4cm} +\frac{\cbeta^2 }{\sbeta v}\delta^{(n)} \thu+\frac{\tan (\phiy-\phiom)}{\cbeta \sbeta^2 v}\delta^{(n)} \tad -\frac{\abslambda c_{2\beta} \vs  \left(\abskappa \vs c_\phiom+\sqrt{2} \ReAlambda\right)}{2 \sbeta^2  c_{\phiom-\phiy} }\delta^{(n)} \tbeta \nonumber \\
&\hspace*{0.4cm}+\frac{\abslambda   \left(2 \abskappa \vs
  c_\phiom+\sqrt{2}
  \ReAlambda\right)}{s_{2\beta}c_{\phiom-\phiy}}\delta^{(n)}
  \vs+\frac{\abslambda \vs^2 c_{\phiom} }{s_{2\beta}c_{\phiom-\phiy}}
  \delta^{(n)} \abskappa+\frac{\sqrt{2}\abslambda \vs }{s_{2\beta}
  c_{\phiom-\phiy}}\delta^{(n)} \ReAlambda \,.
\nonumber 
\end{align}
\endgroup
Next, we give the charged Higgs mass counterterm containing the
product of two one-loop counterterms which contribute at two-loop
order at order $\order$. All one-loop counterterms
$\dvsn,\dabskappan,$ $ \dreakappan, \dphiyn,\dphiksn$ are zero. The
counterterm contribution reads 
\begin{align}
	\Delta^{\tol \tol} M_{H^\pm}^2 &=-\frac{ v^2}{2}(\dabslambdan)^2 -\frac{\abslambda^2 (\deltaone v)^2}{2}+\frac{\abslambda \cbeta  \vs c_{2\beta} \left(\abskappa \vs c_\phiom+\sqrt{2} \ReAlambda\right)}{2 \sbeta^3 c_{\phiom-\phiy}} (\dtanBn)^2
\label{eq:chargedHiggsProductCT}
\\
+& \left(\frac{ \sqrt{2}\vs}{s_{2\beta} c_{\phiom-\phiy}}\deltaone\ReAlambda-2 \abslambda v \deltaone v  -\frac{c_{2\beta}  \vs \left(\abskappa \vs c_\phiom+\sqrt{2} \ReAlambda\right)}{2 \sbeta^2 c_{\phiom-\phiy}}\dtanBn\right)\dabslambdan\crn
-&\fr{1}{v^2}\left(\frac{ \tan (\phiy-\phiom)}{\cbeta \sbeta^2 }\dtadn+\frac{\cbeta^2 }{\sbeta}\dthun+\frac{ \sbeta^2}{\cbeta}\dthdn\right)\deltaone v+\frac{ \sbeta^3}{v}\dtanBn \dthdn
\crn
-&\left(\frac{  \left(\sbeta^2-2 \cbeta^2\right) \tan
  (\phiom-\phiy)}{\sbeta^3 v}\dtadn+\frac{\cbeta^5  }{\sbeta^2
  v}\dthun+\frac{\abslambda c_{2\beta}   \vs}{\sqrt{2} \sbeta^2
  c_{\phiom-\phiy}}\deltaone\ReAlambda\right) \dtanBn \crn
+& \frac{ 2v_s |\kappa | c_{\varphi _\omega } + \sqrt{2} 
\text{Re}\,A_\lambda }{ s_{2\beta} c_{\phi _\omega - \phi _y} } 
\deltaone v_s \deltaone |\lambda |
+ \frac{ \sqrt{2}|\lambda | }{ s_{2\beta} c_{\varphi _\omega - \varphi 
_y} } \deltaone v_s \deltaone \text{Re}\,A_\lambda
+ \frac{ 2v_s |\lambda | c_{\varphi _\omega} }{ s_{2\beta} c_{\varphi 
_\omega - \varphi _y} } \deltaone v_s \deltaone |\kappa | \crn
+& \frac{ |\lambda | c_{2\beta } \left( 2v_s|\kappa | c_{\varphi _\omega 
} + \sqrt{2} \text{Re}\,A_\lambda \right) }{ 2 s_{\beta}^2 c_{\varphi 
_\omega - \varphi _y} } \deltaone v_s \deltaone |\tan \beta |
+ \frac{ v_s^2 c_{\varphi _\omega } }{ s_{2\beta} c_{\varphi _\omega - 
\varphi _y} } \deltaone |\kappa | \deltaone |\lambda | \crn
-& \frac{ 2 v_s^2 |\lambda | c_\beta^2 c_{2\beta} c_{\varphi _\omega } }{ 
s_{2\beta} c_{\varphi _\omega - \varphi _y} } \deltaone |\kappa | 
\deltaone \tan \beta + \frac{\abskappa \abslambda \cos (\phiom)}{s_{2\beta} \cos (\phiom-\phiy)}  (\dvsn )^2  \;. \nonumber
\end{align}

\end{appendix}


\begin{thebibliography}{200}

\bibitem{Golfand:1971iw}
{\relax Yu}.~A. Golfand and E.~P. Likhtman,
\newblock JETP Lett. {\bf 13}, 323 (1971),
\newblock [Pisma Zh. Eksp. Teor. Fiz.13,452(1971)].

\bibitem{Volkov:1973ix}
D.~Volkov and V.~Akulov,
\newblock Phys.Lett. {\bf B46}, 109 (1973).

\bibitem{Wess:1974tw}
J.~Wess and B.~Zumino,
\newblock Nucl.Phys. {\bf B70}, 39 (1974).

\bibitem{Fayet:1974pd}
P.~Fayet,
\newblock Nucl. Phys. {\bf B90}, 104 (1975).

\bibitem{Fayet:1977yc}
P.~Fayet,
\newblock Phys. Lett. {\bf 69B}, 489 (1977).

\bibitem{Fayet:1976cr}
P.~Fayet and S.~Ferrara,
\newblock Phys. Rept. {\bf 32}, 249 (1977).

\bibitem{Nilles:1982dy}
H.~P. Nilles, M.~Srednicki, and D.~Wyler,
\newblock Phys.Lett. {\bf B120}, 346 (1983).

\bibitem{Nilles:1983ge}
H.~P. Nilles,
\newblock Phys.Rept. {\bf 110}, 1 (1984).

\bibitem{Frere:1983ag}
J.~Frere, D.~Jones, and S.~Raby,
\newblock Nucl.Phys. {\bf B222}, 11 (1983).

\bibitem{Derendinger:1983bz}
J.~Derendinger and C.~A. Savoy,
\newblock Nucl.Phys. {\bf B237}, 307 (1984).

\bibitem{Haber:1984rc}
H.~E. Haber and G.~L. Kane,
\newblock Phys.Rept. {\bf 117}, 75 (1985).

\bibitem{Sohnius:1985qm}
M.~Sohnius,
\newblock Phys.Rept. {\bf 128}, 39 (1985).

\bibitem{Gunion:1984yn}
J.~Gunion and H.~E. Haber,
\newblock Nucl.Phys. {\bf B272}, 1 (1986).

\bibitem{Gunion:1986nh}
J.~Gunion and H.~E. Haber,
\newblock Nucl.Phys. {\bf B278}, 449 (1986).

\bibitem{Gunion:1989we}
J.~F. Gunion, H.~E. Haber, G.~L. Kane, and S.~Dawson,
\newblock Front.Phys. {\bf 80}, 1 (2000).

\bibitem{Martin:1997ns}
S.~P. Martin,
\newblock Adv.Ser.Direct.High Energy Phys. {\bf 21}, 1 (2010), hep-ph/9709356.

\bibitem{Dawson:1997tz}
S.~Dawson,
\newblock p. 261 (1997), hep-ph/9712464.

\bibitem{Djouadi:2005gj}
A.~Djouadi,
\newblock Phys.Rept. {\bf 459}, 1 (2008), hep-ph/0503173.

\bibitem{Barbieri:1982eh}
R.~Barbieri, S.~Ferrara, and C.~A. Savoy,
\newblock Phys.Lett. {\bf B119}, 343 (1982).

\bibitem{Dine:1981rt}
M.~Dine, W.~Fischler, and M.~Srednicki,
\newblock Phys.Lett. {\bf B104}, 199 (1981).

\bibitem{Ellis:1988er}
J.~R. Ellis, J.~Gunion, H.~E. Haber, L.~Roszkowski, and F.~Zwirner,
\newblock Phys.Rev. {\bf D39}, 844 (1989).

\bibitem{Drees:1988fc}
M.~Drees,
\newblock Int.J.Mod.Phys. {\bf A4}, 3635 (1989).

\bibitem{Ellwanger:1993xa}
U.~Ellwanger, M.~Rausch~de Traubenberg, and C.~A. Savoy,
\newblock Phys.Lett. {\bf B315}, 331 (1993), hep-ph/9307322.

\bibitem{Ellwanger:1995ru}
U.~Ellwanger, M.~Rausch~de Traubenberg, and C.~A. Savoy,
\newblock Z.Phys. {\bf C67}, 665 (1995), hep-ph/9502206.

\bibitem{Ellwanger:1996gw}
U.~Ellwanger, M.~Rausch~de Traubenberg, and C.~A. Savoy,
\newblock Nucl.Phys. {\bf B492}, 21 (1997), hep-ph/9611251.

\bibitem{Elliott:1994ht}
T.~Elliott, S.~King, and P.~White,
\newblock Phys.Lett. {\bf B351}, 213 (1995), hep-ph/9406303.

\bibitem{King:1995vk}
S.~King and P.~White,
\newblock Phys.Rev. {\bf D52}, 4183 (1995), hep-ph/9505326.

\bibitem{Franke:1995tc}
F.~Franke and H.~Fraas,
\newblock Int.J.Mod.Phys. {\bf A12}, 479 (1997), hep-ph/9512366.

\bibitem{Maniatis:2009re}
M.~Maniatis,
\newblock Int. J. Mod. Phys. {\bf A25}, 3505 (2010), 0906.0777.

\bibitem{Ellwanger:2009dp}
U.~Ellwanger, C.~Hugonie, and A.~M. Teixeira,
\newblock Phys. Rept. {\bf 496}, 1 (2010), 0910.1785.

\bibitem{deFlorian:2016spz}
LHC Higgs Cross Section Working Group, D.~de~Florian {\it et~al.},
\newblock (2016), 1610.07922.

\bibitem{Degrassi:2012ry}
G.~Degrassi {\it et~al.},
\newblock JHEP {\bf 08}, 098 (2012), 1205.6497.

\bibitem{Buttazzo:2013uya}
D.~Buttazzo {\it et~al.},
\newblock JHEP {\bf 12}, 089 (2013), 1307.3536.

\bibitem{Bednyakov:2015sca}
A.~V. Bednyakov, B.~A. Kniehl, A.~F. Pikelner, and O.~L. Veretin,
\newblock Phys. Rev. Lett. {\bf 115}, 201802 (2015), 1507.08833.

\bibitem{Aad:2015zhl}
ATLAS, CMS, G.~Aad {\it et~al.},
\newblock Phys. Rev. Lett. {\bf 114}, 191803 (2015), 1503.07589.

\bibitem{Muhlleitner:2017dkd}
M.~Muhlleitner, M.~O.~P. Sampaio, R.~Santos, and J.~Wittbrodt,
\newblock JHEP {\bf 08}, 132 (2017), 1703.07750.

\bibitem{Ellwanger:1993hn}
U.~Ellwanger,
\newblock Phys.Lett. {\bf B303}, 271 (1993), hep-ph/9302224.

\bibitem{Elliott:1993ex}
T.~Elliott, S.~King, and P.~White,
\newblock Phys.Lett. {\bf B305}, 71 (1993), hep-ph/9302202.

\bibitem{Elliott:1993uc}
T.~Elliott, S.~King, and P.~White,
\newblock Phys.Lett. {\bf B314}, 56 (1993), hep-ph/9305282.

\bibitem{Elliott:1993bs}
T.~Elliott, S.~King, and P.~White,
\newblock Phys.Rev. {\bf D49}, 2435 (1994), hep-ph/9308309.

\bibitem{Pandita:1993tg}
P.~Pandita,
\newblock Z.Phys. {\bf C59}, 575 (1993).

\bibitem{Ellwanger:2005fh}
U.~Ellwanger and C.~Hugonie,
\newblock Phys.Lett. {\bf B623}, 93 (2005), hep-ph/0504269.

\bibitem{Degrassi:2009yq}
G.~Degrassi and P.~Slavich,
\newblock Nucl.Phys. {\bf B825}, 119 (2010), 0907.4682.

\bibitem{Staub:2010ty}
F.~Staub, W.~Porod, and B.~Herrmann,
\newblock JHEP {\bf 1010}, 040 (2010), 1007.4049.

\bibitem{Ender:2011qh}
K.~Ender, T.~Graf, M.~Muhlleitner, and H.~Rzehak,
\newblock Phys.Rev. {\bf D85}, 075024 (2012), 1111.4952.

\bibitem{Drechsel:2016jdg}
P.~Drechsel, L.~Galeta, S.~Heinemeyer, and G.~Weiglein,
\newblock Eur. Phys. J. {\bf C77}, 42 (2017), 1601.08100.

\bibitem{Belanger:2016tqb}
G.~Belanger, V.~Bizouard, F.~Boudjema, and G.~Chalons,
\newblock Phys. Rev. {\bf D93}, 115031 (2016), 1602.05495.

\bibitem{Belanger:2017rgu}
G.~Bélanger, V.~Bizouard, F.~Boudjema, and G.~Chalons,
\newblock Phys. Rev. {\bf D96}, 015040 (2017), 1705.02209.

\bibitem{Goodsell:2014pla}
M.~D. Goodsell, K.~Nickel, and F.~Staub,
\newblock Phys. Rev. {\bf D91}, 035021 (2015), 1411.4665.

\bibitem{Ham:2001kf}
S.~Ham, J.~Kim, S.~Oh, and D.~Son,
\newblock Phys.Rev. {\bf D64}, 035007 (2001), hep-ph/0104144.

\bibitem{Ham:2001wt}
S.~Ham, S.~Oh, and D.~Son,
\newblock Phys.Rev. {\bf D65}, 075004 (2002), hep-ph/0110052.

\bibitem{Ham:2003jf}
S.~Ham, Y.~Jeong, and S.~Oh,
\newblock (2003), hep-ph/0308264.

\bibitem{Funakubo:2004ka}
K.~Funakubo and S.~Tao,
\newblock Prog.Theor.Phys. {\bf 113}, 821 (2005), hep-ph/0409294.

\bibitem{Ham:2007mt}
S.~Ham, S.~Kim, S.~Oh, and D.~Son,
\newblock Phys.Rev. {\bf D76}, 115013 (2007), 0708.2755.

\bibitem{Cheung:2010ba}
K.~Cheung, T.-J. Hou, J.~S. Lee, and E.~Senaha,
\newblock Phys.Rev. {\bf D82}, 075007 (2010), 1006.1458.

\bibitem{Graf:2012hh}
T.~Graf, R.~Grober, M.~Muhlleitner, H.~Rzehak, and K.~Walz,
\newblock JHEP {\bf 10}, 122 (2012), 1206.6806.

\bibitem{Domingo:2017rhb}
F.~Domingo, P.~Drechsel, and S.~Pa{\ss}ehr,
\newblock Eur. Phys. J. {\bf C77}, 562 (2017), 1706.00437.

\bibitem{Muhlleitner:2014vsa}
M.~Muhlleitner, D.~T. Nhung, H.~Rzehak, and K.~Walz,
\newblock JHEP {\bf 1505}, 128 (2015), 1412.0918.

\bibitem{Nhung:2013lpa}
D.~T. Nhung, M.~Muhlleitner, J.~Streicher, and K.~Walz,
\newblock JHEP {\bf 1311}, 181 (2013), 1306.3926.

\bibitem{Muhlleitner:2015dua}
M.~Muhlleitner, D.~T. Nhung, and H.~Ziesche,
\newblock JHEP {\bf 12}, 034 (2015), 1506.03321.

\bibitem{Goodsell:2016udb}
M.~D. Goodsell and F.~Staub,
\newblock Eur. Phys. J. {\bf C77}, 46 (2017), 1604.05335.

\bibitem{Ellwanger:2004xm}
U.~Ellwanger, J.~F. Gunion, and C.~Hugonie,
\newblock JHEP {\bf 0502}, 066 (2005), hep-ph/0406215.

\bibitem{Ellwanger:2005dv}
U.~Ellwanger and C.~Hugonie,
\newblock Comput.Phys.Commun. {\bf 175}, 290 (2006), hep-ph/0508022.

\bibitem{Ellwanger:2006rn}
U.~Ellwanger and C.~Hugonie,
\newblock Comput.Phys.Commun. {\bf 177}, 399 (2007), hep-ph/0612134.

\bibitem{Allanach:2001kg}
B.~Allanach,
\newblock Comput.Phys.Commun. {\bf 143}, 305 (2002), hep-ph/0104145.

\bibitem{Allanach:2013kza}
B.~Allanach, P.~Athron, L.~C. Tunstall, A.~Voigt, and A.~Williams,
\newblock Comput.Phys.Commun. {\bf 185}, 2322 (2014), 1311.7659.

\bibitem{Domingo:2015qaa}
F.~Domingo,
\newblock JHEP {\bf 06}, 052 (2015), 1503.07087.

\bibitem{Domingo:2018uim}
F.~Domingo, S.~Heinemeyer, S.~Pa{\ss}ehr, and G.~Weiglein,
\newblock Eur. Phys. J. {\bf C78}, 942 (2018), 1807.06322.

\bibitem{Staub:2010jh}
F.~Staub,
\newblock Comput.Phys.Commun. {\bf 182}, 808 (2011), 1002.0840.

\bibitem{Staub:2012pb}
F.~Staub,
\newblock Computer Physics Communications {\bf 184}, pp. 1792 (2013),
  1207.0906.

\bibitem{Staub:2013tta}
F.~Staub,
\newblock Comput.Phys.Commun. {\bf 185}, 1773 (2014), 1309.7223.

\bibitem{Goodsell:2014bna}
M.~D. Goodsell, K.~Nickel, and F.~Staub,
\newblock (2014), 1411.0675.

\bibitem{Porod:2003um}
W.~Porod,
\newblock Comput.Phys.Commun. {\bf 153}, 275 (2003), hep-ph/0301101.

\bibitem{Porod:2011nf}
W.~Porod and F.~Staub,
\newblock Comput.Phys.Commun. {\bf 183}, 2458 (2012), 1104.1573.

\bibitem{Athron:2014yba}
P.~Athron, J.-h. Park, D.~Stöckinger, and A.~Voigt,
\newblock Comput. Phys. Commun. {\bf 190}, 139 (2015), 1406.2319.

\bibitem{Athron:2014wta}
P.~Athron, J.-h. Park, D.~Stöckinger, and A.~Voigt,
\newblock Nucl. Part. Phys. Proc. {\bf 273-275}, 2424 (2016), 1410.7385.

\bibitem{Athron:2016fuq}
P.~Athron, J.-h. Park, T.~Steudtner, D.~St{\"o}ckinger, and A.~Voigt,
\newblock JHEP {\bf 01}, 079 (2017), 1609.00371.

\bibitem{Baglio:2013iia}
J.~Baglio {\it et~al.},
\newblock Comput.Phys.Commun. {\bf 185}, 3372 (2014), 1312.4788.

\bibitem{Boudjema:2017ozm}
F.~Boudjema,
\newblock J. Phys. Conf. Ser. {\bf 920}, 012011 (2017).

\bibitem{Goodsell:2017pdq}
M.~D. Goodsell, S.~Liebler, and F.~Staub,
\newblock (2017), 1703.09237.

\bibitem{Staub:2015aea}
F.~Staub {\it et~al.},
\newblock Comput. Phys. Commun. {\bf 202}, 113 (2016), 1507.05093.

\bibitem{Drechsel:2016htw}
P.~Drechsel {\it et~al.},
\newblock Eur. Phys. J. {\bf C77}, 366 (2017), 1612.07681.

\bibitem{Kobayashi:1973fv}
M.~Kobayashi and T.~Maskawa,
\newblock Prog. Theor. Phys. {\bf 49}, 652 (1973).

\bibitem{Skands:2003cj}
P.~Z. Skands {\it et~al.},
\newblock JHEP {\bf 07}, 036 (2004), hep-ph/0311123.

\bibitem{Allanach:2008qq}
B.~C. Allanach {\it et~al.},
\newblock Comput. Phys. Commun. {\bf 180}, 8 (2009), 0801.0045.

\bibitem{Frank20033672003}
M.~Frank,
\newblock {\it Strahlungskorrekturen im Higgs-Sektor des Minimalen
  Supersymmetrischen Standardmodells mit CP-Verletzung},
\newblock PhD thesis, Universit{\"a}t (TH) Karlsruhe, 2003,
\newblock Berlin 2003. Fak. f. Physik, Diss. v. 6.12.2002.

\bibitem{SIEGEL1979193}
W.~Siegel,
\newblock Physics Letters B {\bf 84}, 193  (1979).

\bibitem{1126-6708-2005-03-076}
D.~St{\"o}ckinger,
\newblock Journal of High Energy Physics {\bf 2005}, 076 (2005).

\bibitem{HOLLIK200663}
W.~Hollik and D.~St{\"o}ckinger,
\newblock Physics Letters B {\bf 634}, 63  (2006).

\bibitem{MARTIN2006133}
S.~P. Martin and D.~G. Robertson,
\newblock Computer Physics Communications {\bf 174}, 133  (2006).

\bibitem{THOOFT1979365}
G.~'t~Hooft and M.~Veltman,
\newblock Nuclear Physics B {\bf 153}, 365  (1979).

\bibitem{Nierste1993}
U.~Nierste, D.~M{\"u}ller, and M.~B{\"o}hm,
\newblock Zeitschrift f{\"u}r Physik C Particles and Fields {\bf 57}, 605
  (1993).

\bibitem{Sperling:2013eva}
M.~Sperling, D.~St{\"o}ckinger, and A.~Voigt,
\newblock JHEP {\bf 07}, 132 (2013), 1305.1548.

\bibitem{Sperling:2013xqa}
M.~Sperling, D.~St{\"o}ckinger, and A.~Voigt,
\newblock JHEP {\bf 01}, 068 (2014), 1310.7629.

\bibitem{Brignole:1992uf}
A.~Brignole,
\newblock Phys. Lett. {\bf B281}, 284 (1992).

\bibitem{Chankowski:1992ej}
P.~H. Chankowski, S.~Pokorski, and J.~Rosiek,
\newblock Phys. Lett. {\bf B286}, 307 (1992).

\bibitem{Chankowski:1992er}
P.~H. Chankowski, S.~Pokorski, and J.~Rosiek,
\newblock Nucl. Phys. {\bf B423}, 437 (1994), hep-ph/9303309.

\bibitem{Dabelstein:1994hb}
A.~Dabelstein,
\newblock Z. Phys. {\bf C67}, 495 (1995), hep-ph/9409375.

\bibitem{Dabelstein:1995js}
A.~Dabelstein,
\newblock Nucl. Phys. {\bf B456}, 25 (1995), hep-ph/9503443.

\bibitem{Freitas:2002um}
A.~Freitas and D.~Stockinger,
\newblock Phys. Rev. {\bf D66}, 095014 (2002), hep-ph/0205281.

\bibitem{Degrassi:2014pfa}
G.~Degrassi, S.~Di~Vita, and P.~Slavich,
\newblock Eur. Phys. J. {\bf C75}, 61 (2015), 1410.3432.

\bibitem{Borowka:2015ura}
S.~Borowka, T.~Hahn, S.~Heinemeyer, G.~Heinrich, and W.~Hollik,
\newblock Eur. Phys. J. {\bf C75}, 424 (2015), 1505.03133.

\bibitem{Heinemeyer:2007aq}
S.~Heinemeyer, W.~Hollik, H.~Rzehak, and G.~Weiglein,
\newblock Phys. Lett. {\bf B652}, 300 (2007), 0705.0746.

\bibitem{Heinemeyer:2010mm}
S.~Heinemeyer, H.~Rzehak, and C.~Schappacher,
\newblock Phys. Rev. {\bf D82}, 075010 (2010), 1007.0689.

\bibitem{Staub:2009bi}
F.~Staub,
\newblock Comput.Phys.Commun. {\bf 181}, 1077 (2010), 0909.2863.

\bibitem{Kublbeck:1990xc}
J.~Kublbeck, M.~Bohm, and A.~Denner,
\newblock Comput.Phys.Commun. {\bf 60}, 165 (1990).

\bibitem{Hahn:2000kx}
T.~Hahn,
\newblock Comput.Phys.Commun. {\bf 140}, 418 (2001), hep-ph/0012260.

\bibitem{FeynCalc}
R.~Mertig, M.~B{\"o}hm, and A.~Denner,
\newblock Computer Physics Communications {\bf 64}, 345  (1991).

\bibitem{SHTABOVENKO2016432}
V.~Shtabovenko, R.~Mertig, and F.~Orellana,
\newblock Computer Physics Communications {\bf 207}, 432  (2016).

\bibitem{MERTIG1998265}
R.~Mertig and R.~Scharf,
\newblock Computer Physics Communications {\bf 111}, 265  (1998).

\bibitem{CYROL2017346}
A.~K. Cyrol, M.~Mitter, and N.~Strodthoff,
\newblock Computer Physics Communications {\bf 219}, 346  (2017).

\bibitem{LARIN1993113}
S.~Larin,
\newblock Physics Letters B {\bf 303}, 113  (1993).

\bibitem{Jegerlehner2001}
F.~Jegerlehner,
\newblock The European Physical Journal C - Particles and Fields {\bf 18}, 673
  (2001).

\bibitem{Hollik:2014bua}
W.~Hollik and S.~Pa{\ss}ehr,
\newblock JHEP {\bf 10}, 171 (2014), 1409.1687.

\bibitem{Hollik:2015ema}
W.~Hollik and S.~Pa{\ss}ehr,
\newblock Eur. Phys. J. {\bf C75}, 336 (2015), 1502.02394.

\bibitem{Heinemeyer:1998yj}
S.~Heinemeyer, W.~Hollik, and G.~Weiglein,
\newblock Comput. Phys. Commun. {\bf 124}, 76 (2000), hep-ph/9812320.

\bibitem{Heinemeyer:1998np}
S.~Heinemeyer, W.~Hollik, and G.~Weiglein,
\newblock Eur. Phys. J. {\bf C9}, 343 (1999), hep-ph/9812472.

\bibitem{Degrassi:2002fi}
G.~Degrassi, S.~Heinemeyer, W.~Hollik, P.~Slavich, and G.~Weiglein,
\newblock Eur. Phys. J. {\bf C28}, 133 (2003), hep-ph/0212020.

\bibitem{Frank:2006yh}
M.~Frank {\it et~al.},
\newblock JHEP {\bf 02}, 047 (2007), hep-ph/0611326.

\bibitem{Hahn:2013ria}
T.~Hahn, S.~Heinemeyer, W.~Hollik, H.~Rzehak, and G.~Weiglein,
\newblock Phys. Rev. Lett. {\bf 112}, 141801 (2014), 1312.4937.

\bibitem{Bahl:2016brp}
H.~Bahl and W.~Hollik,
\newblock Eur. Phys. J. {\bf C76}, 499 (2016), 1608.01880.

\bibitem{Bahl:2017aev}
H.~Bahl, S.~Heinemeyer, W.~Hollik, and G.~Weiglein,
\newblock Eur. Phys. J. {\bf C78}, 57 (2018), 1706.00346.

\bibitem{Brignole:2001jy}
A.~Brignole, G.~Degrassi, P.~Slavich, and F.~Zwirner,
\newblock Nucl. Phys. {\bf B631}, 195 (2002), hep-ph/0112177.

\bibitem{Costa:2015llh}
R.~Costa, M.~Muhlleitner, M.~O.~P. Sampaio, and R.~Santos,
\newblock JHEP {\bf 06}, 034 (2016), 1512.05355.

\bibitem{King:2014xwa}
S.~F. King, M.~M{\"u}hlleitner, R.~Nevzorov, and K.~Walz,
\newblock Phys. Rev. {\bf D90}, 095014 (2014), 1408.1120.

\bibitem{Azevedo:2018llq}
D.~Azevedo, P.~Ferreira, M.~Margarete~Mühlleitner, R.~Santos, and
  J.~Wittbrodt,
\newblock (2018), 1808.00755.

\bibitem{PhysRevD.98.030001}
Particle Data Group, M.~Tanabashi {\it et~al.},
\newblock Phys. Rev. D {\bf 98}, 030001 (2018).

\bibitem{Dennerlhcnote}
A.~Denner {\it et~al.},
\newblock (2015), LHCHXSWG-INT-2015-006.

\bibitem{King:2015oxa}
S.~F. King, M.~Muhlleitner, R.~Nevzorov, and K.~Walz,
\newblock Nucl. Phys. {\bf B901}, 526 (2015), 1508.03255.

\bibitem{Bechtle:2008jh}
P.~Bechtle, O.~Brein, S.~Heinemeyer, G.~Weiglein, and K.~E. Williams,
\newblock Comput. Phys. Commun. {\bf 181}, 138 (2010), 0811.4169.

\bibitem{Bechtle:2011sb}
P.~Bechtle, O.~Brein, S.~Heinemeyer, G.~Weiglein, and K.~E. Williams,
\newblock Comput. Phys. Commun. {\bf 182}, 2605 (2011), 1102.1898.

\bibitem{Bechtle:2013wla}
P.~Bechtle {\it et~al.},
\newblock Eur. Phys. J. {\bf C74}, 2693 (2014), 1311.0055.

\bibitem{Bechtle:2013xfa}
P.~Bechtle, S.~Heinemeyer, O.~St{\"o}l, T.~Stefaniak, and G.~Weiglein,
\newblock Eur. Phys. J. {\bf C74}, 2711 (2014), 1305.1933.

\bibitem{Aad:2015iea}
ATLAS, G.~Aad {\it et~al.},
\newblock JHEP {\bf 10}, 054 (2015), 1507.05525.

\bibitem{Aaboud:2016lwz}
ATLAS, M.~Aaboud {\it et~al.},
\newblock Phys. Rev. {\bf D94}, 052009 (2016), 1606.03903.

\bibitem{Inoue:2014nva}
S.~Inoue, M.~J. Ramsey-Musolf, and Y.~Zhang,
\newblock Phys. Rev. {\bf D89}, 115023 (2014), 1403.4257.

\bibitem{Andreev:2018ayy}
ACME, V.~Andreev {\it et~al.},
\newblock Nature {\bf 562}, 355 (2018).

\bibitem{Allanach:2014nba}
B.~C. Allanach, A.~Bednyakov, and R.~Ruiz~de Austri,
\newblock Comput. Phys. Commun. {\bf 189}, 192 (2015), 1407.6130.

\bibitem{SCHARF1994523}
R.~Scharf and J.~Tausk,
\newblock Nuclear Physics B {\bf 412}, 523  (1994).

\bibitem{Chetyrkin:1999qi}
K.~G. Chetyrkin and M.~Steinhauser,
\newblock Nucl. Phys. {\bf B573}, 617 (2000), hep-ph/9911434.

\bibitem{Melnikov:2000qh}
K.~Melnikov and T.~v. Ritbergen,
\newblock Phys. Lett. {\bf B482}, 99 (2000), hep-ph/9912391.

\bibitem{Denner:1991kt}
A.~Denner,
\newblock Fortsch. Phys. {\bf 41}, 307 (1993), 0709.1075.

\bibitem{MACHACEK198383}
M.~E. Machacek and M.~T. Vaughn,
\newblock Nuclear Physics B {\bf 222}, 83  (1983).

\bibitem{MACHACEK1984221}
M.~E. Machacek and M.~T. Vaughn,
\newblock Nuclear Physics B {\bf 236}, 221  (1984).

\bibitem{MACHACEK198570}
M.~E. Machacek and M.~T. Vaughn,
\newblock Nuclear Physics B {\bf 249}, 70  (1985).

\bibitem{PhysRevLett.90.011601}
M.~Luo and Y.~Xiao,
\newblock Phys. Rev. Lett. {\bf 90}, 011601 (2003).

\bibitem{Runge1895}
C.~Runge,
\newblock Mathematische Annalen {\bf 46}, 167 (1895).

\bibitem{Kutta1901}
W.~Kutta,
\newblock Zeit. Math. Phys. {\bf 46}, 435 (1901).

\bibitem{Avdeev:1997sz}
L.~V. Avdeev and M.~{\relax Yu}. Kalmykov,
\newblock Nucl. Phys. {\bf B502}, 419 (1997), hep-ph/9701308.

\bibitem{Harlander:2006rj}
R.~Harlander, P.~Kant, L.~Mihaila, and M.~Steinhauser,
\newblock JHEP {\bf 09}, 053 (2006), hep-ph/0607240.

\bibitem{Harlander:2007wh}
R.~V. Harlander, L.~Mihaila, and M.~Steinhauser,
\newblock Phys. Rev. {\bf D76}, 055002 (2007), 0706.2953.

\bibitem{KING2012207}
S.~King, M.~Mühlleitner, and R.~Nevzorov,
\newblock Nuclear Physics B {\bf 860}, 207  (2012).

\bibitem{PhysRevD.11.1521}
A.~Salam and J.~Strathdee,
\newblock Phys. Rev. D {\bf 11}, 1521 (1975).

\bibitem{GRISARU1979429}
M.~Grisaru, W.~Siegel, and M.~Rocek,
\newblock Nuclear Physics B {\bf 159}, 429  (1979).

\end{thebibliography}


\end{document}